\documentclass[11pt,a4paper]{article}

\usepackage[colorlinks=true, linkcolor=black!50!blue, urlcolor=blue, citecolor=blue, anchorcolor=blue]{hyperref}
\usepackage[font=small,labelfont=bf,margin=0mm,labelsep=period,tableposition=top]{caption}
\usepackage[a4paper,top=1.5cm,bottom=1.5cm,left=2.5cm,right=2.5cm,bindingoffset=0mm]{geometry}

\usepackage[utf8]{inputenc}
\usepackage{graphicx}
\usepackage{epsfig}
\usepackage{cite}
\usepackage{multirow}
\usepackage{amssymb}
\usepackage{amsmath}
\usepackage{dsfont}
\usepackage{url}
\usepackage{xcolor}
\usepackage{afterpage}
\usepackage{url}
\usepackage{hyperref}
\usepackage{booktabs}
\usepackage{tabularx}
\usepackage{enumitem}
\usepackage{cite}

\newcommand{\be}{\begin{equation}}
\newcommand{\ee}{\end{equation}}
\newcommand{\bea}{\begin{eqnarray}}
\newcommand{\eea}{\end{eqnarray}}
\newcommand{\bi}{\begin{itemize}}
\newcommand{\ei}{\end{itemize}}
\newcommand{\ben}{\begin{enumerate}}
\newcommand{\een}{\end{enumerate}}

\def\gsim{\mathrel{\rlap{\lower4pt\hbox{\hskip1pt$\sim$}}
    \raise1pt\hbox{$>$}}}         
\def\lsim{\mathrel{\rlap{\lower4pt\hbox{\hskip1pt$\sim$}}
    \raise1pt\hbox{$<$}}}         
\def\beq{\begin{equation}}
\def\eeq{\end{equation}}
\def\lapprox{\lower .7ex\hbox{$\;\stackrel{\textstyle <}{\sim}\;$}}
\def\gapprox{\lower .7ex\hbox{$\;\stackrel{\textstyle >}{\sim}\;$}}
\newcommand\ptavg{p_{T,\rm avg}}

\numberwithin{equation}{section}
\numberwithin{figure}{section}
\numberwithin{table}{section}

\begin{document}

\begin{flushright}
CERN-TH-2020-073\\
IPPP/20/17\\
Nikhef/2020-001\\
TIF-UNIMI-2020-3\\
ZU-TH 15/20\\
\end{flushright}
\vspace{0.3cm}

\begin{center}
  {\Large \bf Phenomenology of NNLO jet production at the LHC\\[0.2cm] 
  and its impact on parton distributions}
\vspace{1.4cm}

Rabah Abdul Khalek,$^{1,2}$
Stefano Forte,$^{3}$
Thomas Gehrmann,$^{4}$
Aude Gehrmann-De Ridder,$^{4,5}$\\[0.1cm]
Tommaso Giani,$^{6}$
Nigel Glover,$^{7}$
Alexander Huss,$^{7,8}$
Emanuele R. Nocera,$^{2}$
Joao Pires,$^{9}$\\[0.1cm]
Juan Rojo,$^{1,2}$ and
Giovanni Stagnitto$^{3,10,11}$

\vspace{1.0cm}
 
{\it \small 
~$^1$ Department of Physics and Astronomy, VU Amsterdam, 
 1081 HV Amsterdam,\\[0.1cm]
~$^2$ Nikhef Theory Group, Science Park 105, 
 1098 XG Amsterdam, The Netherlands\\[0.1cm]
~$^3$ Tif Lab, Dipartimento di Fisica, Universit\`a di Milano and\\[0.1cm]
 INFN, Sezione di Milano, Via Celoria 16, I-20133 Milano, Italy\\[0.1cm]
~$^4$ Department of Physics, University of Zurich, 
 CH-8057 Zurich, Switzerland\\[0.1cm]
~$^5$ Institute for Theoretical Physics, ETH, 
 CH-8093 Zurich, Switzerland\\[0.1cm]
~$^6$ The Higgs Centre for Theoretical Physics, 
 University of Edinburgh,\\[0.1cm]
 JCMB, KB, Mayfield Rd, Edinburgh EH9 3JZ, Scotland\\[0.1cm]
~$^{7}$ Institute for Particle Physics Phenomenology, 
 Durham University, Durham DH1 3LE, UK\\[0.1cm]
~$^8$ Theoretical Physics Department, CERN, 
 CH-1211 Geneva 23, Switzerland\\[0.1cm]
~$^{9}$LIP, Avenida Professor Gama Pinto 2, P-1649-003 Lisboa, Portugal\\[0.1cm]
~$^{10}$ Sorbonne Universit\'e, CNRS, Laboratoire de Physique 
 Th\'orique et Hautes \'Energies, F-75005 Paris, France\\[0.1cm]
~$^{11}$ Universit\'e Paris Diderot, F-75013 Paris, France 
}

\vspace{1.0cm}

{\bf \large Abstract}

\end{center}
We present a systematic investigation of jet production at hadron
colliders from a phenomenological point of view, with the dual aim of
providing a validation of theoretical calculations and guidance to future
determinations of parton distributions (PDFs).
We account for all available inclusive jet and dijet production
measurements 
from ATLAS and CMS at 7 and 8 TeV by including them
in a global PDF determination, and comparing to
theoretical predictions at NNLO QCD supplemented by electroweak (EW) corrections.
We assess the compatibility of the PDFs,
specifically the gluon, obtained before and after inclusion of the
jet data.
We compare the single-inclusive jet and dijet
observables  in terms of perturbative behaviour upon inclusion of QCD
and EW corrections, impact
on the PDFs, and 
global fit quality.  In the single-inclusive case,  we also
investigate the
role played by different scale choices and the
stability of the results upon changes in  modelling
of the correlated experimental systematics.

\clearpage

\tableofcontents

\section{Introduction}
\label{sec:introduction}

The inclusive jet cross-section is the simplest hadron collider
observable with a purely strongly interacting final state.
The computation of next-to-next-to-leading order (NNLO) QCD corrections to it
was completed recently~\cite{Ridder:2013mf,Currie:2013dwa,Currie:2016bfm} 
(see also Ref.~\cite{Czakon:2019tmo}), and opens
up the possibility of doing precision phenomenology with jet
observables. Whereas single-inclusive
jets have been used for the determination of 
the parton distributions (PDFs) of the proton~\cite{Gao:2017yyd}
for over thirty years~\cite{Martin:1987vw}, there is a number of
unsettled
theoretical
issues related to the definition of the observable which is most
promising and appropriate for precision QCD studies, such as the
determination of the PDFs and of the strong coupling constant $\alpha_s$.

The simplest inclusive observable, the single-inclusive jet cross-section~\cite{Aversa:1988fv,Ellis:1988hv},
has the undesirable feature
of being non-unitary: each event is counted more than once, so 
the integral of the differential cross-section is not equal to the
total cross-section. The dijet cross-section is free of this issue and
it appears to be especially well-suited for PDF 
determination~\cite{Giele:1994xd}. However, for 
this observable  several 
scale choices are possible, because the more complex nature of the final state 
offers a wide choice of dimensionful kinematic variables;
consequently, the significant scale dependence of NLO results has so far 
effectively prevented the use of this observable for PDF determination.

The availability of  NNLO calculations has opened up the possibility of
settling these issues, though their full understanding has posed a
theoretical challenge, with the single-inclusive jet and dijet
observables presenting different features. On the one hand, the issue
of scale choice for the dijet observable has been essentially settled by
the NNLO computation, with the scale dependence being under control at
NNLO and the dijet invariant mass $m_{jj}$ emerging as the preferred choice.
On the other hand, the single-inclusive jet cross-section has shown a 
dependence on the choice of scale which is not significantly reduced from NLO 
to NNLO~\cite{Currie:2017ctp}, so that the understanding of the
perturbative behavior, the scale dependence~\cite{Currie:2018xkj},
and even the appropriate definition~\cite{Cacciari:2019qjx} of this 
observable are non-trivial. A careful analysis reveals that the apparent lack 
of improvement of scale stability from NLO to NNLO is due to an accidental 
NLO scale cancellation which occurs for particular values of the jet 
radius~\cite{Dasgupta:2016bnd,Cacciari:2019qjx}. The
persistence of a dependence on the central scale choice at
NNLO can in turn be understood as a consequence of infrared sensitivity, which
is aggravated by particular scale choices~\cite{Currie:2018xkj}.
It then appears that the non-unitary definition of the observable is in
fact necessary for perturbative stability, with dijets offering
essentially the only viable unitary stable alternative~\cite{Cacciari:2019qjx}.
From these studies the partonic transverse energy $\widehat{H}_T$ emerges as the 
optimal scale choice~\cite{Currie:2018xkj} for the calculation
of single-inclusive jet cross-sections.

In this work, we address these issues from a phenomenological point of
view, specifically within the context of a global PDF determination:
we study the effect of adding jet cross-sections to a global dataset, 
with various choices of the observable (single-inclusive jet, or dijet) 
and of the scale. In each case, we assess the fit quality and the impact
of the jet data on the PDFs, at various perturbative orders.
This allows us to achieve two main goals. First, we can test 
phenomenologically the conclusions of past theoretical
studies~\cite{Ridder:2013mf,Currie:2013dwa,Currie:2016bfm,Currie:2017ctp,
Currie:2018xkj,Cacciari:2019qjx}, by checking which observable and which scale 
choice leads to better perturbative stability, better PDF compatibility with 
other data and better fit quality, and more stringent constraints on the PDFs.
Secondly, these results make it possible to optimize the choice of 
jet observables in view of their inclusion in future global PDF fits, 
and assess their impact as a means of PDF determination.

We will consider the complete inclusive jet~\cite{Aad:2014vwa,Aaboud:2017dvo,
Chatrchyan:2012bja,Khachatryan:2016mlc}
and dijet\cite{Aad:2013tea,Chatrchyan:2012bja,Sirunyan:2017skj} dataset
from ATLAS and CMS at $\sqrt{s}=7$ and 8 TeV.
Whereas most recent global determinations of the proton
PDFs~\cite{Ball:2017nwa,Harland-Lang:2014zoa,Hou:2019efy} include 
some of these jet datasets (for instance, NNPDF3.1 included
the ATLAS and CMS single-inclusive data with  $\sqrt{s}=2.76$ and 7~TeV),
and other studies have assessed the impact of some jet measurements on smaller
datasets~\cite{Khachatryan:2014waa,Sirunyan:2017skj}, this is the
first time that the full LHC-Run I jet dataset is being considered, and
specifically the first time dijet data are included in a modern global PDF
determination.

Thanks to the availability of such a wide dataset, we will be able
to pursue the two main goals discussed above, by including these jet
data in the NNPDF3.1 dataset, while keeping the rest of the global
dataset and adopting the same general PDF fitting methodology.
In addition, we will look into  two further 
secondary issues. First, we will study the impact of the inclusion of
electroweak (EW) corrections to jet predictions.
Second, we will assess the sensitivity of results to the 
treatment of experimental correlated systematic uncertainties, thus
addressing the issue, recently raised  {\it e.g.} in
Ref.~\cite{Harland-Lang:2017ytb}, of the sensitivity of  
some LHC jet datasets to variations 
in the experimental correlation model, which may lead to  substantial 
differences in fit quality.

The outline of this paper is as follows.
In Sect.~\ref{sec:expdata} we discuss the experimental data
for single-inclusive jet and inclusive dijet production.
In Sect.~\ref{sec:theory} we present the theory, and in particular
discuss  NNLO QCD and  EW
corrections and scale choices.
Our results for the  global PDF analyses that we performed  are
presented in
Sect.~\ref{sec:results}, where we also discuss their implications and
summarize our findings. Future implications and avenues for further
research are briefly
addressed in Sect.~\ref{sec:summary}.

\section{Experimental data}
\label{sec:expdata}

We now discuss the single-inclusive jet and dijet data. We first
summarize available inclusive  jet production data from the LHC.
We then
review the jet cross-sections included in the NNPDF3.1 PDF determination and
their treatment. We
finally provide details on the treatment and kinematic coverage
of the single-inclusive jet and
dijet datasets 
that we will use in this paper.

\subsection{Jet production at the LHC}

The ATLAS and CMS collaborations have performed a number of measurements of
the   single-inclusive  and dijet cross-sections  at
different center of mass energies,
ranging from $\sqrt{s}=2.76$~TeV to 13~TeV.
In this work, we will focus on the 7 and 8 TeV data, for which
single-inclusive and dijet data corresponding to the same underlying
dataset and  integrated luminosity can be compared.

The $\sqrt{s}=7$ and
8 TeV data are summarized in Table~\ref{tab:input_datasets}, where for
each dataset we indicate the experiment, the measured quantity,
  the center of mass energy $\sqrt{s}$, the integrated luminosity $\mathcal{L}$,
  the number of datapoints $n_{\rm dat}$, and  the published
  reference. All measurements are performed using the anti-$k_t$
  algorithm~\cite{Cacciari:2008gp} in the four-momentum recombination scheme, which leads to
  jets with non-vanishing invariant mass.
  The relevant kinematic variables are defined as follows.
  For single-inclusive  jets, $p_T$ and $y$ are the jet transverse momentum
  and rapidity.
  For dijets, $m_{jj}$ is the dijet invariant mass,   $y^*=|y_1-y_2|/2$
  and $|y_{\rm max}|=\max(|y_1|,|y_2|)$ are respectively
  the absolute rapidity difference and maximum absolute rapidity
  of the two leading jets of the event.
 Finally, for dijet triple-differential
 distributions,  $\ptavg=(p_{T,1}+p_{T,2})/2$ is
 the average transverse momentum of the two leading jets,
  and $y_b=|y_1+y_2|/2$ is the boost of the dijet system.

\begin{table}[!t]
\centering
\scriptsize
\renewcommand{\arraystretch}{1.90}
\begin{tabularx}{\textwidth}{XXcccccc}
\toprule
  Experiment 
& Measurement   
& $\sqrt{s}$ [TeV]
& $\mathcal{L}$ [fb$^{-1}$] 
& $R$
& Distribution  
& $n_{\rm dat}$ 
& Reference   \\
\midrule
  ATLAS  
& Inclusive jets  
& 7 
& 4.5
& 0.6
& $d^2\sigma/dp_Td|y|$  
& 140  
& \cite{Aad:2014vwa}  \\
  CMS  
& Inclusive jets  
& 7
& 4.5
& 0.7
& $d^2\sigma/dp_Td|y|$  
& 133 
& \cite{Chatrchyan:2012bja}  \\
  ATLAS  
& Inclusive jets  
& 8
& 20.2
& 0.6
& $d^2\sigma/dp_Td|y|$  
& 171  
& \cite{Aaboud:2017dvo}  \\
  CMS  
& Inclusive jets  
& 8 
& 19.7
& 0.7
& $d^2\sigma/dp_Td|y|$  
& 185  
& \cite{Khachatryan:2016mlc}  \\
\midrule
  ATLAS  
& Dijets  
& 7 
& 4.5
& 0.6
& $d^2\sigma/dm_{jj}d|y^{*}|$  
& 90 
& \cite{Aad:2013tea}  \\
  CMS  
& Dijets  
& 7 
& 4.5
& 0.7
& $d^2\sigma/dm_{jj}d|y_{\rm max}|$  
& 54  
& \cite{Chatrchyan:2012bja}  \\
  CMS  
& Dijets  
& 8 
& 19.7
& 0.7
& $d^3\sigma/dp_{T,\rm avg}dy_b dy^{*}$  
& 122  
& \cite{Sirunyan:2017skj}  \\
\bottomrule
\end{tabularx}

\vspace{0.3cm}
\caption{\small The LHC single-inclusive jet and dijet cross-section data
   that will be used  in this study. For each dataset we indicate the experiment,
   the measurement, the center of mass energy $\sqrt{s}$, the luminosity 
   $\mathcal{L}$, the jet radius $R$, the measured distribution, the number of 
   datapoints $n_{\rm dat}$ and the reference.}
\label{tab:input_datasets}
\end{table}

In addition to the data listed in  Table~\ref{tab:input_datasets},
ATLAS and CMS have also performed
measurements at $\sqrt{s}=13$~TeV, though so far with smaller 
integrated luminosities than for their Run I counterparts:
at Run II, the single-inclusive jet measurements 
from ATLAS~\cite{Aaboud:2017wsi} and CMS~\cite{Khachatryan:2016wdh}
have $\mathcal{L}=3.2$ fb$^{-1}$ and
$\mathcal{L}=71$ pb$^{-1}$ respectively, while the dijet measurements from
ATLAS~\cite{Aaboud:2017wsi} and CMS~\cite{Sirunyan:2018xdh} have
$\mathcal{L}=3.2$ fb$^{-1}$ and $\mathcal{L}=2.3$ fb$^{-1}$.
For this reason, we do not include  these datasets.
Very recently, CMS has presented a single-inclusive jet measurement at
$\sqrt{s}=13$~TeV, based on a luminosity of $\mathcal{L}=35.9$
fb$^{-1}$~\cite{Sirunyan:2020uoj}.

We will also not include single-inclusive jet data
at $\sqrt{s}=2.76$~TeV~\cite{Aad:2013lpa,Khachatryan:2015luy}
and 5.02~TeV~\cite{Sirunyan:2018qel}. The main motivation for these
measurements was to provide a baseline for proton-lead
and lead-lead data taken at the same center of mass energy.
A possible exception could be  the 5.02 TeV CMS double-differential
cross-section data,  based on an integrated
luminosity of $\mathcal{L}=27.4$ pb$^{-1}$: indeed,
a recent study~\cite{Eskola:2019dui} claims that they might also
impact the proton PDFs.
We will  investigate this dataset in a follow-up study
based on an update of the nNNPDF1.0 analysis~\cite{AbdulKhalek:2019mzd} of
nuclear parton distribution functions.

In addition,
ATLAS and CMS have also presented several measurements of multijet
($\ge 3$ jets) production. 
For example, ATLAS has provided  measurements of three
  jet cross-sections at 7 TeV~\cite{Aad:2014rma}, differential
 in three-jet mass $m_{jjj}$ and the sum of the
 absolute rapidity separations between the three leading jets, $|y_*|$;
 and of four-jet cross-sections at  8 TeV~\cite{Aad:2015nda},
 differential in the $p_T$ of the four leading jets in the event.
 CMS also has a measurement of the
 3-jet production cross-section at 7 TeV~\cite{CMS:2014mna}
 differential in the invariant mass of the three jets $m_{jjj}$.
Because theoretical predictions are currently only available up 
to NLO for these observables, they will not be considered here, though
they are important for other applications such as the validation
of Monte Carlo event generators and searches for physics
beyond the Standard Model.
    
\subsection{Jet data in NNPDF3.1}
\label{sec:data31}

The present study will be based on the PDF fitting
framework adopted for the
NNPDF3.1 global
PDF determination~\cite{Ball:2017nwa}.
As already mentioned, the NNPDF3.1 dataset includes
several single-inclusive jet data.
Specifically, for ATLAS 
the $\sqrt{s}=7$ TeV data  from  2010~\cite{Aad:2011fc} and
2011~\cite{Aad:2014vwa} and the  $\sqrt{s}=2.76$
TeV~\cite{Aad:2013lpa} data (including cross-correlations
between the 2.76 TeV and the 7 TeV data).
For the 2011 7~TeV data
only the central rapidity bin ($y_{\rm jet}\le 0.5$)
was included, due to the difficulty in achieving
a satisfactory description of the complete set of rapidity bins using the
default experimental covariance matrix.
From the CMS experiment, NNPDF3.1 included the measurements
at $\sqrt{s}=7$~TeV~\cite{Chatrchyan:2012bja} and 
$2.76$ TeV~\cite{Khachatryan:2015luy}, with their cross-correlations.
Finally the CDF Run II data
with the $k_T$ algorithm~\cite{Abulencia:2007ez} was also included.
Note that the value of the jet radius $R$ is different for each
of these measurements:  $R=0.4$ for the ATLAS 7 TeV 2010 and the 
2.76~TeV measurements; $R=0.6$ for the ATLAS 7 TeV 2011 measurement;
and $R=0.7$ for the CDF and CMS measurements.

In the default NNPDF3.1 PDF determination, theory predictions for all
these data were obtained by combining NLO coefficient functions with
NNLO perturbative evolution, because full NNLO results were not
available then. In order to account for the 
 missing NNLO corrections, a missing higher order uncertainty,
 estimated from scale variations,  was
 added to jet data, as a fully correlated systematics.
A variant PDF set was also produced by 
only including the two datasets for which the NNLO corrections were
available at the time, namely  the ATLAS and CMS 7 TeV 2011 data, with all the
remaining jet data removed, and now using full NNLO theory. 
This reduced, but fully NNLO, dataset was also used for the
determination of the strong coupling in Ref.~\cite{Ball:2018iqk}, for
the  PDFs with QED corrections~\cite{Bertone:2017bme} and the PDFs
with small-$x$ resummation~\cite{Ball:2017otu}, and for the recent studies 
of theoretical uncertainties on
PDFs~\cite{AbdulKhalek:2019ihb,AbdulKhalek:2019bux}. 
In all these previous studies, the renormalization and factorization
scales were set equal to the jet transverse momentum,
$\mu_{F}=\mu_{R}=p_{T,\rm jet}$.

\subsection{Jet data in this analysis}
\label{sec:datahere}

The single-inclusive jet data from ATLAS and CMS  used
in this work are the double-differential ($y$, $p_T$) distributions
listed in Table~\ref{tab:input_datasets}.
The ATLAS 7~TeV data cover
the  range  100~GeV~$\le p_T\le$~1.992~TeV
and $0\le |y|\le 3$, while the ATLAS 8~TeV data cover the
same rapidity range, but with an extended range of transverse momenta, namely
70~GeV~$\le p_T\le$~2.5~TeV. In our default fit we include only the central
rapidity bin ($y_{\rm jet}\le 0.5$) of the ATLAS 7~TeV, for ease of
comparison with NNPDF3.1. This is
not expected to affect results, as in Ref.~\cite{Ball:2017nwa} it was
shown that PDFs fitted to the central rapidity bin provide an equally
good fit to all other rapidity bins, and in
Ref.~\cite{Nocera:2017zge}  it was checked explicitly that PDFs
determined including each rapidity bin from this data in turn
are indistinguishable. We  will revisit this issue in
Sect.~\ref{sec:corr}, where we will discuss
variant fits in which all rapidity bins are included, and we will
consider alternative correlation models both for these data and for
their 8~TeV counterpart, as suggested in
Refs.~\cite{Harland-Lang:2017ytb,Aaboud:2017dvo}. 

The CMS 7 TeV data cover
the  range  100~GeV~$\le p_T\le$~2.0~TeV
and  $0\le |y|\le 2.5$, and 
the CMS 8 TeV data the extended range  74~GeV~$\le p_T\le$~2.5~TeV  and
$0\le |y|\le 3.0$.
We note that in the case of the CMS 8 TeV single-inclusive jets,
measurements for $p_T < 74$~GeV are also available, but these
are excluded from the fit because non-perturbative and resummation corrections,
not accounted for by fixed-order computations, are large at small $p_T$.
We therefore retain only 185 points out of a total of 239. 

For the dijet cross-sections
we consider three  Run I measurements from ATLAS and CMS, specifically
the ATLAS and CMS 7~TeV~\cite{Aad:2013tea,Chatrchyan:2012bja} double-differential distributions
and the CMS 8~TeV triple-differential distributions~\cite{Sirunyan:2017skj}.
Note that currently ATLAS dijet measurements are only available
 at 7 and 13~TeV, but not at 8~TeV.
The ATLAS 
data are  double-differential in  $m_{jj}$ and $|y^{*}|$.
The corresponding ranges are 260~GeV~$\le m_{jj}\le$~4.27~TeV
and $0\le y^*\le 3.0$.
The CMS 7 TeV data~\cite{Chatrchyan:2012bja} are instead
double-differential in  $m_{jj}$ and $|y_{\rm max}|$. 
The ranges are  $200~{\rm GeV} \le m_{jj} \le 5$ TeV
  and $0 \le |y|_{\rm max} \le 2.5$.
The CMS  8~TeV~\cite{Sirunyan:2017skj} data are  triple
  differential in
  $\ptavg$, $y_b$, and $|y^{*}|$.
  The ranges are 133~GeV~$\le \ptavg \le$~1.78~TeV and $0 \le
  y_b, y^* \le 3$. 

For all these measurements, we will use the complete set of systematic 
uncertainties and correlations available from {\sc HepData}. 
Various correlation models, whereby specific systematic uncertainties are  
decorrelated to a different extent, have been proposed, depending on the dataset.
As a representative example, we will study some of these models in the 
case of the ATLAS 7~TeV and 8~TeV single-inclusive jet cross-sections.

\section{Theoretical calculations and implementation}
\label{sec:theory}

In this section we present the main aspects of the 
theoretical computations on which our phenomenological studies are
based. First we address QCD corrections, discuss the scale
choice, and assess the size of NNLO corrections. Then we discuss
EW corrections, assess their size, and explain how they are combined 
with QCD corrections for the purpose of PDF determination.

\subsection{QCD corrections}
\label{sec:QCD}

Single-inclusive and dijet observables display a somewhat different 
perturbative behavior. We discuss the two observables in turn: for each 
of them, we present the dependence of results on the central scale,  
its optimal choice, and the NNLO corrections.

The single-inclusive jet cross-section is in general rather sensitive
to the choice of central scale, even at NNLO. A
detailed study of the scale dependence of the NNLO QCD predictions
for single-inclusive jet production was carried out in~\cite{Currie:2018xkj},
where three different scales (and their multiples) were discussed in
detail: the individual jet transverse momentum $p_T$, the leading 
jet transverse momentum $p_{T,1}$, and the scalar sum of the 
transverse momenta of all partons in the event
\be
\widehat{H}_T = \sum_{i\in {\rm partons}} p_{T,i} \, .
\ee
Note that $p_{T,1}$ and $\widehat{H}_T$ are event-based choices,  i.e.
all jets in the event have the same scale, while $p_T$ is a jet-based choice, 
 i.e. it is a property of the individual jet within a given event.

The commonly used scale choices $\mu=p_T$ or $\mu=p_{T,1}$ lead to
predictions which even at NNLO may differ by an amount which is
comparable to, or larger than, their scale dependence~\cite{Currie:2017ctp}, 
a behavior which was traced in Ref.~\cite{Currie:2018xkj} to the infrared 
sensitivity of the second-jet contribution, and which is aggravated by the 
choice $\mu=p_{T,1}$. In Ref.~\cite{Currie:2018xkj} scale
choices were thus compared according to a number of criteria:
perturbative convergence; scale uncertainty as error estimate; perturbative 
convergence of the individual jet spectra; and stability of the second jet 
distribution. The event-based scale 
$\mu=\widehat{H}_T$  and the jet-based scale $\mu=2p_T$ were singled
out as optimal choices. Here we will adopt  $\mu=\widehat{H}_T$ as central 
scale choice; results obtained with this scale choice will be compared in
Sect.~\ref{sec:results} to those found using  $\mu=p_T$, which was the baseline 
choice adopted in previous NNPDF determinations, specifically NNPDF3.1.

NNLO QCD corrections computed with NNLOJET~\cite{Gehrmann-DeRidder:2019ibf}
will be included by supplementing theoretical predictions
accurate to NLO QCD with $K$-factors defined as
\be
\label{eq:kfactor}
K_{\rm NNLO}^{\rm QCD}
\equiv 
\frac{\sum_{ij}\widetilde{\sigma}^{\rm NNLO}_{ij} \otimes \mathcal{L}^{\rm NNLO}_{ij}}{\sum_{ij}
\widetilde{\sigma}^{\rm NLO}_{ij}\otimes \mathcal{L}^{\rm NNLO}_{ij}} \, ,
\ee
where the sum runs over partonic subchannels,
$\widetilde{\sigma}_{ij}$ are  partonic cross-sections, and 
$\mathcal{L}_{ij}$ the corresponding parton luminosities, computed
both in the
numerator and the denominator using NNPDF3.1 NNLO as a fixed input PDF
set.  

\begin{figure}[!t]
\centering
\includegraphics[scale=0.47]{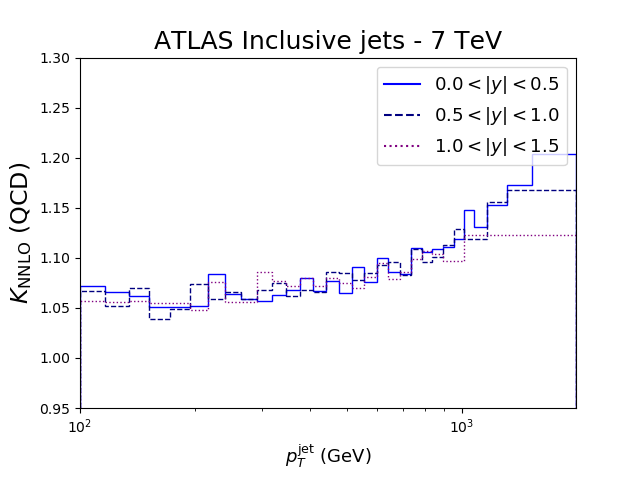}
\includegraphics[scale=0.47]{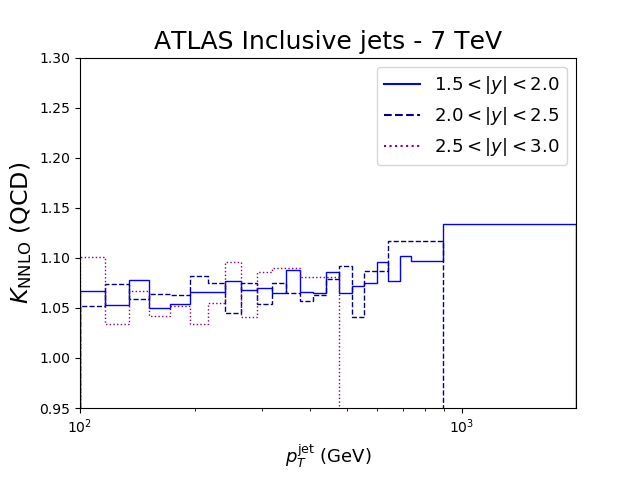}\\
\includegraphics[scale=0.47]{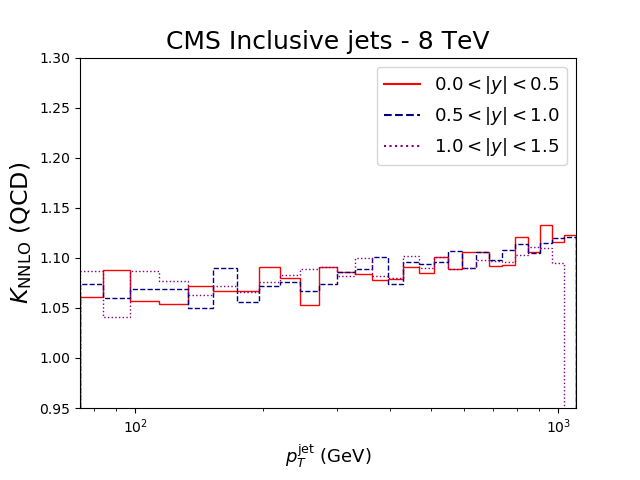}
\includegraphics[scale=0.47]{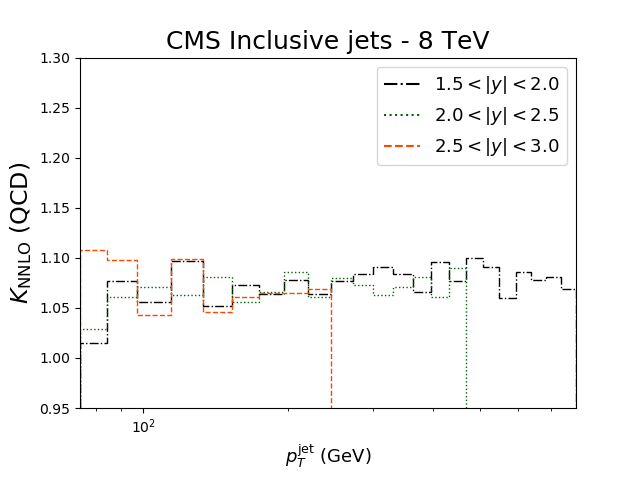}\\
\caption{The NNLO QCD $K$-factors, Eq.~(\ref{eq:kfactor}),
   for  the ATLAS 7 TeV (top) and CMS 8 TeV (bottom) single-inclusive
   jet cross-sections evaluated using NNPDF3.1 PDFs and scale 
   $\mu=\widehat{H}_T$. Results are shown as a function of the jet $p_T$ in 
   different jet rapidity bins, with the central (forward) bins shown in the 
   left (right) plot.}
\label{fig:kfactqcd_incljets_atlas7_HTp_R06}
\end{figure}

In Fig.~\ref{fig:kfactqcd_incljets_atlas7_HTp_R06} we show the NNLO QCD 
$K$-factors, Eq.~(\ref{eq:kfactor}), corresponding to the ATLAS 7 TeV  
and CMS 8 TeV single-inclusive jet cross-sections evaluated with the NNPDF3.1 NNLO
PDF set and $\mu=\widehat{H}_T$ as central scale. Results are shown as a 
function of the jet $p_T$ in different jet rapidity bins, with the central 
(forward) bins in the left (right) plot. At central rapidities, the NNLO 
$K$-factor increases monotonically with $p_T$ from about 5\% to about 20--25\%.
This growth with $p_T$ becomes less marked as the jet rapidity increases: 
in fact at 8~TeV for $|y|\ge 1.5$ the $K$-factor depends only mildly on the 
jet $p_T$. The $K$-factors display moderate point-to-point
fluctuations, especially 
in the forward rapidity bins.

We now turn to dijets. A variety of scale choices is possible: 
two popular choices are the dijet invariant mass $m_{jj}$ and the average
transverse momentum  $\ptavg$ of the two leading jets. Theoretical 
predictions computed with either of these scale choices differ significantly 
at NLO. This difference is substantially reduced at NNLO, with $\mu=m_{jj}$ 
emerging as a preferred choice, based on the criteria of perturbative 
convergence, and residual scale dependence of the NNLO
prediction~\cite{Currie:2017eqf,Currie:2018oxh}. This is the scale
choice which we will adopt in the sequel.

\begin{figure}[!t]
\centering
\includegraphics[scale=0.47]{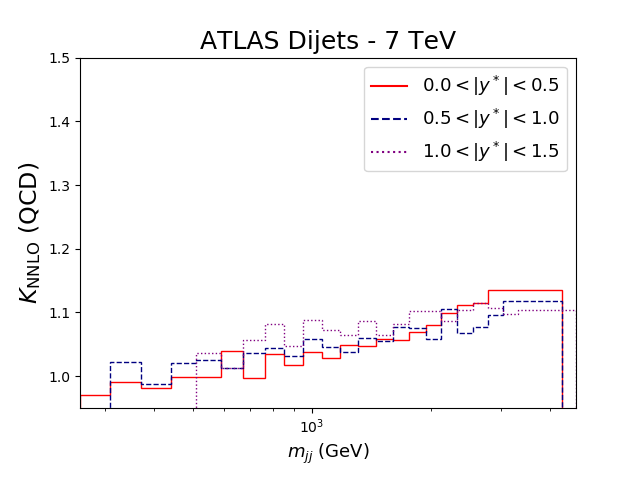}
\includegraphics[scale=0.47]{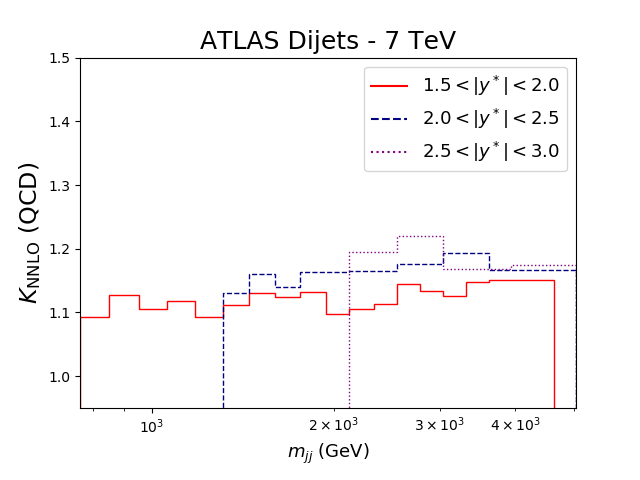}\\
\includegraphics[scale=0.47]{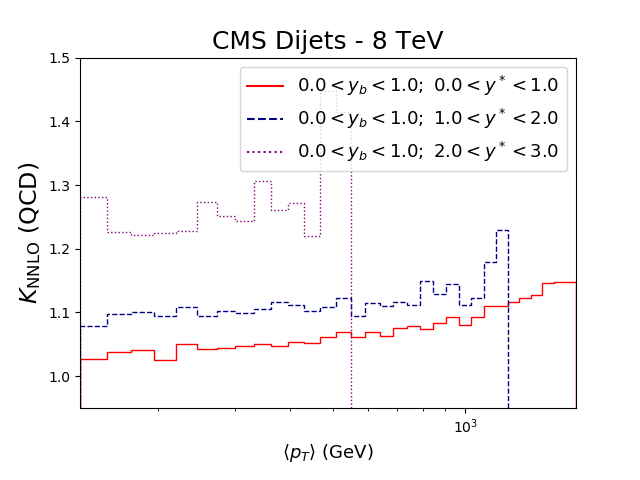}
\includegraphics[scale=0.47]{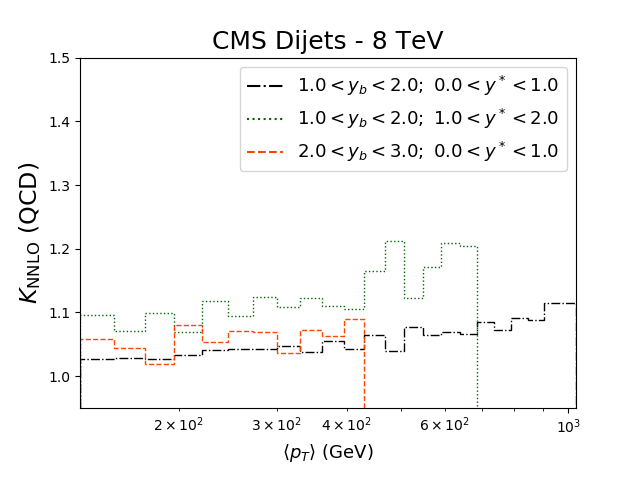}\\
\caption{The NNLO QCD $K$-factors, Eq.~(\ref{eq:kfactor}),
   corresponding to the ATLAS 7 TeV (top) and CMS 8 TeV (bottom) dijet
   cross-sections evaluated with NNPDF3.1 PDF and scale  $\mu=m_{jj}$.
   Results are shown as function of the jet $p_T$ in different jet rapidity 
   bins, with the most central (forward) bins in the left (right) plot.}
\label{fig:kfactqcd_dijets} 
\end{figure}

In Fig.~\ref{fig:kfactqcd_dijets} we display the NNLO QCD $K$-factors, 
Eq.~(\ref{eq:kfactor}), computed  with this scale choice and the NNDPF3.1 NNLO 
PDF set, for the ATLAS 7~TeV and CMS 8~TeV dijet cross-sections. 
For ATLAS, the $K$-factors at small rapidity separations are somewhat
below
unity for low invariant masses, then  grow monotonically with $m_{jj}$ up
to about $K\sim1.15$ at the highest $m_{jj}\sim4$~TeV. For larger rapidity separations, $1.5 \le |y^*| \le 3.0$, 
the $K$-factors are less sensitive to $m_{jj}$, and their value corresponds to 
corrections between 10\% and 20\%. For CMS, as previously mentioned, the measurement
is presented as a triple-differential distribution in $\ptavg$, $y^*$,
and $y_b$. As seen in Fig.~\ref{fig:kfactqcd_dijets}, the qualitative behavior
of the  $K$-factors is similar in all rapidity bins, and shows a monotonic
growth with $\ptavg$. However, the value depends strongly on the rapidity 
difference, with the $K$-factor larger at larger $y^*$.
For example, in the $0 \le y_b,y^* \le 1$ bin the $K$-factor ranges
from a few percent at low $\ptavg$ to up to 15\%, while in the
$0 \le y_b \le 1$ and $2 \le y^* \le 3$ bin it goes up to 25\%.
These $K$-factors display sizable point-to-point fluctuations.

\subsection{Electroweak corrections}
\label{sec:ew}

We have determined EW corrections for all of the single-inclusive jet and dijet
datasets considered in this work by using the calculation of
Ref.~\cite{Dittmaier:2012kx}, suitably extended to the case of
single-inclusive jets. The EW corrections computed in 
Ref.~\cite{Dittmaier:2012kx} include
$O(\alpha \alpha_s)$ and $O(\alpha^2)$ tree level contributions
(where $\alpha$ and $\alpha_s$ are the electromagnetic and strong couplings,
respectively), and the weak radiative corrections of $O(\alpha \alpha_s^2)$.
In particular, they include the virtual exchange of weak bosons that give
rise to the dominant EW Sudakov logarithms, suitably combined with the
respective hard QCD emissions to cancel infrared singularities.
This is what will be
referred to as EW corrections in the remainder of this paper.

We include EW corrections through a $K$-factor defined as
\be
\label{eq:kfactorEWK}
K^{\rm EW}
\equiv 
\frac{\sum_{ij}\widetilde{\sigma}^{\rm LO\> QCD+EW}_{ij}\otimes \mathcal{L}^{\rm NNLO}_{ij}}{
\sum_{ij}\widetilde{\sigma}^{\rm LO,QCD}_{ij}\otimes \mathcal{L}^{\rm NNLO}_{ij}} \, ,
\ee
where the partonic cross-sections in the numerator are
obtained by combining the contributions
computed in Ref.~\cite{Dittmaier:2012kx} with the LO QCD computation.
The $K$-factor defined in Eq.~\eqref{eq:kfactorEWK} has been computed
using a proprietary code~\cite{Dittmaier:2012kx}.
Electroweak $K$-factors have been evaluated  using consistently  the
NNPDF3.1 NNLO PDF set, and the same  scale choice as that of the corresponding 
NNLO QCD predictions. Note that because of cancellations between (negative) 
Sudakov logarithms and (positive) subleading Born contributions, 
the $K$-factors are quite sensitive to the underlying
parton decomposition, and it is  consequently important to make a
consistent choice of PDFs in the computation of QCD and EW $K$-factors.

\begin{figure}[!t]
\centering
\includegraphics[scale=0.47]{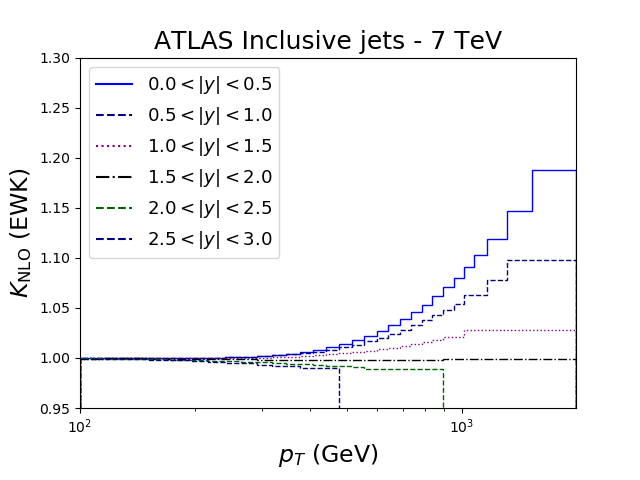}
\includegraphics[scale=0.47]{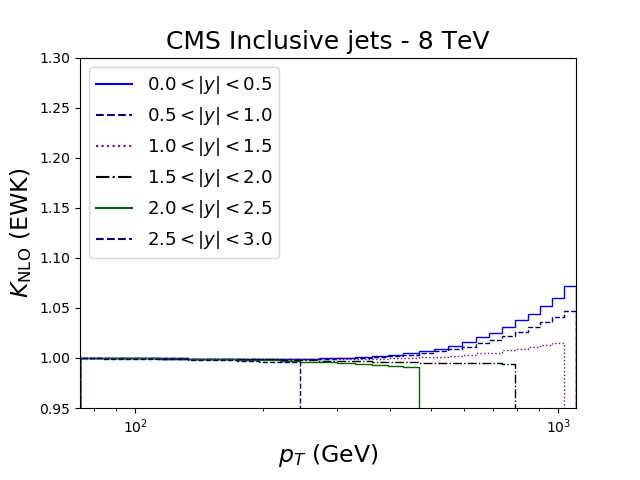}\\
\includegraphics[scale=0.47]{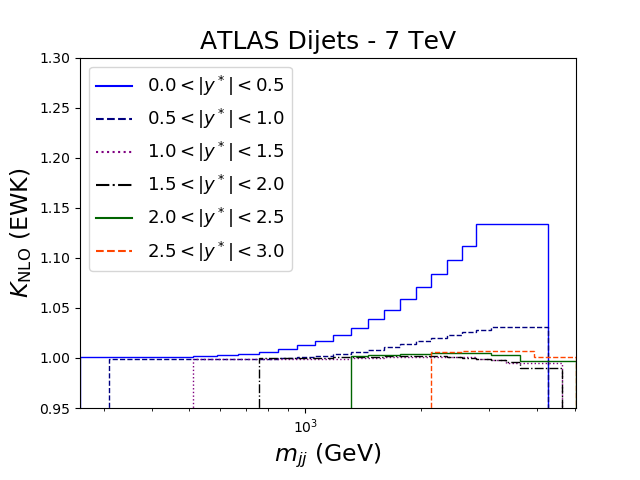}
\includegraphics[scale=0.47]{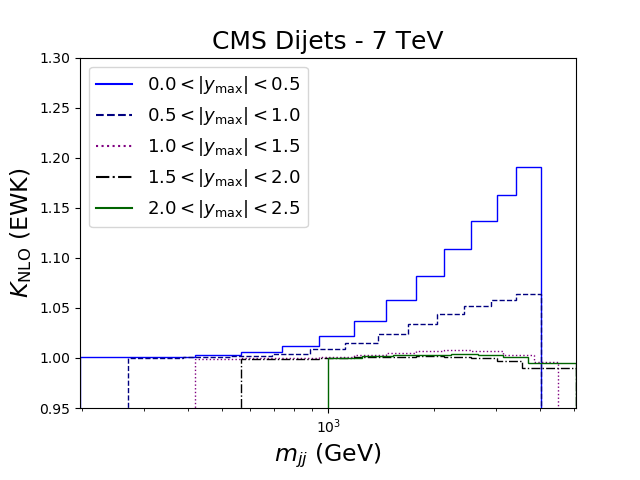}\\
\caption{The EW $K$-factors, Eq.~\eqref{eq:kfactorEWK}, for the ATLAS and CMS
   single-inclusive (top) and dijet (bottom)  measurements. For 
   single-inclusive jets the $K$-factors are shown as a function of jet $p_T$ 
   in six different rapidity bins. For dijets they are shown as a function of 
   the dijet invariant mass $m_{jj}$ for different  $y^*$ bins for ATLAS (left)  
   or $y_{\rm max}$ bins for  CMS (right).}
\label{fig:kfactewk_dijets7}
\end{figure}

The $K$-factors thus computed are shown
in Fig.~\ref{fig:kfactewk_dijets7} for the ATLAS 7~TeV and CMS 8~TeV 
single-inclusive jet cross-sections and for the ATLAS and CMS 7~TeV dijet
cross-sections. Results are shown as a function of $p_T$ for single-inclusive 
jets and as a function of $m_{jj}$ for dijets, in bins of rapidity $y$
(single-inclusive), absolute rapidity difference  $y^*$ (ATLAS dijets)
or maximum absolute rapidity $y_{\rm max}$ (CMS dijets).
In all cases the qualitative behavior is similar: the $K$-factor is
close to unity for small values of $p_T$ or $m_{jj}$; it is flat (in
fact slightly decreasing) for large values of the rapidity variable;
and it grows with respectively $p_T$ or $m_{jj}$ at central rapidity, 
the growth being stronger at smaller rapidity.
The largest EW correction can reach 20\% or more for transverse
momenta or invariant masses in the TeV range and smaller rapidity.

\subsection{Implementation}
\label{sec:implementation}

For each dataset, we produce fast interpolation grids, accurate to NLO in 
QCD, whereby partonic matrix elements are precomputed in such a way that the 
numerical convolution with generic input PDFs can be efficiently approximated 
by means of interpolation techniques. To this purpose, we use 
{\sc NLOJET++}~\cite{Nagy:2001fj} interfaced to
{\sc FastNLO}~\cite{Wobisch:2011ij}. The computation is performed with 
the scale choices discussed in Sect.~\ref{sec:QCD}, and it is 
benchmarked against the {\sc NNLOJET} computation. These fast interpolation 
grids are then combined with PDF evolution kernels, in a format compliant with 
the NNPDF framework, using {\sc APFELgrid}~\cite{Bertone:2016lga}. Such a 
combination is required to speed up the computation 
of hadronic observables when the fit is performed.

Fast interpolation grids accurate to NNLO, for instance in the 
{\sc APPLfast} format, are not yet publicly available: indeed, the 
{\sc NNLOJET}+{\sc APPLfast} fast interpolation tables with NNLO QCD
corrections  
are so far only available for jet production in deep-inelastic 
scattering~\cite{Britzger:2019kkb}.  We therefore implement NNLO and EW
corrections by
supplementing our NLO grids with the 
QCD and EW $K$-factors defined above, which we combine through
the multiplicative prescription
\be
\label{eq:qcdewk}
\frac{d^2\sigma}{dp_T dy}\Bigg|_{\rm NNLO_{QCD}+EW} = \frac{d^2\sigma}{dp_T dy}
\Bigg|_{\rm NLO_{QCD}} \times K_{\rm NNLO}^{\rm QCD}(p_T,y,\sqrt{s})
\times K^{\rm EW}(p_T,y,\sqrt{s}) \, .
\ee
The first term on the right-hand side of the equation is the output
of the NLO computation, while the second and third terms are the bin-by-bin 
QCD and EW $K$-factors defined in Eqs.~\eqref{eq:kfactor} 
and~\eqref{eq:kfactorEWK}, respectively. If the EW $K$-factor is not included, 
Eq.~\eqref{eq:qcdewk} exactly reproduces
the NNLO results obtained with {\sc NNLOJET}.

\begin{figure}[!t]
\centering
\includegraphics[scale=0.47]{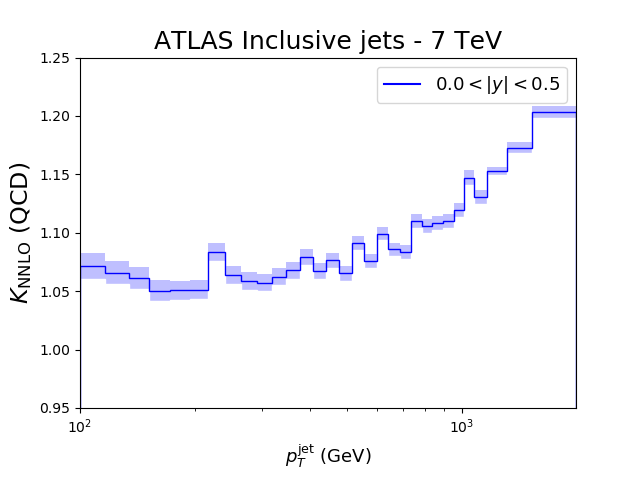}
\includegraphics[scale=0.47]{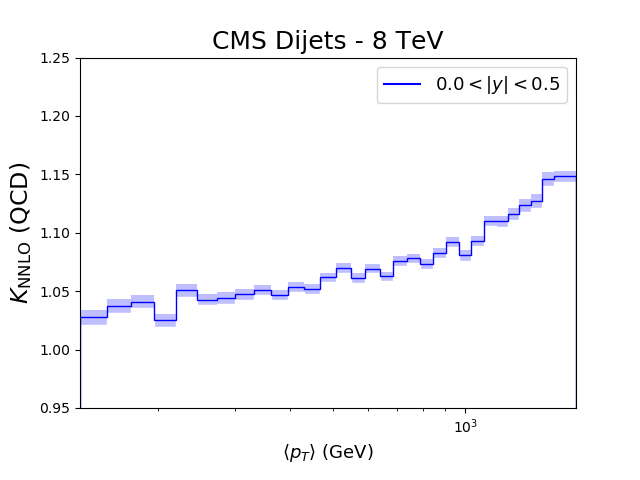}\\
\caption{The NNLO QCD $K$-factors for the central rapidity bins of the ATLAS 
   7~TeV single-inclusive jets (left) and CMS 8~TeV dijets (right), with
   the Monte Carlo numerical uncertainties shown as filled bands around the 
   central result.}
\label{fig:kfactqcd_werrors} 
\end{figure}

As observed in Sects.~\ref{sec:QCD}-\ref{sec:ew}, QCD $K$-factors are 
affected by point-to-point fluctuations which reveal an underlying
numerical uncertainty. For illustration purposes, this uncertainty is 
displayed in Fig.~\ref{fig:kfactqcd_werrors} for the central rapidity bins 
of the ATLAS 7 TeV single-inclusive jet and of the CMS 8 TeV dijet distributions. 
We have estimated this uncertainty
through the procedure for the suppression of outliers as described in Ref.~\cite{Ridder:2016rzm}.
When performing PDF fits, this numerical uncertainty is added in
quadrature to the experimental uncertainty,
fully uncorrelated datapoint by datapoint. An alternative possibility
would be to perform a smooth interpolation of the $K$-factor, see
Ref.~\cite{Carrazza:2017bjw}.
     
Finally, we note that the theoretical computations of single-inclusive and dijet 
observables are subject to non-perturbative corrections and to missing higher 
order uncertainties (MHOU). The former arise from the underlying event and 
multiple parton interactions, and are estimated by the experimental 
collaborations by comparing predictions obtained from different Monte Carlo 
parton shower generators. In the case of all of the CMS measurements, they are 
provided in terms of point-by-point rescaling factors, which we apply 
to the data together with an additional, fully correlated, systematic 
uncertainty, which we estimate as the difference between the value at
each datapoint before and after rescaling.
The estimate of MHOUs requires some care, especially for single-inclusive jets. 
This is due to the fact that there are accidental cancellations which occur for 
values of the jet radius $R\sim0.5$ which are close to the values adopted by 
ATLAS and CMS, where the NLO scale dependence evaluated in a standard 
way is artificially small~\cite{Dasgupta:2016bnd,Cacciari:2019qjx}, and thus
is not a good estimator of the MHOU. A  more reliable estimate of the MHOU 
requires performing uncorrelated scale 
variation~\cite{Dasgupta:2016bnd,Bellm:2019yyh}.
The inclusion of MHOU in PDF fits, though in principle possible using 
the formalism of Refs.~\cite{AbdulKhalek:2019bux,AbdulKhalek:2019ihb}, 
goes beyond the scope of this paper, and we will not consider it further.

\section{Results}
\label{sec:results}

We now present our main results. They consist of a set of global PDF
determinations, in which the NNPDF3.1
global dataset  is supplemented by
the  single-inclusive jet and inclusive dijet
data presented in Sect.~\ref{sec:datahere}: by comparing fit results,
we study the impact of varying
the jet observable,
the data, and the theory settings.
Specifically, we have performed fits including either single-inclusive
or dijet data, in each case using either the full data set, or
7~TeV data or  8~TeV data only,
and with theory at pure NLO QCD, pure NNLO QCD, or NNLO QCD
supplemented by EW corrections as discussed in Sect.~\ref{sec:ew}. 
For the single-inclusive 7~TeV data we have also performed fits with
alternative choices of central scale. Finally, for the ATLAS 7~TeV
and 8~TeV
single-inclusive jet data we have studied the
effect of the treatment of correlated systematics .
We will first present in Sect.~\ref{sec:resincljets} all PDF sets based on
single-inclusive data, including variations of scale choice and
decorrelation model, then  in Sect.~\ref{sec:resdijets} PDF sets based on inclusive
dijet data, and finally in Sect.~\ref{sec:comparisonjets} draw general comparative
conclusions on the behavior of different observables at different
perturbative orders. 

\subsection{PDF sets}
\label{subsec:fit_settings}
\begin{figure}[!t]
\centering
\includegraphics[scale=0.48]{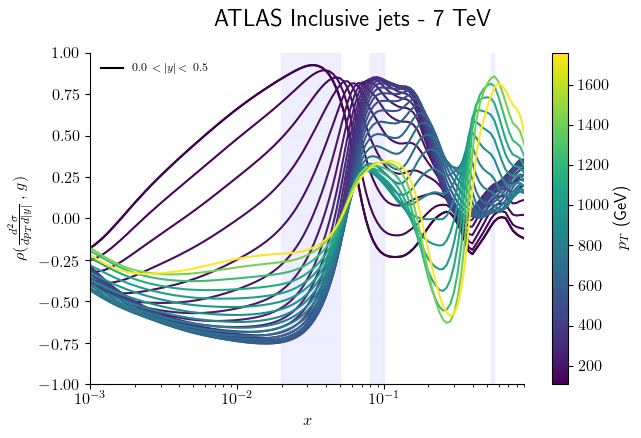}
\includegraphics[scale=0.48]{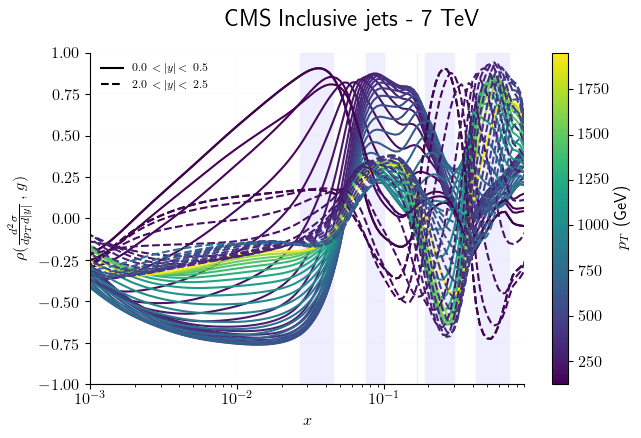}\\
\includegraphics[scale=0.48]{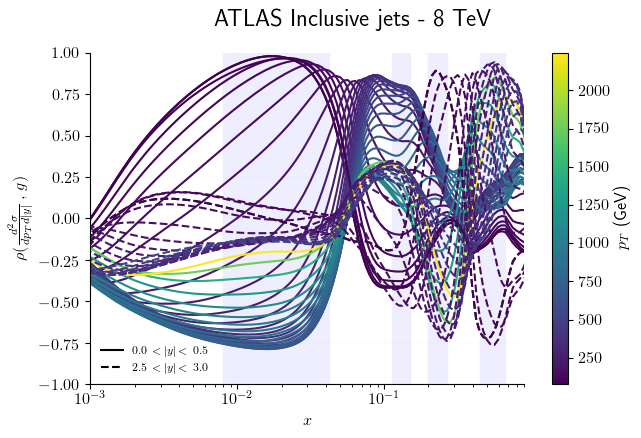}
\includegraphics[scale=0.48]{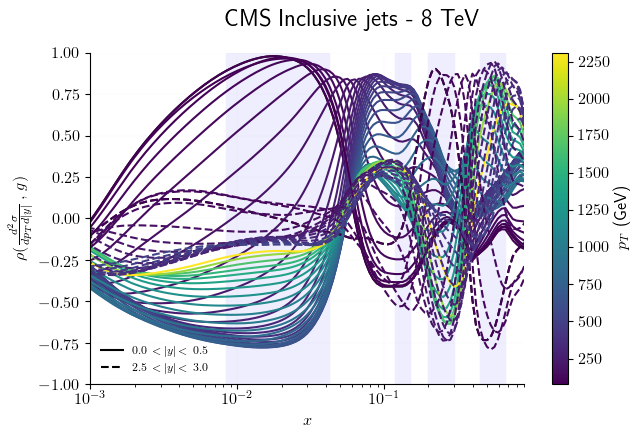}\\
\includegraphics[scale=0.48]{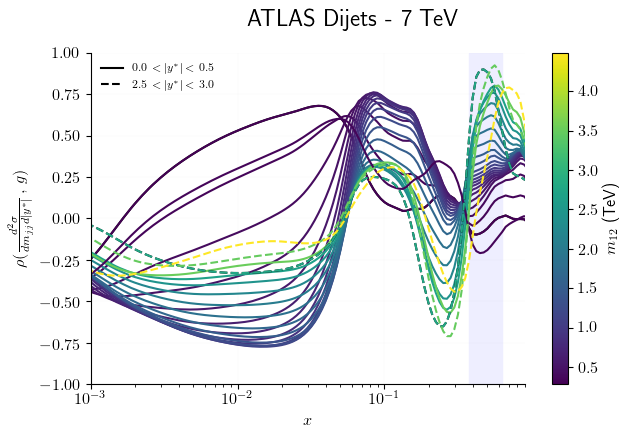}
\includegraphics[scale=0.48]{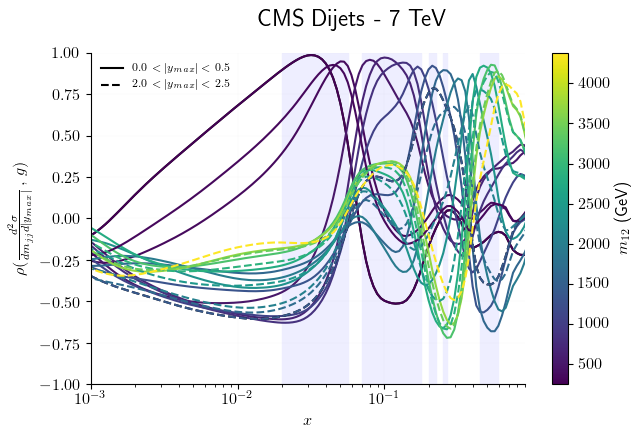}\\
\includegraphics[scale=0.48]{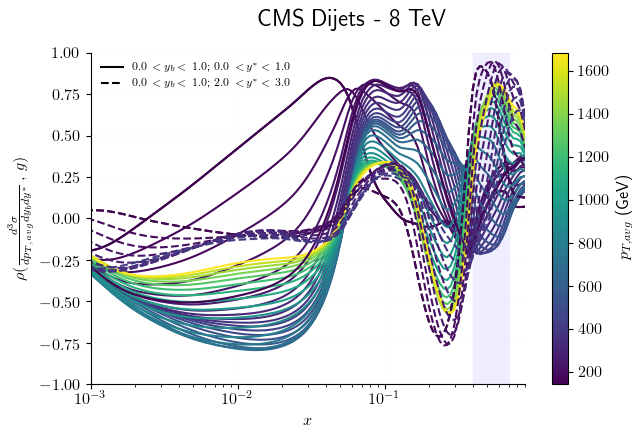}\\
\caption{Correlation coefficients between the data of Table~\ref{tab:input_datasets} and
  the  gluon
  PDF $g(x)$. Each curve corresponds to a different datapoint, with the
  value of $p_T$ corresponding to the color code on the right of the
  plot, and only curves for the points in the
   largest and smallest rapidity bins are shown. The shaded bands
   denote regions in which the maximum correlation is greater than
   90\% of the maximum correlation in the whole plot.}
\label{Fig:correlations}
\end{figure}

The inclusion of jet data in a global NNPDF3.1-like PDF determination
essentially impacts only the gluon PDF, as was shown in
Ref.~\cite{Ball:2017nwa}, while leaving other PDFs essentially
unchanged. The impact of the jet data on the gluon PDF can be assessed
by computing the correlation coefficient (as defined in
Ref.~\cite{Ball:2008by}, see also Ref.~\cite{Alekhin:2011sk}) between
each datapoint and the gluon. The correlations for the largest and
smallest rapidity bins of the datasets of
Table~\ref{tab:input_datasets} are shown in
Figure~\ref{Fig:correlations}, computed using the default baseline
NNLO PDF set (before inclusion of the jet data, \#bn, see
Table~\ref{tab:listfits} below). Correlations are seen to be large or very
large (up to almost one) for all $x\gtrsim 10^{-2}$. Interestingly,
the choice of the rapidity variable $y_{\rm max}$ appears to maximize
the range in which a large correlation with the gluon is observed.

The full list of PDF determinations that we will discuss is given in
Table~\ref{tab:listfits}, together with an ID that will be used to
identify them. In this and all subsequent tables and plots ``jets'' is
short for single-inclusive jets. Each row corresponds to a different choice
of dataset or methodological settings, while columns correspond to the
theory adopted: QCD at NLO or NNLO, without or with EW corrections
included.
By NLO or NNLO we mean that jets have been consistently added
with NLO or NNLO theory  to the respective NLO or NNLO global
fit; note that EW corrections instead are only included for the jet
observable, with all other observables in the global fit computed
using pure QCD theory. 

\begin{table}[!t]
\renewcommand*{\arraystretch}{1.60}
\scriptsize
\centering
\begin{tabularx}{\textwidth}{Xlll}
\toprule
& NNLO$_{\rm QCD}$+EW
& NNLO$_{\rm QCD}$
& NLO$_{\rm QCD}$\\
\midrule
baseline (see text)                           &  ---      &  bn     & b    \\
\midrule
ATLAS \& CMS jets   7-8~TeV                   & janw      & ---    & ---   \\
ATLAS \& CMS jets   7~TeV                     & j7nw      & j7n    & j7    \\
ATLAS \& CMS jets   7~TeV ($\mu=p_T^{\rm jet}$) &  ---      & j7n-pt & j7-pt \\
ATLAS \& CMS jets   8~TeV                     & j8nw      & j8n    & j8    \\
ATLAS \& CMS jets   7-8~TeV
(all ATLAS bins, full corr.)                  & janw-7cor & ---    & ---   \\
ATLAS \& CMS jets   7-8~TeV
(all ATLAS bins 7 TeV, full decorr.)          & janw-7dec & ---    & ---   \\
ATLAS \& CMS jets   7-8~TeV
(all ATLAS bins 7 TeV, part. corr.)           & janw-7pcor& ---    & ---   \\
ATLAS \& CMS jets   7-8~TeV
(all ATLAS bins 8 TeV, full decorr.)          & janw-8dec & ---    & ---   \\
ATLAS \& CMS jets   7-8~TeV
(all ATLAS bins 8 TeV, part. corr.)           & janw-8pcor& ---    & ---   \\
\midrule
ATLAS \& CMS dijets 7-8~TeV                   & danw      & ---    & ---   \\
ATLAS \& CMS dijets 7~TeV                     & d7nw      & d7n    & d7    \\
CMS          dijets 8~TeV                     & d8nw      & d8n    & d8    \\
\bottomrule
\end{tabularx}

\vspace{0.3cm}
\caption{The PDF determinations discussed in this study and their
  IDs. Each row corresponds to a different choice of input jet dataset or fit
  settings (listed in the first column), and each column corresponds
  to a different theory accuracy (listed in the first row).
  The ID encodes the process used (j for single-inclusive
  jets and d for dijets); the data used (a for all, 7 or 8 for the
  7~TeV or 8~TeV datasets); the perturbative accuracy (n for QCD NNLO,
  w if EW corrections included); and the choice of scale (pt when
  $\mu=p_T^{\rm jet}$); special treatment of the ATLAS 7~TeV data (cor
  for all bins included, correlated; dec, decorrelated; pcor,
  partially correlated); and special treatment of the ATLAS 8~TeV data
  (dec for all bins included, decorrelated; pcor, partially correlated).
  In this and subsequent tables and plots
  ``jets'' is short for single-inclusive jets.}
\label{tab:listfits}
\end{table}

The jet data of Table~\ref{tab:input_datasets} are added to a baseline
dataset, which essentially coincides with the  NNPDF3.1 dataset. This
dataset includes:
fixed-target neutral-current 
(NC) DIS structure function data from NMC~\cite{Arneodo:1996kd,Arneodo:1996qe}, 
SLAC~\cite{Whitlow:1991uw} and BCDMS~\cite{Benvenuti:1989rh}; charged-current 
(CC) DIS structure function data from CHORUS~\cite{Onengut:2005kv} and 
NuTeV~\cite{Goncharov:2001qe,Mason:2006qa}; HERA data from their combined 
measurements~\cite{Abramowicz:2015mha}, including charm-production cross 
sections~\cite{Abramowicz:1900rp} and $b$-tagged structure 
functions~\cite{Aaron:2009af,Abramowicz:2014zub}; fixed-target Drell-Yan data 
from E866~\cite{Webb:2003ps,Webb:2003bj,Towell:2001nh} and 
E605~\cite{Moreno:1990sf}; collider Drell-Yan data from 
CDF~\cite{Aaltonen:2010zza} and D0~\cite{Abazov:2007jy,
Abazov:2013rja,D0:2014kma}; and Drell-Yan, inclusive gauge boson, and top-pair
production data from 
ATLAS~\cite{Aad:2013iua,Aad:2014qja,Aad:2011dm,Aaboud:2016btc,Aad:2015auj,
Aad:2014kva,Aaboud:2016pbd,Aad:2015mbv}, CMS~\cite{Chatrchyan:2012xt,
Chatrchyan:2013mza,Chatrchyan:2013tia,Khachatryan:2016pev,Khachatryan:2015oaa,
Khachatryan:2016mqs,Khachatryan:2015uqb,Khachatryan:2015oqa} 
and LHCb~\cite{Aaij:2012vn,Aaij:2012mda,Aaij:2015gna,Aaij:2015zlq}. 
In total this baseline dataset contains $n_{\rm dat}=3813$ datapoints,
see Ref.~\cite{Ball:2017nwa} for more details. The number of
datapoints corresponding to the  jet data included in the various fits
of Table~\ref{tab:listfits} is given in Table~\ref{tab:chi2s} and in Table~\ref{tab:chi2sD} below.

In all of these fits, experimental systematic uncertainties
are fully correlated across bins of different kinematic variables,
while statistical uncertainties coming from the unfolding are correlated only
across bins of transverse momentum (for jets) or invariant mass (for dijets),
but not across rapidity bins. The possibility of removing some or all
of these correlations will be discussed in Sect.~\ref{sec:corr} below.
Multiplicative uncertainties are treated with the 
$t_0$-method~\cite{Ball:2009qv}, and all fits in Table~\ref{tab:listfits}
are iterated once to ensure  convergence of the $t_0$ method and
preprocessing (see Ref.~\cite{Ball:2017nwa} for more details).

For jet or dijet data, non-perturbative corrections are included
by default, as are Monte Carlo uncertainties due to finite numerical
precision of NNLO QCD $K$-factor computations (see Sect.~\ref{sec:theory} for 
details). The factorization and renormalization scales are by default taken 
to be  $\mu=\widehat{H}_T$ for single-inclusive jets, and $\mu=m_{jj}$ for dijets (see the discussion
in Sect.~\ref{sec:theory}). An alternative choice of scale for
single-inclusive jets will be considered in Sect.~\ref{sec:scale} below.

All the fits listed in Table~\ref{tab:listfits} otherwise closely follow
the NNPDF3.1 analysis~\cite{Ball:2017nwa}. Specifically, the same
settings and codes are used for the 
computation of physical observables in the baseline dataset, and the
same choice of 
kinematic cuts, of values of physical parameters, and of fitting
methodology are adopted. 
All PDF sets include  $N_{\rm rep}$=100 Monte Carlo replicas. 
The {\sc ReportEngine} software~\cite{zahari_kassabov_2019_2571601} is used in 
the sequel to analyze each fit and compute various fit
metrics. Specifically, we consider
the $\chi^2$ of the theory prediction for  each 
dataset or combinations of datasets, defined according to  Eqs.~(7)-(8)
of Ref.~\cite{Ball:2012wy}, and the 
distance $d$ between pairs of fits (see e.g.
Eq.~(48) of Ref.~\cite{Ball:2014uwa} for its definition).  

The values of the $\chi^2$ per
datapoint for all fits with default settings at NLO and NNLO with or
without EW corrections and 
single-inclusive jet or dijet data are collected in
Tab.~\ref{tab:chi2s} and in Tab.~\ref{tab:chi2sD} respectively; $\chi^2$ values are shown
for all data in the global dataset, grouped by process type  (DIS NC,
DIS CC, Drell-Yan, $Z$ $p_T$, top pair) and for
all jet data, both those which are and those which are not included in
each fit. 
The values of $\chi^2$ per datapoint for all jet data (included or not
included) for all fits performed with alternative choices of central
scale or alternative decorrelation models are collected in
Tab.~\ref{tab:chi2_suppl}. In these tables, $\chi^2$ values
corresponding to data not included in each fit are enclosed in square
brackets. 

\begin{table}[t]
\renewcommand*{\arraystretch}{1.60}
\scriptsize
\centering
\begin{tabularx}{\textwidth}{Xrccccccccc}
\toprule
 Dataset                    & $n_{\rm dat}$ &     b  &   bn   &  janw  &    j7  &   j7n  &  j7nw  &    j8  &   j8n  &  j8nw  \\
\midrule
 DIS NC                     &       2103  &  1.17  &  1.17  &  1.18  &  1.17  &  1.18  &  1.17  &  1.17  &  1.17  &  1.18  \\
 DIS CC                     &        989  &  1.06  &  1.10  &  1.11  &  1.06  &  1.11  &  1.10  &  1.08  &  1.11  &  1.11  \\
 Drell-Yan                  &        577  &  1.35  &  1.33  &  1.30  &  1.35  &  1.31  &  1.31  &  1.34  &  1.31  &  1.31  \\
 $Z$ $p_T$                  &        120  &  1.84  &  1.01  &  1.02  &  1.85  &  1.02  &  1.02  &  1.89  &  1.03  &  1.03  \\
 Top pair                   &         24  &  1.10  &  1.05  &  1.25  &  1.09  &  1.06  &  1.02  &  2.00  &  1.61  &  1.24  \\
 \ \ ATLAS $\sigma_{t\bar t}$ &          3  &  2.02  &  0.90  &  0.70  &  1.68  &  0.74  &  0.72  &  1.70  &  0.79  &  0.78  \\
 \ \ ATLAS $t\bar{t}$ rap   &          9  &  1.12  &  1.22  &  2.01  &  1.25  &  1.38  &  1.31  &  2.93  &  2.78  &  1.96  \\
 \ \ CMS $\sigma_{t\bar t}$   &          3  &  0.53  &  0.22  &  0.21  &  0.42  &  0.24  &  0.31  &  0.34  &  0.17  &  0.19  \\
 \ \ CMS $t\bar{t}$ rap     &          9  &  0.98  &  1.17  &  0.98  &  0.96  &  1.09  &  1.04  &  1.65  &  1.12  &  0.99  \\
 Jets (all)                 &        520  & [1.48] & [2.60] &  1.88  & [1.86] & [2.45] & [2.53] & [1.20] & [1.75] & [1.89] \\
 \ \ Jets (fitted)          &             &  ---   &  ---   &  1.88  &  0.79  &  1.15  &  1.12  &  1.40  &  2.05  &  2.20  \\
 \ \ ATLAS 7 TeV            &         31  & [1.26] & [1.87] &  1.59  &  1.12  &  1.73  &  1.15  & [1.07] & [1.69] & [1.62] \\
 \ \ ATLAS 8 TeV            &        171  & [2.60] & [5.01] &  3.22  & [3.55] & [4.76] & [4.58] &  2.03  &  3.18  &  3.25  \\
 \ \ CMS   7 TeV            &        133  & [0.60] & [1.06] &  1.09  &  0.71  &  1.01  &  1.11  & [0.72] & [0.94] & [1.14] \\
 \ \ CMS   8 TeV            &        185  & [1.10] & [1.59] &  1.25  & [1.24] & [1.47] & [1.80] &  0.81  &  1.01  &  1.23  \\
 Dijets (all)               &        266  & [3.49] & [3.07] & [2.10] & [4.16] & [2.96] & [2.56] & [3.34] & [2.21] & [2.22] \\
 \ \ Dijets (fitted)        &             &  ---   &  ---   &  ---   &  ---   &  ---   &  ---   &   ---  &  ---   &  ---   \\
 \ \ ATLAS 7 TeV            &         90  & [1.49] & [2.47] & [1.95] & [1.77] & [2.46] & [1.97] & [1.43] & [2.28] & [2.01] \\
 \ \ CMS   7 TeV            &         54  & [2.06] & [2.40] & [2.08] & [2.43] & [2.50] & [2.12] & [1.65] & [2.00] & [2.15] \\
 \ \ CMS   8 TeV            &        122  & [5.60] & [3.81] & [2.21] & [6.70] & [3.53] & [3.20] & [5.48] & [2.26] & [2.39] \\
\midrule
 Total                      &             &  1.20  & 1.18   &  1.28  &  1.17  &  1.17  & 1.17  &   1.39  &  1.27  &  1.27  \\
\bottomrule
\end{tabularx}

\vspace{0.3cm}
\caption{The $\chi^2$ per datapoint for all fits of
  Table~\ref{tab:listfits} including single-inclusive jet data, with default settings.
  Results are shown
  for all datasets, aggregated by process type. For jets data, results are
  shown both for the sets included in each fit, and also for those not
  included, enclosed in square brackets. Combined results are also shown
  for all single-inclusive jet and for all dijet data, both for
  the full set, and for those included in each fit.
  The number of datapoints in each
  dataset is also shown.}
\label{tab:chi2s}
\end{table}
\begin{table}[t]
\renewcommand*{\arraystretch}{1.60}
\scriptsize
\centering
\begin{tabularx}{\textwidth}{Xrccccccccc}
\toprule
 Dataset                    & $n_{\rm dat}$ &     b  &   bn   &  danw  &    d7  &   d7n  &  d7nw  &    d8  &   d8n  &  d8nw  \\
\midrule
 DIS NC                     &       2103  &  1.17  &  1.17  &  1.18  &  1.17  &  1.17  &  1.17  &  1.21  &  1.18  &  1.18  \\
 DIS CC                     &        989  &  1.06  &  1.10  &  1.12  &  1.07  &  1.09  &  1.09  &  1.11  &  1.11  &  1.12  \\
 Drell-Yan                  &        577  &  1.35  &  1.33  &  1.29  &  1.36  &  1.33  &  1.32  &  1.32  &  1.28  &  1.28  \\
 $Z$ $p_T$                  &        120  &  1.84  &  1.01  &  1.07  &  1.85  &  1.03  &  1.03  &  2.06  &  1.07  &  1.08  \\
 Top pair                   &         24  &  1.10  &  1.05  &  1.14  &  1.16  &  1.06  &  1.04  &  1.57  &  1.34  &  1.26  \\
 \ \ ATLAS $\sigma_{t\bar t}$ &          3  &  2.02  &  0.90  &  0.66  &  1.79  &  0.74  &  0.73  &  0.80  &  0.68  &  0.69  \\
 \ \ ATLAS $t\bar{t}$ rap   &          9  &  1.12  &  1.22  &  1.57  &  1.26  &  1.34  &  1.32  &  2.41  &  2.02  &  1.82  \\
 \ \ CMS $\sigma_{t\bar t}$   &          3  &  0.53  &  0.22  &  0.53  &  0.48  &  0.29  &  0.28  &  0.01  &  0.74  &  0.67  \\  
 \ \ CMS $t\bar{t}$ rap     &          9  &  0.98  &  1.17  &  1.04  &  1.07  &  1.09  &  1.07  &  1.42  &  1.04  &  1.04  \\
 Jets (all)                 &        520  & [1.48] & [2.60] & [2.06] & [1.62] & [2.75] & [2.70] & [1.42] & [1.94] & [2.14] \\
 \ \ Jets (fitted)          &             &  ---   &  ---   &  ---   &  ---   &  ---   &  ---   &  ---   &  ---   &  ---   \\
 \ \ ATLAS 7 TeV            &         31  & [1.26] & [1.87] & [1.63] & [1.26] & [1.86] & [1.74] & [1.00] & [1.70] & [1.61] \\
 \ \ ATLAS 8 TeV            &        171  & [2.60] & [5.01] & [3.36] & [2.62] & [4.80] & [4.65] & [2.18] & [3.30] & [3.55] \\
 \ \ CMS   7 TeV            &        133  & [0.60] & [1.06] & [1.06] & [0.71] & [1.13] & [1.14] & [0.77] & [0.97] & [1.07] \\
 \ \ CMS   8 TeV            &        185  & [1.10] & [1.59] & [1.64] & [1.42] & [2.16] & [2.17] & [1.27] & [1.41] & [1.68] \\
 Dijets (all)               &        266  & [3.49] & [3.07] &  1.65  & [3.03] & [2.21] & [2.16] & [2.38] & [1.74] & [1.71] \\
 \ \ Dijets (fitted)        &             &  ---   &  ---   &  1.65  &  1.33  &  1.79  &  1.72  &  3.69  &  1.59  &  1.68  \\
 \ \ ATLAS 7 TeV            &         90  & [1.49] & [2.47] &  1.76  &  1.20  &  1.94  &  1.78  & [1.04] & [1.96] & [1.78] \\
 \ \ CMS   7 TeV            &         54  & [2.06] & [2.40] &  1.60  &  1.54  &  1.55  &  1.63  & [1.67] & [1.70] & [1.66] \\
 \ \ CMS   8 TeV            &        122  & [5.60] & [3.81] &  1.58  & [5.03] & [2.70] & [2.67] &  3.69  &  1.59  &  1.68  \\
\midrule
 Total                      &             &  1.20  &  1.18  &  1.22  &  1.33  &  1.20  &  1.19  &  1.33  &  1.20  &  1.20  \\
\bottomrule
\end{tabularx}

\vspace{0.3cm}
\caption{Same as Table~\ref{tab:chi2s}, but now for dijets. The
  baseline is repeated for ease of reference.}
\label{tab:chi2sD}
\end{table}

\subsection{Single-inclusive jets}
\label{sec:resincljets}

We first present PDF sets obtained by including single-inclusive jet data.
We discuss in turn
the impact and consistency of individual  datasets; perturbative QCD
stability and the impact of EW corrections; the choice of central
scale; and alternative data treatment and
decorrelation models for the ATLAS 7~TeV data.

\subsubsection{Impact and consistency of datasets}
\label{sec:impcondata}

We provide a general comparative assessment of the
impact of single-inclusive jet data on PDFs by 
comparing fits performed with the default theory settings
of Sect.~\ref{sec:theory} and the highest theory accuracy, i.e. NNLO
QCD theory for jet data and the rest of the global fit,
and EW corrections included in
the jet predictions only. According to the data included, these correspond
to the fits \#bn, \#janw, \#j7nw, and \#j8nw of Table~\ref{tab:listfits}.

First, we compare fit \#janw, that contains all of the single-inclusive
jet data, to the baseline \#bn, which does not include any jet
data. Note that, as discussed in Sect.~\ref{sec:datahere}, in our
default global dataset only the central rapidity bin of the ATLAS
7~TeV data is included. Fits in which the full 7~TeV ATLAS dataset is
included will be discussed in Sect.~\ref{sec:corr} below. 
In Fig.~\ref{fig:jet_data_total} we display the distance between 
the PDF central values  for the two fits, and the gluon PDF in both
fits, normalized to the baseline, both at  $Q=100$~GeV.
 Recall that the 
distance $d$ is the difference in units of the standard deviation $\sigma$ 
of the mean, so for a sample of 100 replicas $d\sim 1$ corresponds to 
statistically identical PDFs (replicas extracted from the same underlying 
distribution) and $d\sim 10$ corresponds to PDFs that differ by one sigma. 
From 
Table~\ref{tab:chi2s}, we note  that  individual jet datasets are  
well described (with $\chi^2$ per datapoint of order one), 
except the  8~TeV ATLAS data ($\chi^2=3.22$), to be
investigated in greater detail below. In comparison 
to the baseline fit, the inclusion of the single-inclusive jet data leads to a slight 
deterioration in the description of the ATLAS top pair rapidity distributions,
whose  $\chi^2$ per datapoint increases from 1.22 to 2.01. On the
other hand, it leads to an improvement in the description of the dijet
data, especially the 8~TeV CMS data, which are not included in any of
these fits. This suggests that the inclusion of single-inclusive and dijet data have a
similar impact on PDFs, as we shall also see in Sect.~\ref{sec:impcondatad} and
discuss in greater detail in
Sect.~\ref{sec:comparisonjets} below. 

As mentioned above, and as it is clear from the distance plot in
Fig.~\ref{fig:jet_data_total}, single-inclusive
jet data only have an impact on the gluon. The regions which are most
affected are $x\simeq 0.05$, $0.1\lesssim x \lesssim 0.2$, and
$0.3\lesssim x\lesssim 0.5$,
consistently with the correlation plots of Fig.~\ref{Fig:correlations}:
in these regions  the gluon PDF changes by up to slightly more than
half sigma.
In comparison to the baseline, the central gluon PDF 
is suppressed by about 2\% in the small $x$ region and enhanced by about 4\% in 
the large $x$ regions, though it always remains within the uncertainty band
of the baseline.

\begin{figure}[!t]
\centering
\includegraphics[scale=0.49]{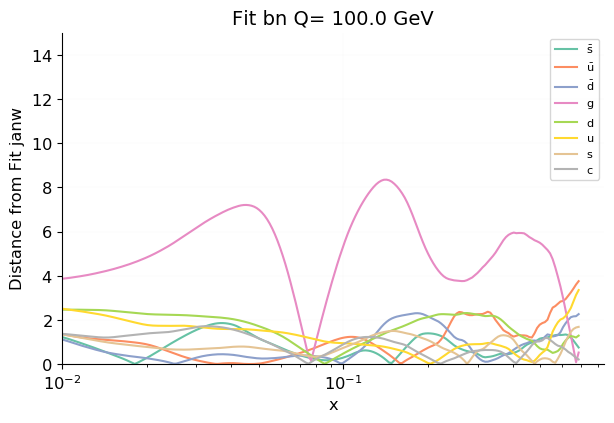}
\includegraphics[scale=0.49]{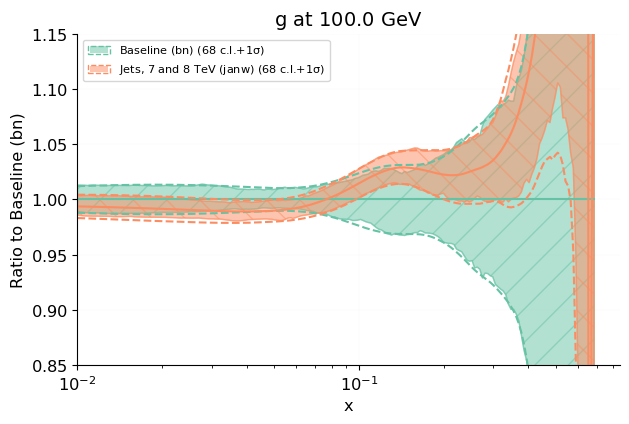}\\
\caption{ Comparison between the baseline fit with no jet data  (\#bn)
  and the fit with all single-inclusive jet data included (\#janw), both with
  default settings and the most accurate theory (NNLO QCD, including EW
  corrections for jets). The distance (see text) between all PDFs
  (left) and the ratio of the gluon PDF to the baseline (right) are shown at the scale
  $Q=100$~GeV. The shaded band is the 68\% confidence interval,
  while the dashed lines are the edge of one sigma interval.}
\label{fig:jet_data_total}
\end{figure}
\begin{figure}[!t]
\centering
\includegraphics[scale=0.49]{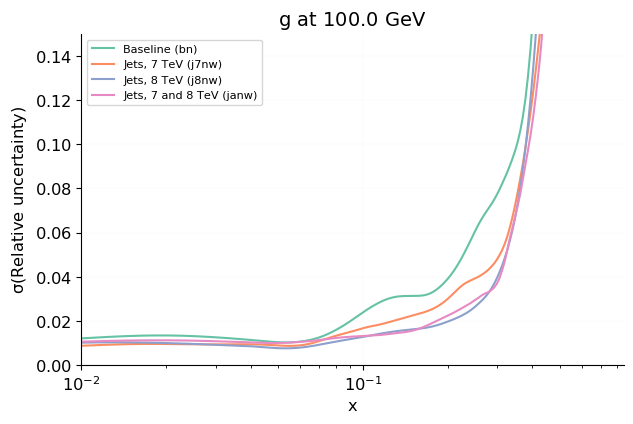}
\includegraphics[scale=0.49]{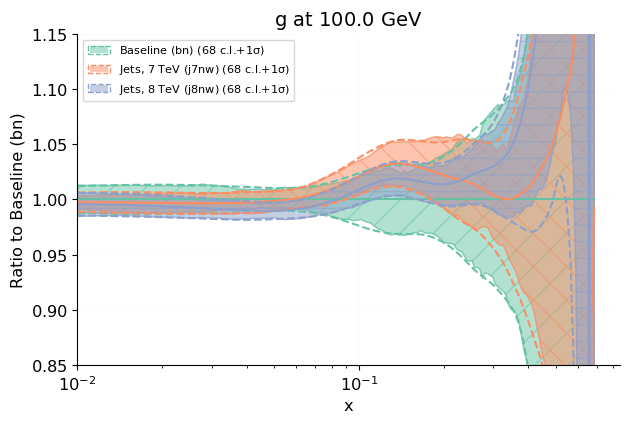}
\caption{ Comparison between the baseline fit with no jet data
  (\#bn), and the fits with only 7~TeV (\#j7nw) or only 8~TeV (\#j8nw)
  jet data included. The relative uncertainty on the gluon PDF (left)
  and the ratio of the gluon PDF to the baseline (right) are shown at
  $Q=100$ GeV. All results are shown as ratios to the baseline.}
\label{fig:jet_data_partial}
\end{figure}

We next assess the relative impact of different jet datasets, by
adding to the comparison of the baseline (\#bn) and the fit with all single-inclusive jet
data (\#janw) also fits in  which only  7~TeV  (\#j7nw) or 8~TeV
(\#j8nw) jet data are included, all with the same settings (NNLO
QCD+EW). The comparison is shown  in Fig.~\ref{fig:jet_data_partial},
where we compare the gluon and its relative uncertainty. Here and
henceforth, when comparing relative uncertainties, the uncertainties
shown are computed as a ratio to a common baseline, i.e. the plot
displays all uncertainties as a percentage of the same reference fit. 
From Table~\ref{tab:chi2s}, 
we note  that the unsatisfactory description of the ATLAS 8~TeV data
persists even when the 7~TeV data are not included in the fit, and the
deterioration in fit quality for the ATLAS top data in the global
fit is also similar. On the other hand, the fit in which only 7~TeV jets
are included shows excellent fit quality both for the jet data and the
global dataset. A significant difference between these two datasets is
that for the 7~TeV data only the central rapidity bin is included,
while for the 8~TeV data all rapidity bins are included: this suggests
that the 8~TeV data may also be affected by similar issues in the
treatment of correlations between rapidity bins. We will see in
Sect.~\ref{sec:corr} that this is indeed the case.

The relative pull of the jet datasets at 7~TeV and 8~TeV can be inferred from 
Fig.~\ref{fig:jet_data_partial}. They both lead to a comparable suppression of
the gluon PDF of about 1\% in the region $0.3\lesssim x\lesssim 0.5$, while 
they respectively enhance it by 4\% and 2\% in the region 
$0.1\lesssim x \lesssim 0.2$.
However, the decrease in gluon uncertainty is rather more marked upon
inclusion of the 8~TeV data, and in fact, results obtained including
all jet data, or only 8~TeV are almost identical.
Specifically, in comparison to the
baseline, inclusion of the 8~TeV data results in a reduction of the
relative gluon uncertainty at $x\simeq 0.2$ from 
4\% to 1.5\%, to be compared to the reduction  4\% to 3\% when the
7~TeV data are included. A similar behavior was observed in the recent
CT18 global PDF determination~\cite{Hou:2019efy}, which includes the ATLAS and
CMS jet datasets at 7~TeV and the CMS jet dataset at 8~TeV. 

\subsubsection{Impact of higher-order QCD and EW corrections}
\label{sec:hoimp}

\begin{figure}[!t]
\centering
\includegraphics[scale=0.49]{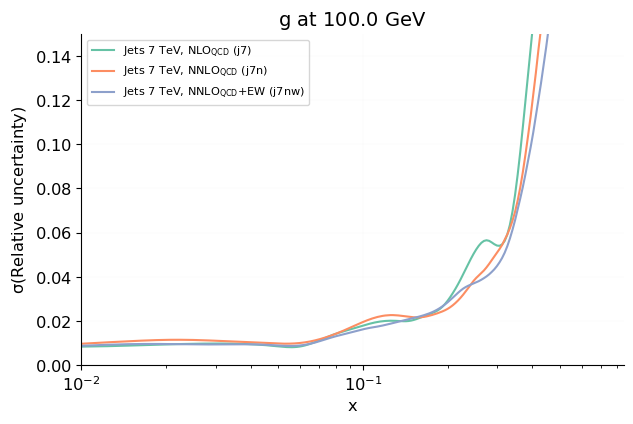}
\includegraphics[scale=0.49]{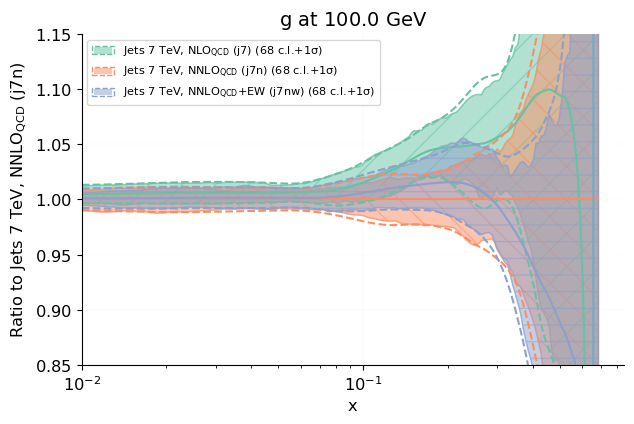}
\\
\includegraphics[scale=0.49]{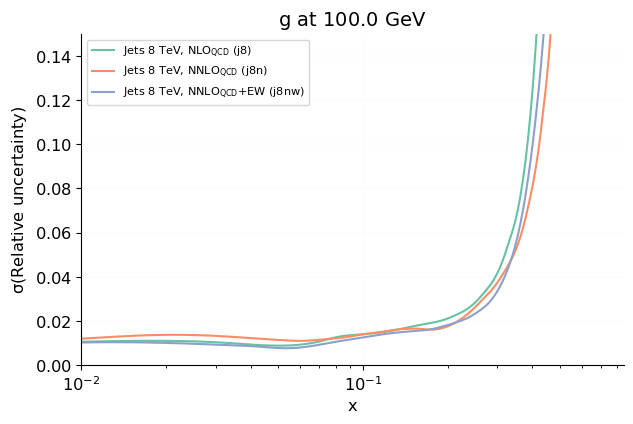}
\includegraphics[scale=0.49]{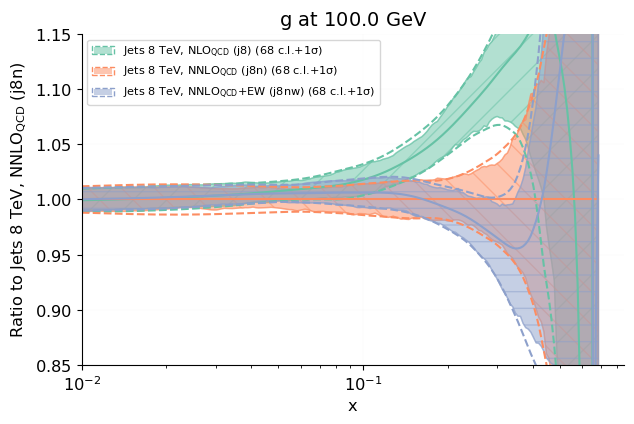}\\
\caption{Same as Fig.~\ref{fig:jet_data_partial}, but now comparing
  fits with
  NLO, NNLO and NNLO+EW theory for 7~TeV (top: fits \#j7, \#j7n,
  \#j7nw respectively) and 8~TeV (bottom: fits \#j8, \#j8n,
  \#j8nw). All results are now shown as ratios to the NNLO fit.}
\label{fig:jet_corrs}
\end{figure}
Having assessed the impact of various single-inclusive
jet data on PDFs with optimal theory settings,
we now turn to the assessment of the perturbative stability of
results. To this purpose, we compare fits at NLO, NNLO and with EW
corrections (included for jet data only). The comparison is performed
separately for the 7~TeV and  8~TeV data (fits
\#j7, \#j7n,  \#j7nw; and  \#j8, \#j8n,  \#j8nw respectively) both in
oder to check consistency and to get a more detailed picture of the
impact of different datasets. The gluon PDFs at $Q=100$~GeV and their
uncertainty for these fits are compared in  Fig.~\ref{fig:jet_corrs}.

It is clear from the figure that, both for 7~TeV and 8~TeV data, at NLO
the gluon undergoes a significant distortion in the region $0.1\lesssim
x \lesssim 0.5$ in comparison to the NNLO results  shown in
Fig.~\ref{fig:jet_data_partial}. Specifically, at the peak,
$x\sim 0.3$ the NLO gluon turns out to be by 30-40\% larger than the
baseline. This effect is driven by the jet data: we have verified that
in the baseline (without jet data) the NLO gluon does show some
distortion in comparison to the NNLO baseline,
but by a much smaller amount, with the largest enhancement
of order 5\%. This is thus evidence for large missing NNLO corrections
to the single-inclusive jet cross section in
the NLO fit.  The effect is more
pronounced for the  8~TeV data, which can be understood as a
consequence of their greater precision.

The effect of EW
corrections is rather more moderate, with the shift of the central
value always within the NNLO uncertainty band. Also, EW
corrections seem to have an opposite effect when added to the fit to
the 7~TeV or the 8~TeV data, leading to a slight enhancement of the gluon
in the former case
and a significant suppression for
$x\gtrsim 0.2$ in the latter case.
  For both
datasets,  the uncertainty on the gluon for $x\gtrsim0.1$, where the
jet data have an impact, is reduced by a non-negligible amount by the
inclusion of NNLO corrections.
On the other hand the impact of the EW corrections is less clear. All
this suggests that NNLO corrections have a significant impact, by
affecting the best-fit large-$x$ gluon shape and improving its
precision, while the impact of EW corrections is minor, and not clear-cut.

The effect of the inclusion of the NNLO and EW corrections on fit
quality is less clear. Indeed, from Table~\ref{tab:chi2s}, we observe
that generally the fit quality to jet data deteriorates somewhat upon
inclusion of NNLO corrections, and a little more upon inclusion of EW
corrections. On the other hand, the global fit quality, as measured by
the total $\chi^2$, is unchanged for the 7~TeV data, and it improves
significantly, from 1.39 to 1.27, for the more precise 8~TeV data, with
the improvement mostly driven by the top and $Z$ $p_T$ data which are
most sensitive to the gluon. However, as already noted in
Sect.~\ref{sec:impcondata}, the $\chi^2$ of the top data deteriorates
when adding the jet data to the baseline, and the fit quality to the
ATLAS 8~TeV data remains unchanged. This suggests that, for the more precise
 8~TeV data,
the NNLO corrections reduce a tension between top and jets (especially
ATLAS).

In summary, we conclude that, consistently with previous theoretical
investigations~\cite{Currie:2018xkj,Cacciari:2019qjx} NNLO
corrections have a sizable impact on single-inclusive jets, and in
particular their inclusion leads to a reduction of the uncertainty on
the large-$x$ gluon PDF and an improved consistency of the jet data
with the rest of the global dataset, demonstrated by a reduction of
the shift of the gluon central value upon inclusion of jets, and as an
improvement of the global $\chi^2$ (for the more precise 8~TeV jet
data), when  going from NLO to NNLO. Electroweak corrections do not
appear to lead to improvements either in terms of fit quality or PDF
uncertainty.
\begin{table}[!t]
\renewcommand*{\arraystretch}{1.60}
\scriptsize
\centering
\begin{tabularx}{\textwidth}{Xlcccccccccc}
Dataset              & $n_{\rm dat}$ & \rotatebox{90}{j7}  & \rotatebox{90}{j7-pt}
                     & \rotatebox{90}{j7n}  & \rotatebox{90}{j7n-pt}
                     & \rotatebox{90}{janw} & \rotatebox{90}{janw-7cor}
                     & \rotatebox{90}{janw-7dec} & \rotatebox{90}{janw-7pcor}
                     & \rotatebox{90}{janw-8dec} & \rotatebox{90}{janw-8pcor} \\
\toprule
ATLAS jets 7 TeV     & 31~(140)    &  1.12  &  1.13  &  1.13  &  1.13   &  1.59  &  2.44     &  1.22     &  1.22      &  1.59     &  1.61      \\     
ATLAS jets 8 TeV     & 171         & [3.55] & [3.93] & [4.76] & [4.99]  &  3.22  &  3.16     &  3.19     &  3.20      &  0.83     &  0.98      \\ 
CMS jets 7 TeV       & 133         &  0.71  &  0.91  &  0.95  &  0.94   &  1.09  &  1.04     &  1.04     &  1.04      &  1.12     &  1.12      \\
CMS jets 8 TeV       & 185         & [1.24] & [1.16] & [1.47] & [1.81]  &  1.25  &  1.52     &  1.53     &  1.54      &  1.42     &  1.42      \\    
\midrule
ATLAS dijets 7 TeV   &  90         & [1.77] & [1.98] & [2.46] & [2.55]  & [1.95] & [1.86]    & [1.86]    & [1.88]     & [1.98]    & [1.98]     \\    
CMS dijets 7 TeV     &  54         & [2.43] & [2.52] & [2.50] & [2.57]  & [2.08] & [1.90]    & [1.96]    & [1.89]     & [2.19]    & [2.17]     \\    
CMS dijets 8 TeV     & 122         & [6.70] & [7.48] & [3.53] & [3.89]  & [2.21] & [1.95]    & [2.46]    & [2.47]     & [2.96]    & [3.04]     \\       
\bottomrule
\end{tabularx}

\vspace{0.3cm}
\caption{Same as Table~\ref{tab:chi2s} for
  fits performed with alternative choices of central
  scale or alternative decorrelation models. Now only $\chi^2$ values
  for jet data are shown. Results for the fits with default settings
  \#j7, \#j7n and \#janw, already shown  in Table~\ref{tab:chi2s} are included
  for ease of reference. Note that for the fits with alternative decorrelation
  models for the ATLAS 7~TeV data
  (\#janw-7cor, \#janw-7dec and \#janw-7pcor) the number of ATLAS 7~TeV
  data is  $n_{\rm dat}=140$ instead of  $n_{\rm dat}=30$ as for all other fits.}
\label{tab:chi2_suppl}
\end{table}
\begin{figure}[!t]
\centering
\includegraphics[scale=0.49]{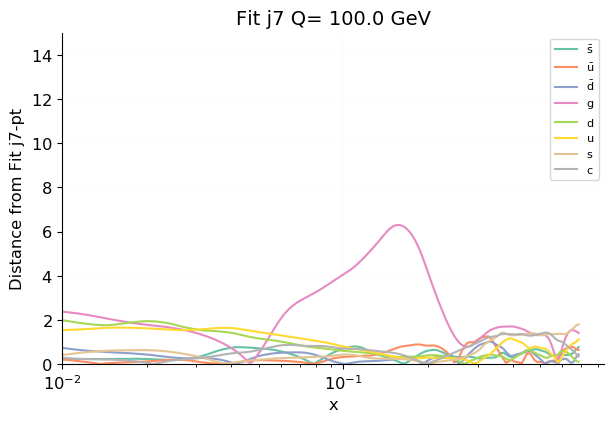}
\includegraphics[scale=0.49]{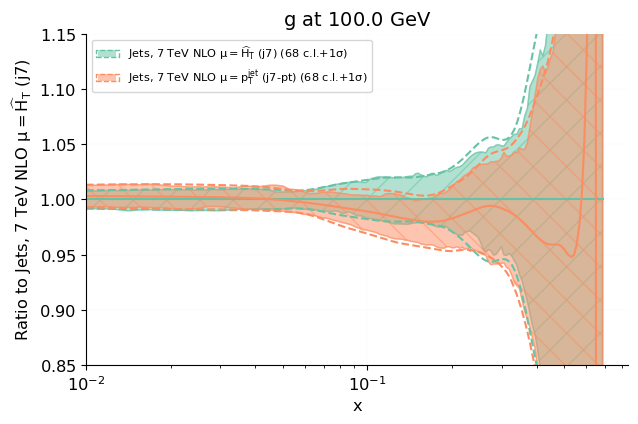}\\
\includegraphics[scale=0.49]{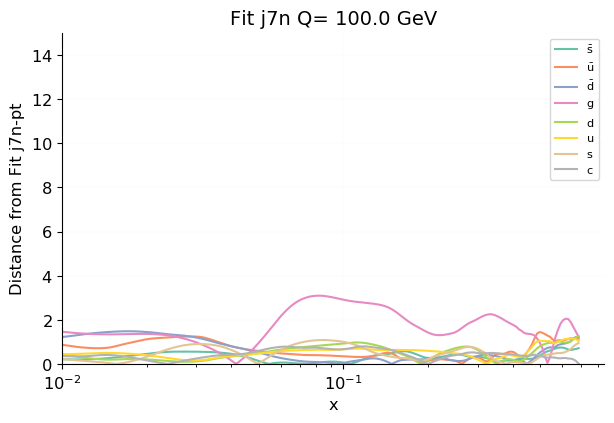}
\includegraphics[scale=0.49]{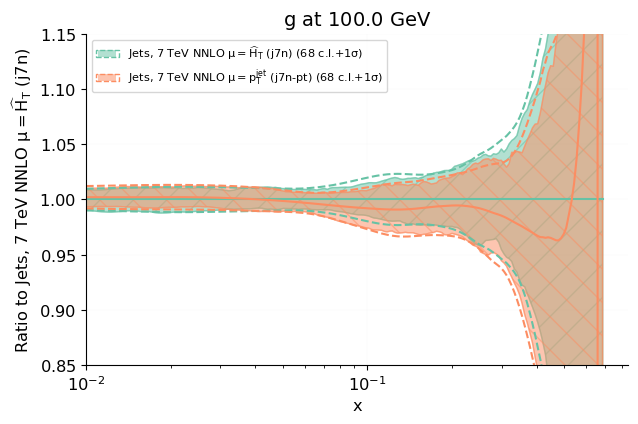}\\
\caption{Same as Fig.~\ref{fig:jet_data_total}, but now comparing
  fits to the 7~TeV data with the choices of central renormalization and
  factorization scale scale $\mu=\widehat{H}_T$ (as shown in
  Fig.~\ref{fig:jet_corrs}, top) and
  $\mu=p_T^{\rm  jet}$ at NLO 
   (fits \#j7 and \#j7-pt) and NNLO (fits  \#j7n and \#j7n-pt).  The gluon
is  shown as ratio to the fits with $\mu=\widehat{H}_T$.}
\label{fig:jet_data_scale}
\end{figure}

\subsubsection{Impact of the choice of scale}
\label{sec:scale}

We now turn to an assessment of the impact of the choice of central
scale: specifically, we compare results obtained by fitting with our default
scale choice $\mu=\widehat{H}_T$, chosen as optimal based on the
studies of Ref.~\cite{Currie:2018xkj},  and with the scale choice
$\mu=p_T^{\rm   jet}$ used in NNPDF3.1~\cite{Ball:2017nwa} and
previous NNPDF studies. For ease of comparison to
Ref.~\cite{Ball:2017nwa}, the comparison is performed for fits to the 7~TeV
data (fits  \#j7 and \#j7-pt at NLO, and  \#j7n and \#j7n-pt at
NNLO).
In Fig.~\ref{fig:jet_data_scale} we show the distance 
between PDF central values of the two pairs of fits, at NLO 
and NNLO, and compare the corresponding gluon PDFs.


Inspection of Table~\ref{tab:chi2_suppl} 
 shows that at NLO the scale choice
$\mu=\widehat{H}_T$ leads to a better description of the jet data,
 both included and not  included in the fits, with respect to
$\mu=p_T^{\rm jet}$.  However, the
effect of the scale choice on the PDFs  is very mild (see
Fig.~\ref{fig:jet_data_scale}), with a localized
modification of the gluon below the half sigma level for
$x\simeq 0.2$  and no effect on the other PDFs. On the other hand, at
NNLO the two scale choices lead to almost indistinguishable results, both in
terms of fit quality and PDF shape, with the scale choice
$\mu=\widehat{H}_T$ leading to a slightly better description of data
not included in the fit, and a difference in gluon central values
barely above statistical indistinguishability.

We conclude that the scale choice $\mu=\widehat{H}_T$  is
perturbatively more stable, in that it leads to a better NLO fit, but
that at NNLO the choice of central scale is not an issue. Both
conclusions are in agreement with the findings of  Ref.~\cite{Currie:2018xkj}.

\subsubsection{Impact of the choice of correlation models}
\label{sec:corr}

We finally discuss the impact of different correlation models on the
ATLAS single-inclusive jet data.
As repeatedly mentioned, only the central rapidity bin of
the ATLAS 7~TeV  data was  included in NNPDF3.1 and thus
in our default fit because it was not possible
to obtain a good fit when all rapidity bins were included, yet PDFs
fitted to each rapidity bin turned out to
be very close to each other~\cite{Ball:2017nwa,Nocera:2017zge}: this
suggests issues in
the covariance matrix for these data, as extensively discussed in
Ref.~\cite{Harland-Lang:2017ytb}.
Further, as shown in
Sect.~\ref{sec:impcondata}, the corresponding ATLAS 8~TeV data appear to be fully
consistent with the 7~TeV data, yet lead to a poor $\chi^2$ when
included in the global fit, which suggests that they may suffer from a
similar problem.

Here we will first check that indeed the inclusion of all
rapidity bins from the 7~TeV ATLAS data does not change the results
for the PDFs, 
as argued in Refs.~\cite{Ball:2017nwa,Nocera:2017zge}, but now by
fitting all rapidity bins simultaneously, rather than one at a time as
in Ref.~\cite{Nocera:2017zge}, and with the new scale choice and
jet dataset adopted here. We will then  address the
issue of the impact of the choice of correlation model, in particular
by decorrelating different rapidity bins as suggested in
Ref.~\cite{Harland-Lang:2017ytb} for the 7~TeV data and in
Ref.~\cite{Aaboud:2017dvo} for the 8~TeV data.

To this purpose, we have performed five variant fits of our most accurate
fit with default settings (\#janw) in which alternative treatments of 
the 7~TeV or 8~TeV ATLAS data are considered in turn (see
Table~\ref{tab:chi2_suppl}). Concerning the 7~TeV data,  in a first
fit (\#janw-7cor) all ATLAS 
rapidity bins are included: so in this fit the 7~TeV and 8~TeV are
treated on an equal footing, with all bins included and correlated
systematics treated using the published covariance matrix. The
correlation pattern is then modified: 
in fit \#janw-7uncor systematics are assumed to be
uncorrelated across rapidity bins, and in fit  \#janw-7pcor systematics
are partially decorrelated, following the prescription suggested in
Ref.~\cite{Harland-Lang:2017ytb}. Concerning the 8~TeV data, we start
with the default fit (\#janw), and we obtain from it two variants by
modifying, as suggested in Ref.~\cite{Aaboud:2017dvo},
the treatment of   three (out of 659) correlated 
systematic uncertainties, related to the jet energy scale, specifically
  to the flavour response, the fragmentation and the pile-up. In 
  fit \#janw-8dec these
  three uncertainties are completely decorrelated; in 
  fit \#janw-8pcor they are partly decorrelated by splitting each
  uncertainty into three components and decorrelating one of them (see
  Tab.~6 or Ref.~\cite{Aaboud:2017dvo}).

\begin{figure}[!t]
\centering
\includegraphics[scale=0.49]{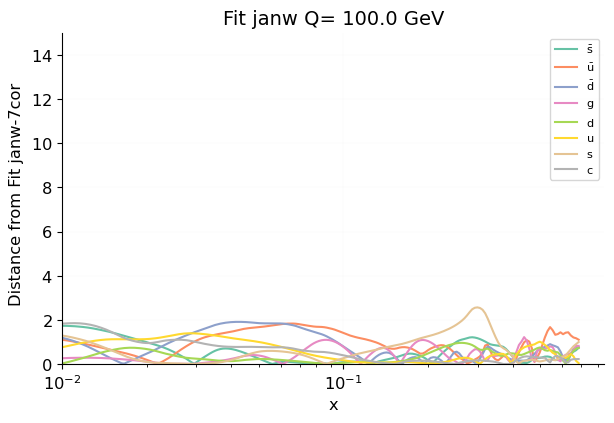}
\includegraphics[scale=0.49]{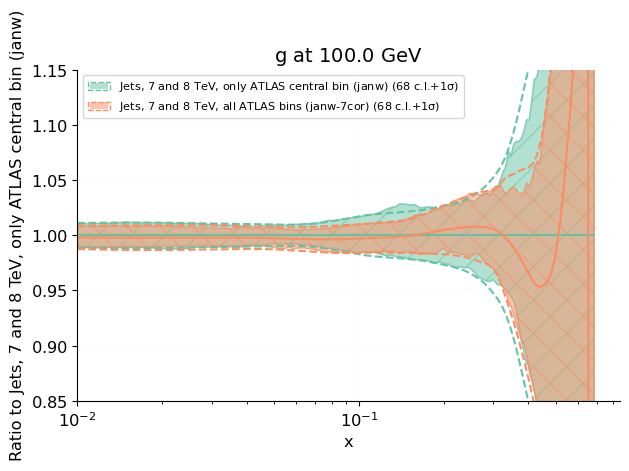}\\
\caption{Same as Fig.~\ref{fig:jet_data_total}, but now comparing
  the default most accurate fit to single-inclusive jet data (all datasets,
  NNLO QCD+EW, fit \#janw), in which only the central rapidity bin of
  the ATLAS 7~TeV data is included, to a fit in which all rapidity
  bins are included (\#janw-7cor). The gluon
  is  shown as ratio to the former fit.}
\label{fig:jet_data_model} 
\end{figure}

The fit with only the central rapidity bin of the ATLAS
7~TeV data (\#janw), and the fit in which all ATLAS data are included,
with fully correlated systematics (\#janw-cor), are compared in 
Fig.~\ref{fig:jet_data_model}, where we show distances
between the two sets of PDFs, and we compare directly the gluon PDFs,
shown as a ratio to the default optimal fit. It is clear that all PDFs
including the gluon are essentially unchanged. On the other hand, the
$\chi^2$ now increases very substantially. However, the
$\chi^2$ to all other jet and dijet data (both fitted and not fitted) is
essentially unchanged, consistently with the fact that the gluon is
very stable. In short, we confirm the previous
conclusion~\cite{Ball:2017nwa,Nocera:2017zge}  that including all
rapidity bins of 
the ATLAS 7~TeV data has almost no impact on the PDFs, despite the
considerable deterioration of the $\chi^2$ for this data.

\begin{figure}[!t]
\centering
\includegraphics[scale=0.49]{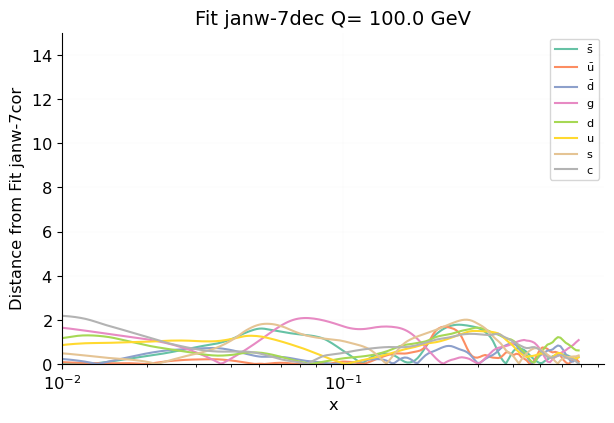}
\includegraphics[scale=0.49]{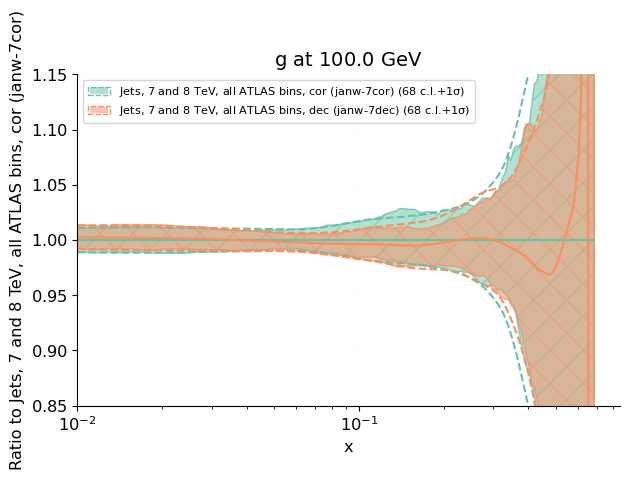}\\
\caption{Same as Fig.~\ref{fig:jet_data_total}, but now comparing the fits to
  single-inclusive jet data (all datasets, NNLO QCD+EW), in which all rapidity
  bins of the ATLAS 7~TeV data are included, either with (fit \#janw-7cor) or 
  without (fit \#janw-7dec) experimental correlations on selected systematic
  uncertainties. The gluon is  shown as ratio to the former fit.}
\label{fig:jet_data_model_7TeV} 
\end{figure}

\begin{figure}[!t]
\centering
\includegraphics[scale=0.49]{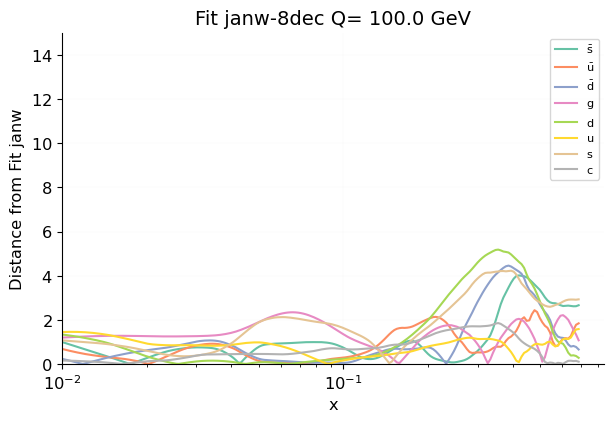}
\includegraphics[scale=0.49]{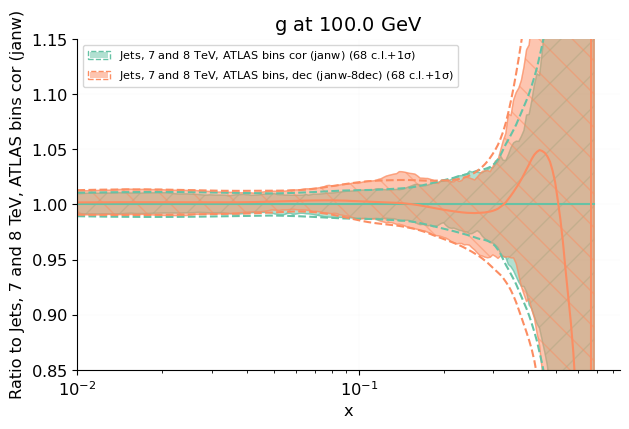}\\
\caption{Same as Fig.~\ref{fig:jet_data_total}, but now comparing
  the default most accurate fit to single-inclusive jet data (all datasets,
  NNLO QCD+EW, fit \#janw), to a fit in which selected systematic uncertainties
  are decorrelated in the ATLAS 8~TeV data (fit \#janw-8dec). The gluon
  is  shown as ratio to the former fit.}
\label{fig:jet_data_model_8TeV} 
\end{figure}

Inspection of the $\chi^2$ values from Table~\ref{tab:chi2_suppl}
further reveals that as soon as systematics are decorrelated the
$\chi^2$ value improves considerably. However,
once again the
$\chi^2$ to all other jet and dijet data changes very little. Hence
once again we conclude that the PDFs are stable upon decorrelation. In
fact, all $\chi^2$ values remain essentially the same regardless of
whether correlations are completely removed, or only partially
removed following the suggestion of
Ref.~\cite{Harland-Lang:2017ytb}, thereby validating the conclusion
that these correlations are the problematic ones.
The stability of PDFs in general and the gluon in particular
upon decorrelation is confirmed by a direct comparison, shown in
Fig.~\ref{fig:jet_data_model}. Distances between all PDFs before and after
decorrelation  is
seen to be compatible with statistical fluctuations.

Turning now to the 8~TeV data, a similar pattern is found. Namely,
upon decorrelation the $\chi^2$ for the ATLAS data improves
considerably, but $\chi^2$ values for all other datasets are almost
unaffected, with very similar results obtained when fully or partially
decorrelating the relevant sources of systematics following
Ref.~\cite{Aaboud:2017dvo}, thus validating the prescription of this
reference. Also in this case, the stability of the PDFs is confirmed
by direct comparison in Fig.~\ref{fig:jet_data_model_8TeV}.

In summary, firstly,  we confirm the conclusion of
Refs.~\cite{Ball:2017nwa,Nocera:2017zge} that inclusion of all of the
ATLAS 7~TeV jet data with full correlations has a significant impact on the
fit quality but not on the PDFs. Furthermore, we confirm
that the correlation model suggested in
Ref.~\cite{Harland-Lang:2017ytb} leads to a good description of this
data, without any significant change in the PDFs when the
decorrelation is performed. And finally, we find that the ATLAS 8~TeV
data behave in a very similar way, and in particular that the
correlation model suggested in Ref.~\cite{Aaboud:2017dvo} leads to
good fit quality without significant change in PDFs.

\subsection{PDF fits with dijet data}
\label{sec:resdijets}

We now turn to PDF fits in which dijet data rather that
single-inclusive jet data are included. Also in this case, we first
discuss the impact and compatibility of these data, and then the
perturbative stability of results.

\subsubsection{Impact and consistency of datasets}
\label{sec:impcondatad}

We assess the impact of dijet data on PDFs by
comparing fits with optimal settings, i.e. with NNLO QCD theory, and
EW corrections included (for jets only), and either the
full dataset (\#danw), or the 7~TeV (\#d7nw) or 8~TeV (\#d8nw) data included in turn.

\begin{figure}[!t]
\centering
\includegraphics[scale=0.49]{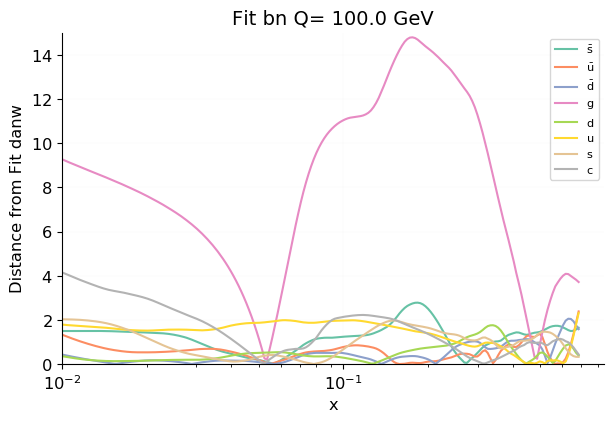}
\includegraphics[scale=0.49]{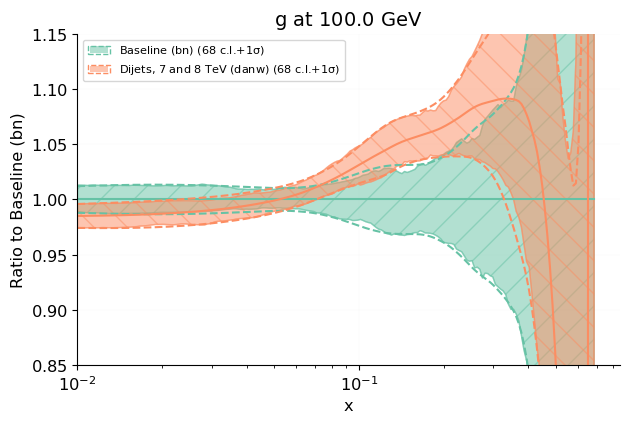}\\
\caption{Same as Fig.~\ref{fig:jet_data_total}, but now for dijets.}
\label{fig:dijet_data_total}
\end{figure}

We start by comparing to the baseline \#bn, with no jet data, fit
\#danw in which all dijet data are included; PDFs are compared in Fig.~\ref{fig:dijet_data_total}.
From Table~\ref{tab:chi2sD}, we see that  
individual dijet datasets are overall fairly well described (the $\chi^2$
per datapoint is around 1.5 for each of them). Inclusion of the dijet
datasets in the baseline leads to an improved description of
single-inclusive jet data, just like (see Sect.~\ref{sec:resincljets})
inclusion of single-inclusive jet data leads to an improved
description of dijets. This confirms consistency of the
single-inclusive and dijet data. Unlike in the case of
single-inclusive jet data, no tension is observed between dijet data
and the rest of the global dataset (specifically top rapidity
distributions), whose $\chi^2$ is left almost unchanged.

As in the case of single-inclusive jets, only the gluon PDF is
affected by the inclusion of dijet data, with the strongest impact
observed in the regions  $x\simeq 0.01$ and  $0.06\lesssim x \lesssim 0.4$
(see  Fig.~\ref{fig:dijet_data_total}). In the former region the gluon
is suppressed by about  2\%, corresponding to a shift in
central value by about one sigma; in the latter it is enhanced by up
to 10\% around $x\sim 0.3$ , corresponding
to a  shift by about
one and a half sigma, hence outside the error band of the baseline.
These shifts are qualitatively similar to those observed upon
inclusion of the single-inclusive jet data, but somewhat more
pronounced and in a somewhat wider kinematic region.

\begin{figure}[!t]
\centering
\includegraphics[scale=0.49]{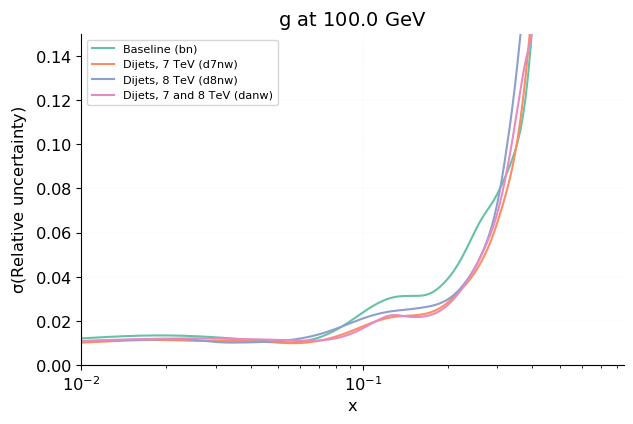}
\includegraphics[scale=0.49]{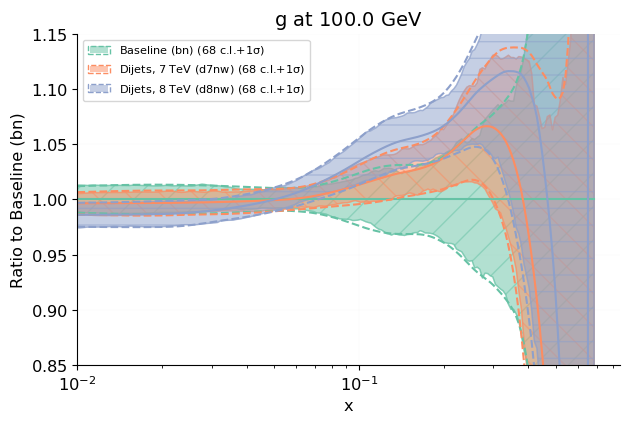}
\caption{Same as Fig.~\ref{fig:jet_data_partial}, but now for dijets.}
\label{fig:dijet_data_partial}
\end{figure}

We then turn to the assessment of the relative impact of different
datasets, by comparing to  the baseline (\#bn) the fits in which
 only  7~TeV  (\#d7nw) or only 8~TeV  (\#d8nw) dijet data  are
 included, see
 Fig.~\ref{fig:dijet_data_partial}.
 From Table~\ref{tab:chi2sD}, we see that the fit quality is equally
 good for  7~TeV or 8~TeV data, however
the fit to the 8~TeV dijet data is closer to the fit in which all
dijet data are included, in that it leads to a similar description of all of the
jet and dijet data, including those that are not included in either fit.
Such a description is better in both fits than in the fit to the 7~TeV dijet 
datasets only, and is accompanied by a similar change in the description of the
ATLAS top pair differential rapidity distributions. This suggests that
among the dijet data, the 8~TeV  data provide the dominant contribution.

The relative impact of the 7~TeV and 8~TeV data on the gluon central values
and uncertainty can be directly inferred from
Fig.~\ref{fig:dijet_data_partial}. The impact of the two datasets on
the gluon central value is
qualitatively the same, and thus also the same as that of the full
dijet dataset, but with the 8~TeV data having a stronger impact,
almost equivalent to the impact of the full dataset.
The reduction in uncertainty in comparison to the baseline
due to either dataset is almost the same, with a slightly stronger
reduction observed for the 7~TeV data,
by about  3-4\% to 3\% at $x\simeq 0.2$. Consequently,
the gluon PDF determined when including all of the dijet data is very
close to that found when including only the 8~TeV data, thus
confirming that the 8~TeV data have a dominant impact on the gluon
central value.

\begin{figure}[!t]
\centering
\includegraphics[scale=0.49]{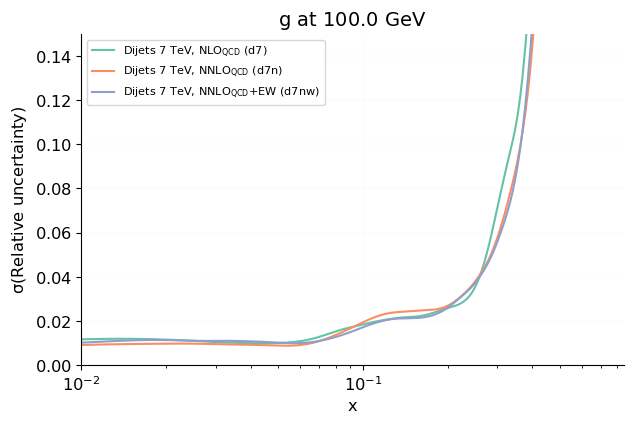}
\includegraphics[scale=0.49]{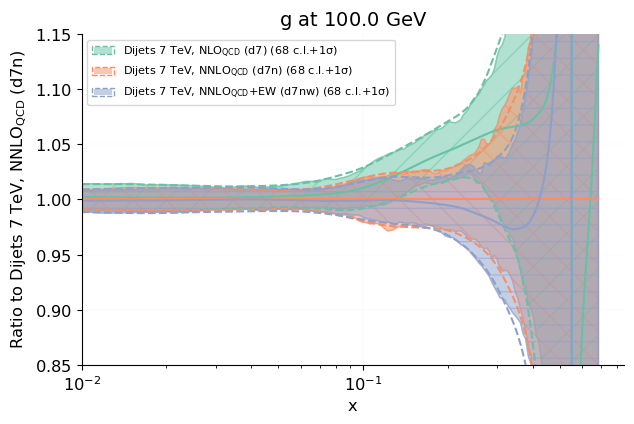}\\
\includegraphics[scale=0.49]{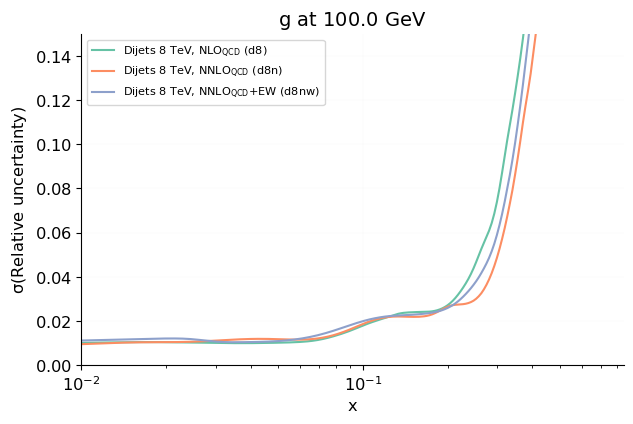}
\includegraphics[scale=0.49]{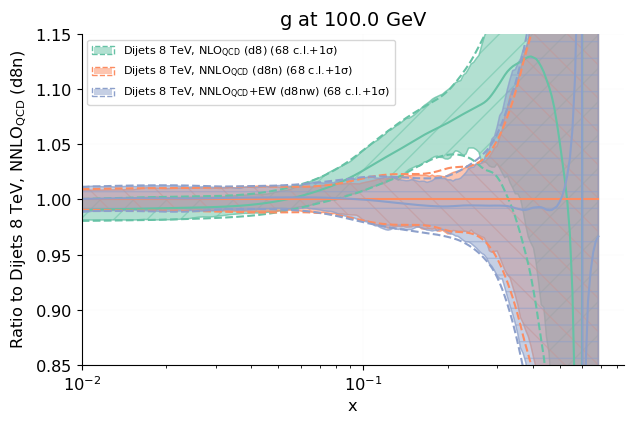} \\
\caption{Same as Fig.~\ref{fig:jet_corrs}, but now for dijets.}
\label{fig:dijet_corrs}
\end{figure}
\subsubsection{Impact of higher order QCD and EW corrections}
\label{sec:hoimpd}

As for single-inclusive jets, we assess the perturbative stability of
fits with dijet data by comparing fits at NLO, NNLO and with EW
corrections, separately for the 7~TeV and  8~TeV data, i.e.,
respectively, fits
\#d7, \#d7n,  \#d7nw; and  \#d8, \#d8n,  \#d8nw. The gluon PDFs for
these fits are compared in Fig.~\ref{fig:dijet_corrs}.

The figure shows that the perturbative behavior of the gluon upon
inclusion of the dijet data is very similar to what observed when
including single-inclusive jets. Namely,
at NLO the gluon is distorted in the region
$0.1\lesssim x \lesssim 0.5$ in comparison to the NNLO results already shown in
Fig.~\ref{fig:dijet_data_partial}, with the effect more pronounced for the
dominant and more precise 8~TeV data, again providing evidence for
large missing NNLO corrections. 
The effect of the EW  corrections is even less marked than in the case
of single-inclusive jets: in fact, their inclusion leaves the gluon
PDF almost unchanged.
For both datasets, inclusion of the NNLO corrections leads to a
reduction in uncertainty, more marked for 8~TeV data, while inclusion
of the EW corrections has no clear effect; in fact, for the 8~TeV data
it leads to a slight increase of the uncertainty. As in the case of
single-inclusive jets, we conclude that NNLO corrections have a strong
impact by modifying the gluon shape and reducing its uncertainty,
while EW corrections have essentially no impact.

Unlike in the case of single-inclusive jets, where the inclusion of
NNLO corrections did not have a clear impact on fit quality, for dijets
at NNLO there is a  clear improvement in $\chi^2$ values (see
Table~\ref{tab:chi2sD}). Specifically, when all dijet data are included
at NLO, the $\chi^2$ of the global fit deteriorates significantly in
comparison to the baseline, with the largest effect seen in data which
are most sensitive to the gluon, such as the $Z$ $p_T$ distribution
and the top rapidity distribution. This deterioration goes away upon
inclusion of NNLO corrections.
Indeed, when
NNLO corrections are included, the quality of the global fit including
dijets improves considerably, corresponding now  to a fit quality which is
essentially the same for the fits with or without the dijet data.
Accordingly, the fit quality to
the dijet data is significantly better at NNLO than at NLO.
 The effect is driven by the more precise 8~TeV data. Indeed, the same
 pattern is observed 
 when only 8~TeV data are included, while with 7~TeV data only
fit quality to the dijet data
at NLO and NNLO is essentially the same, and so is the fit quality
with or without dijet data.

This means that  inclusion of NNLO corrections is crucial in order to
ensure compatibility of the dijet data with the rest of the global
dataset. Interestingly, when fitting dijet data
no clear improvement in the fit quality of single-inclusive jet data
(not fitted) is seen when going from NLO to
NNLO  .
Inclusion of EW corrections has no
significant effect on fit quality.

We conclude that for dijets NNLO corrections have a significant impact
on both fit quality, the central value of the gluon PDF and its
uncertainty, with a clear pattern of improvement when going from NLO
to NNLO.

\subsection{Single-inclusive jets vs. dijets: a comparative assessment}
\label{sec:comparisonjets}

\begin{figure}[!t]
\centering
\includegraphics[scale=0.49]{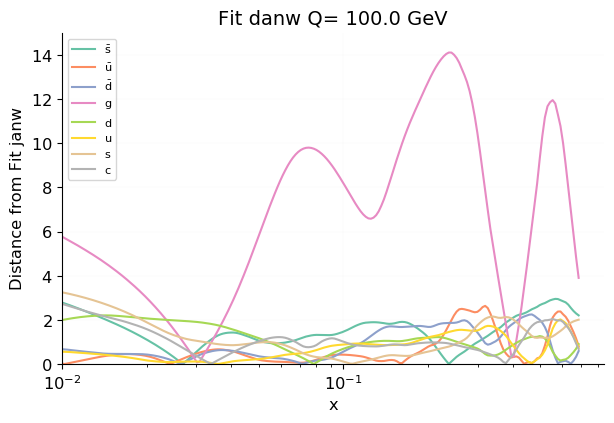}
\includegraphics[scale=0.49]{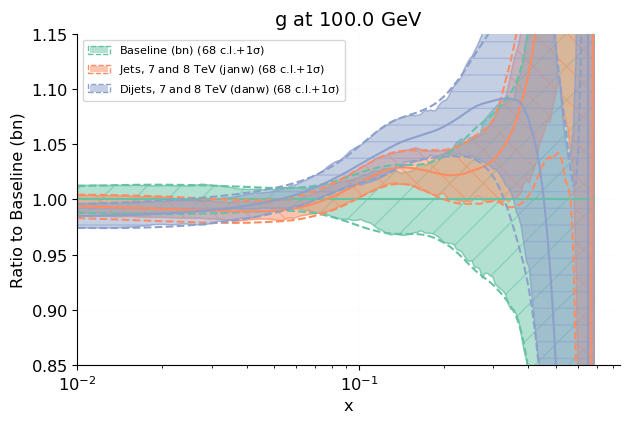}\\
\caption{Same as Fig.~\ref{fig:jet_data_total}, but now comparing the
  fits with  all single-inclusive jet data (\#janw), and that with all
  dijet data (\#danw) and highest theory accuracy (NNLO QCD+ EW) and
  default settings. In the gluon comparison (right) results are
  displayed as a ratio to the baseline with no jet data included (also
  shown for reference).}
\label{fig:jet_dijet_1}
\end{figure}

\begin{figure}[!t]
\centering
\includegraphics[scale=0.49]{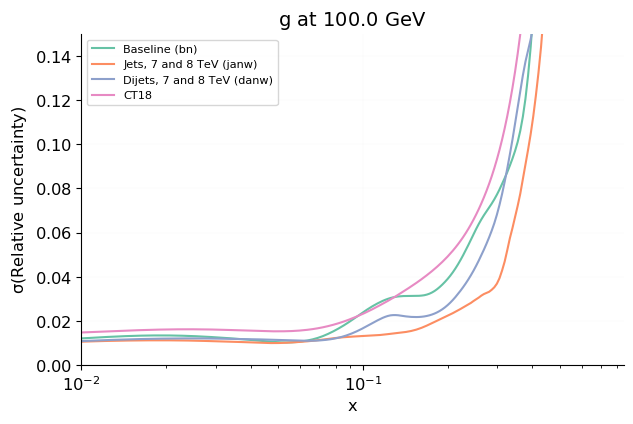}
\includegraphics[scale=0.49]{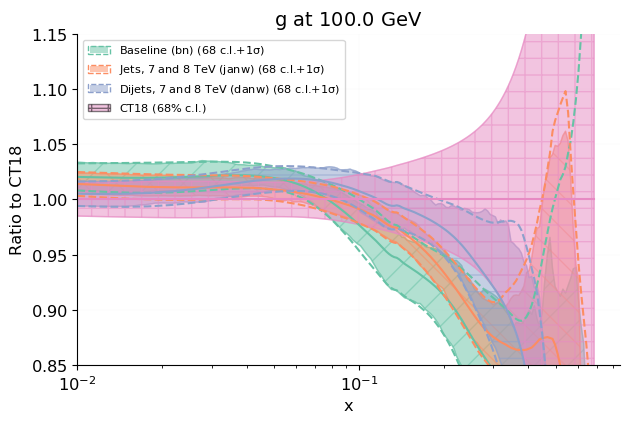}\\
\caption{Same as Fig.~\ref{fig:jet_data_partial},
but now comparing the baseline (\#bn) to the fits with all
 single-inclusive jet (\#janw) and 
  dijet data (\#danw) of Fig.~\ref{fig:jet_dijet_1}. All results are
  shown as a ratio to the CT18 fit (also
  shown for reference).}
\label{fig:jet_dijet_2}
\end{figure}

\begin{figure}[!t]
\centering
\includegraphics[scale=0.46]{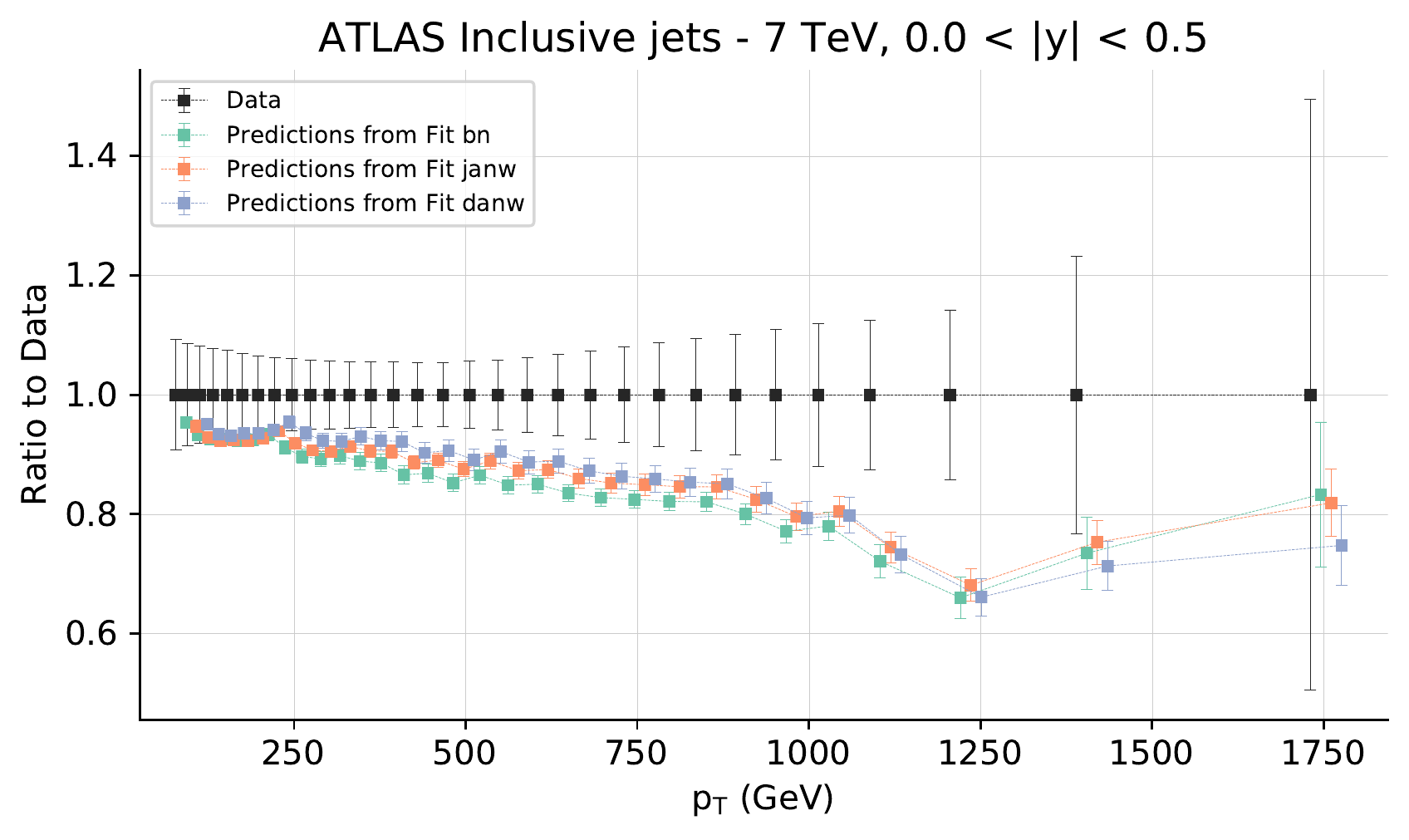}
\includegraphics[scale=0.46]{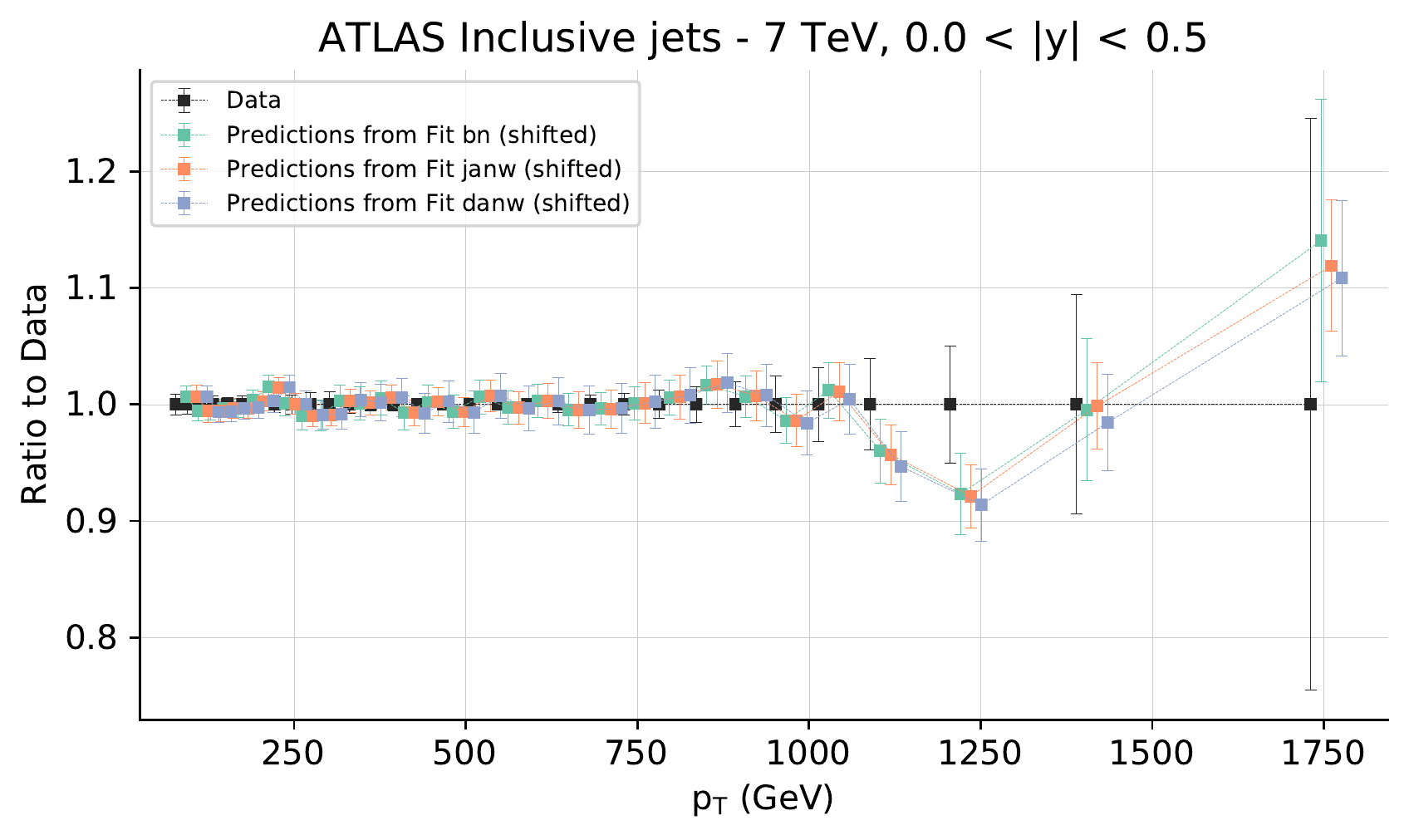}\\
\includegraphics[scale=0.46]{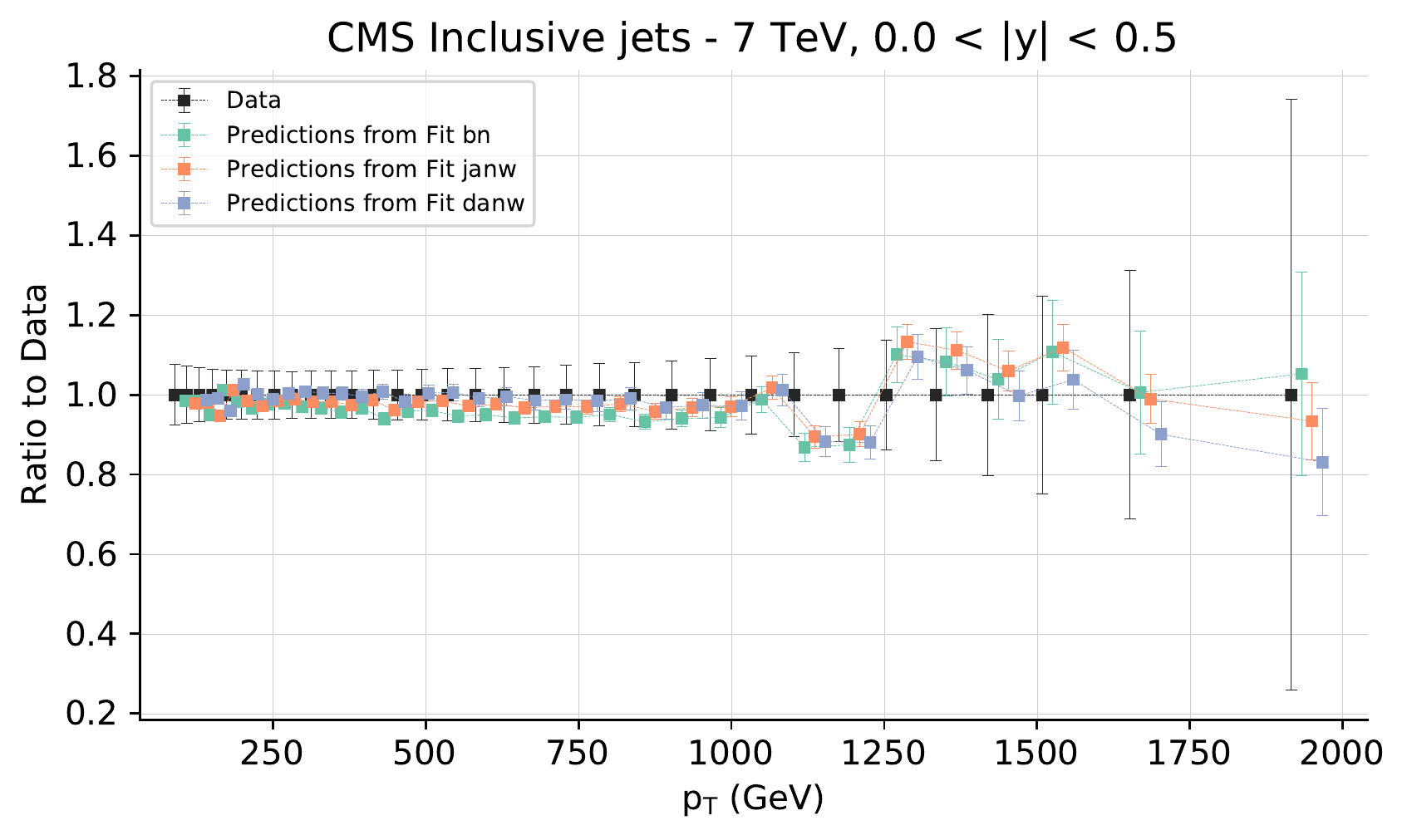}
\includegraphics[scale=0.46]{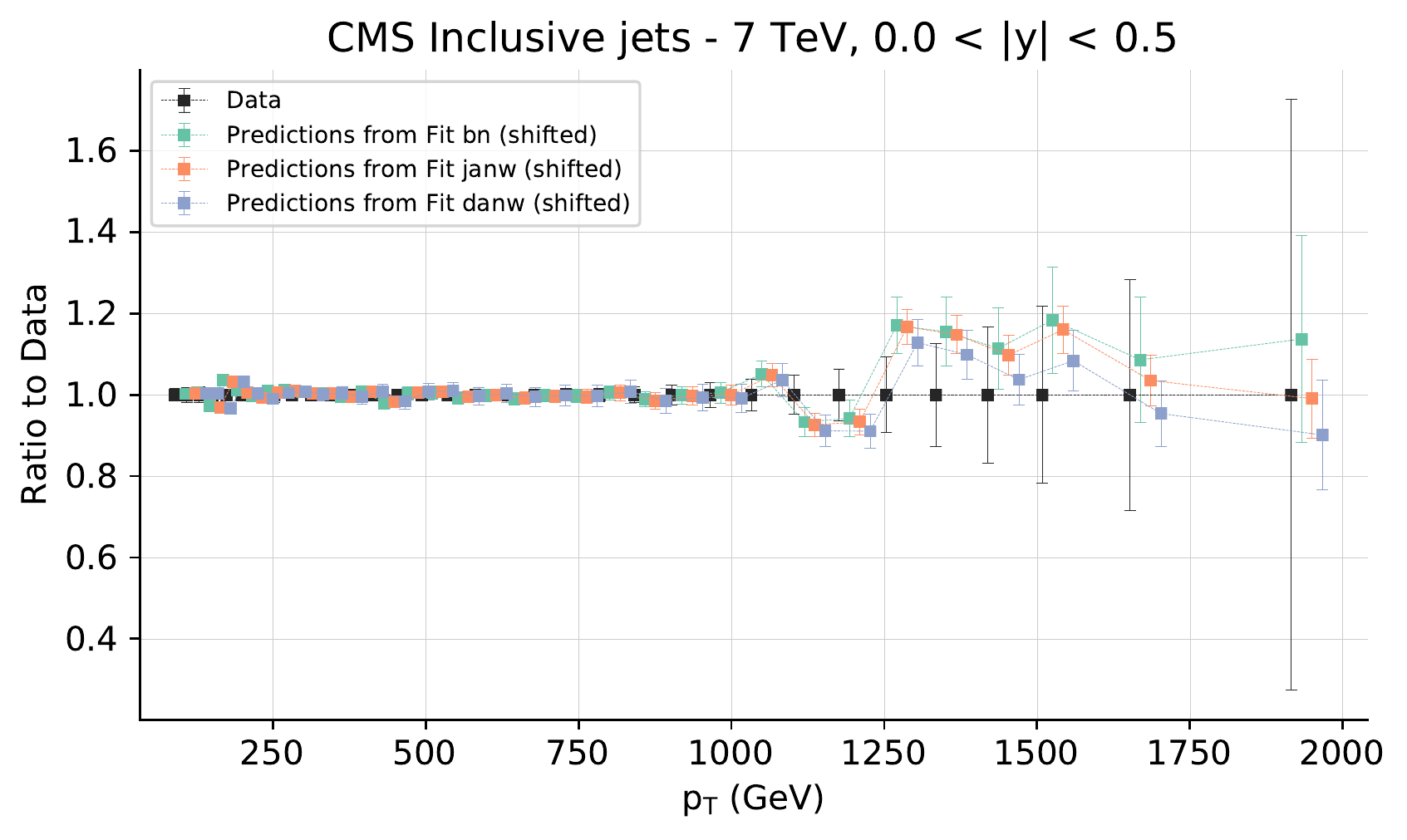}\\
\includegraphics[scale=0.46]{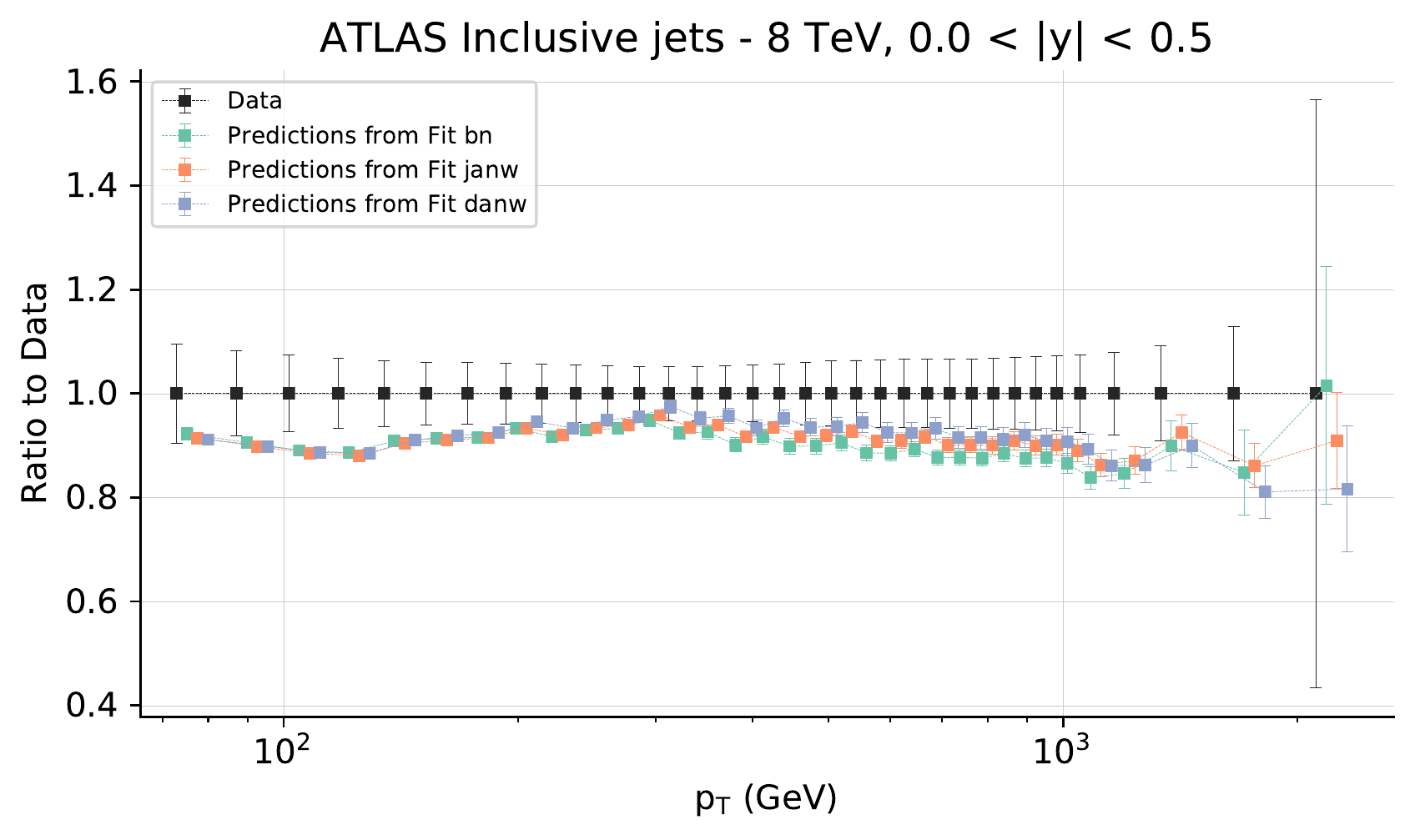}
\includegraphics[scale=0.46]{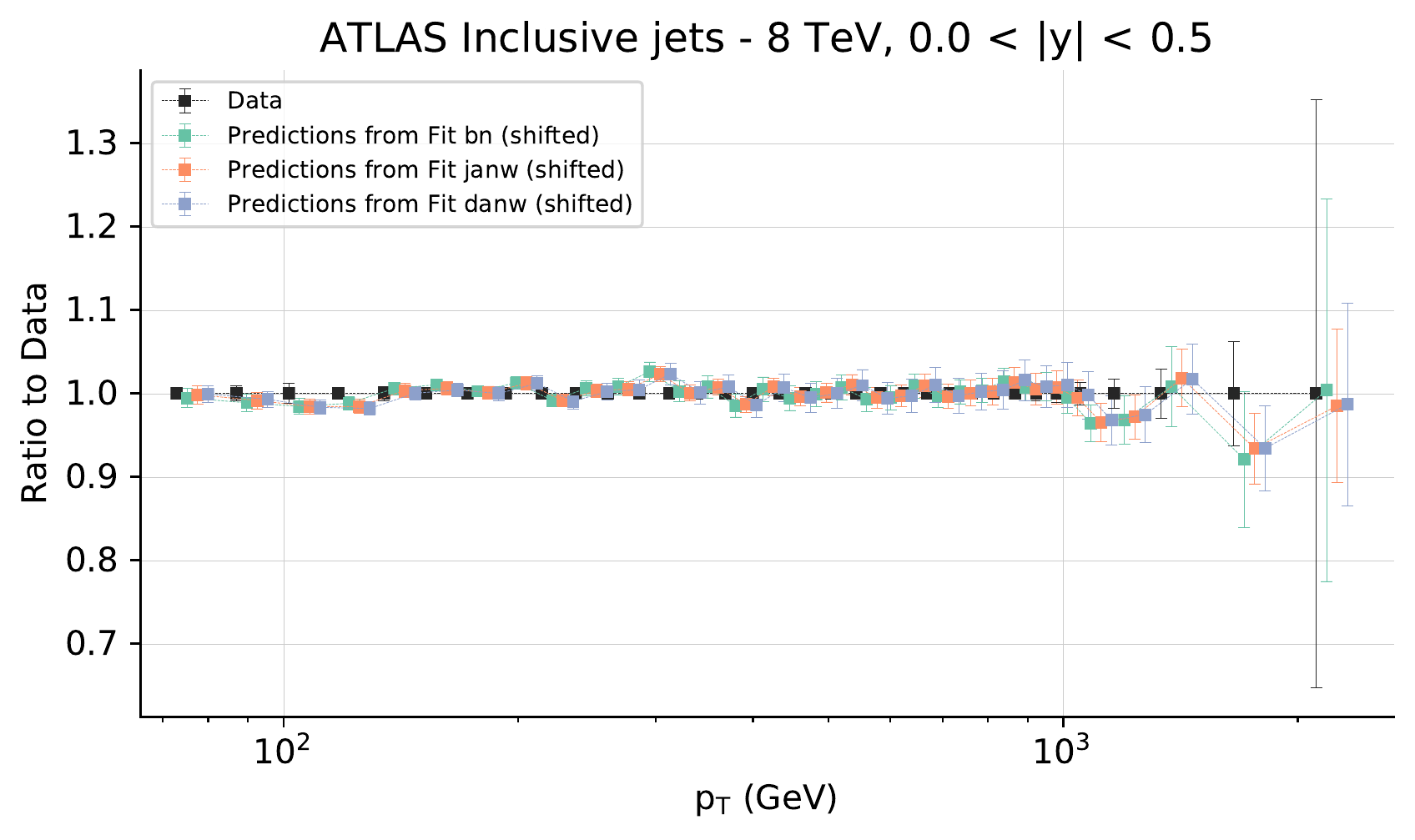}\\
\includegraphics[scale=0.46]{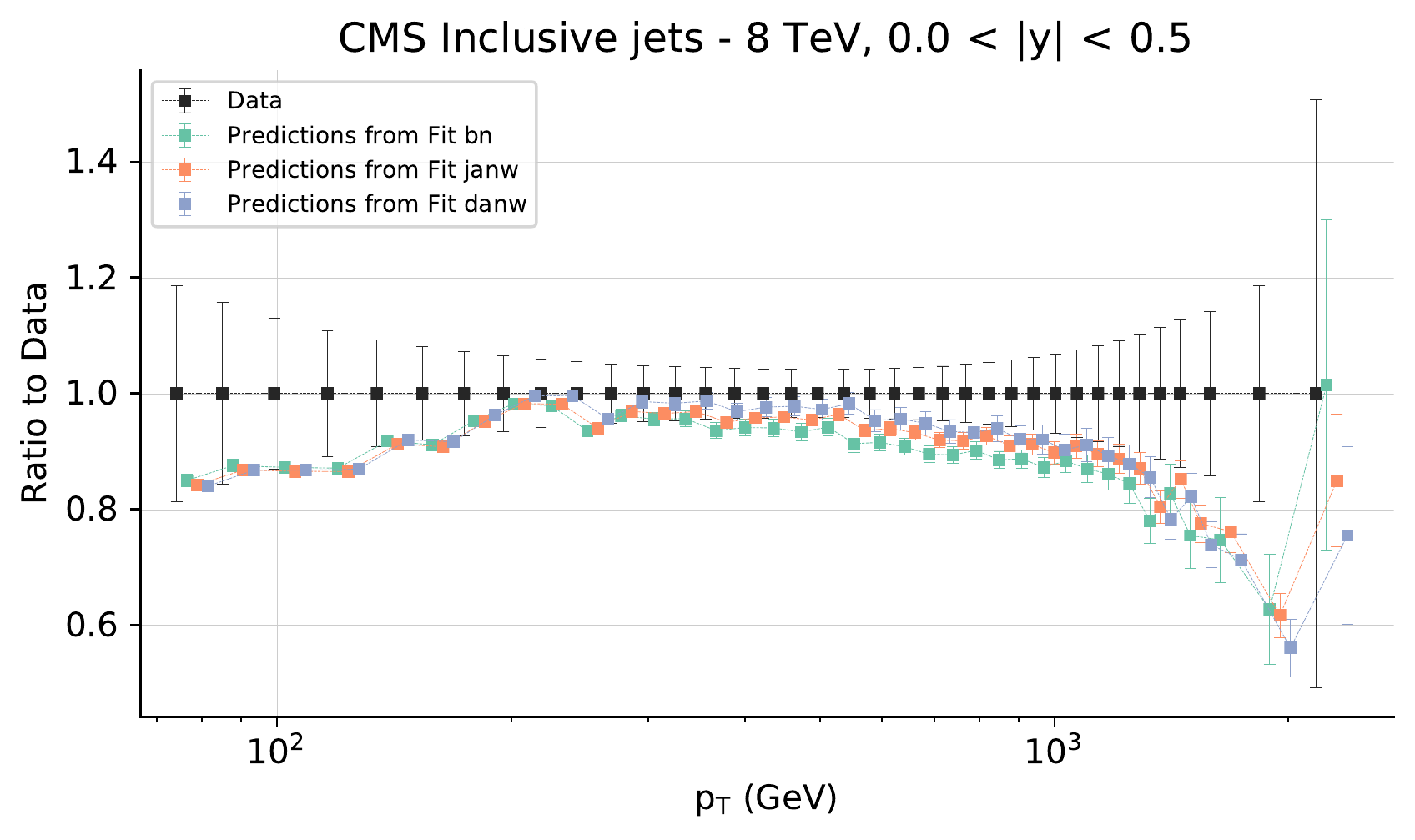}
\includegraphics[scale=0.46]{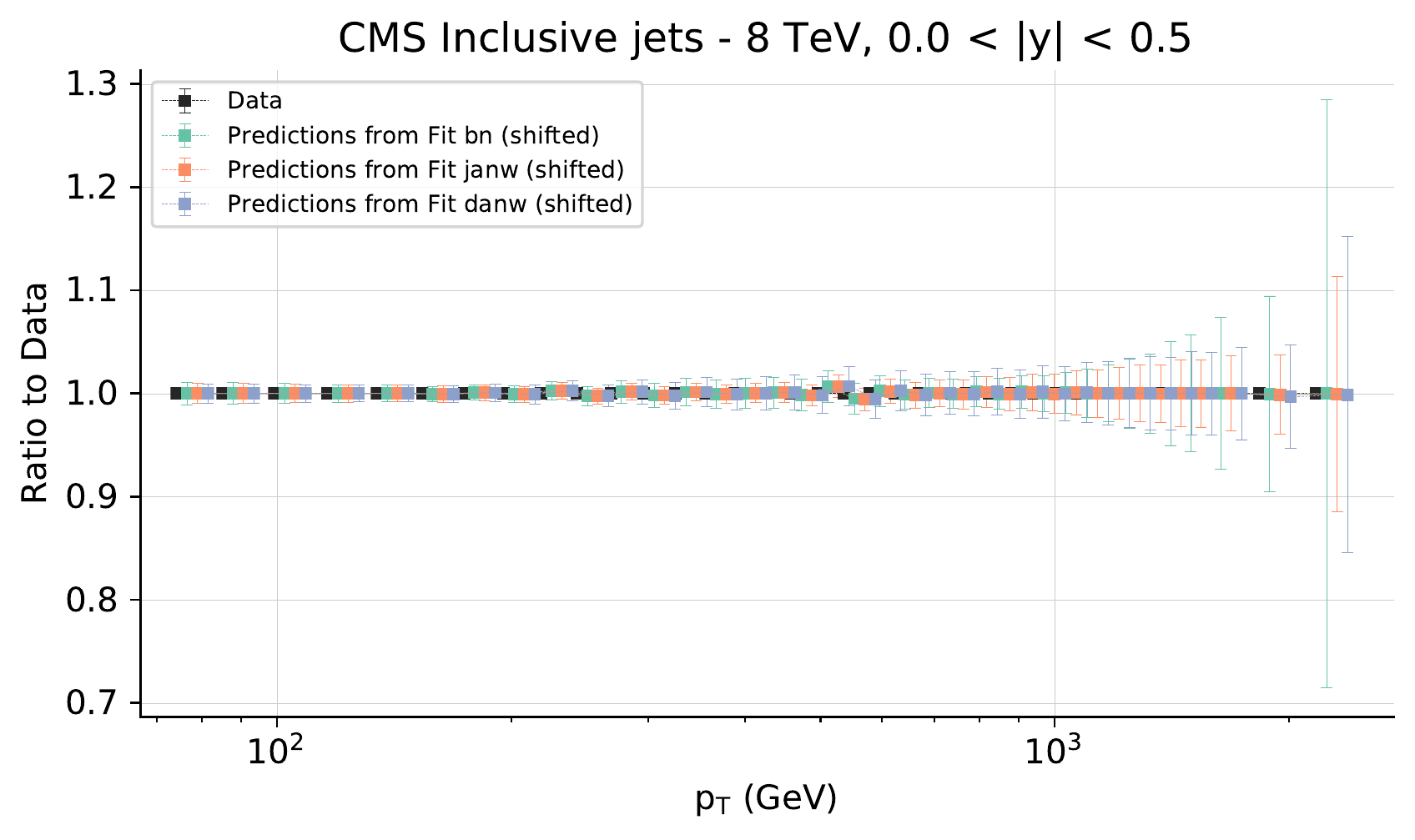}\\
\caption{The theory to data ratio for a representative set of points from
each single-inclusive  jet dataset introduced in Sect.~\ref{sec:expdata}.
  Specifically, we show the central rapidity bins for all the ATLAS and CMS jet
  datasets at 7 and 8 TeV. Theoretical predictions are computed from fits \#bn,
  \#janw and \#danw with corresponding theoretical accuracy, and shown
  as a ratio to the experimental data. For the data, either the full
  uncertainty is shown (left) or only the uncorrelated uncertainty,
  with the correlated uncertainty kept into account as a shift of the
  datapoint (right).}
\label{fig:datatheory_jets}
\end{figure}

\begin{figure}[!t]
\centering
\includegraphics[scale=0.46]{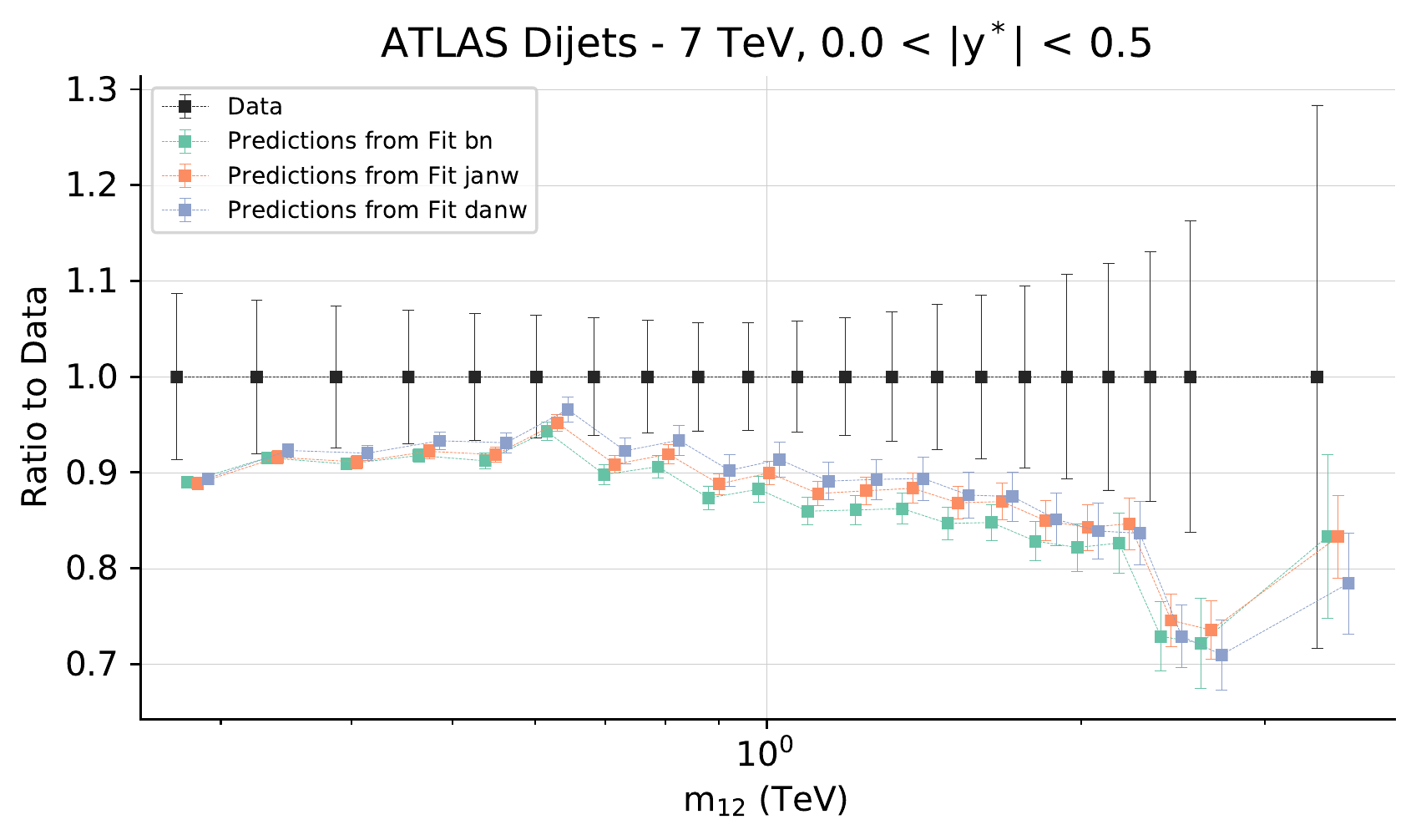}\\
\includegraphics[scale=0.46]{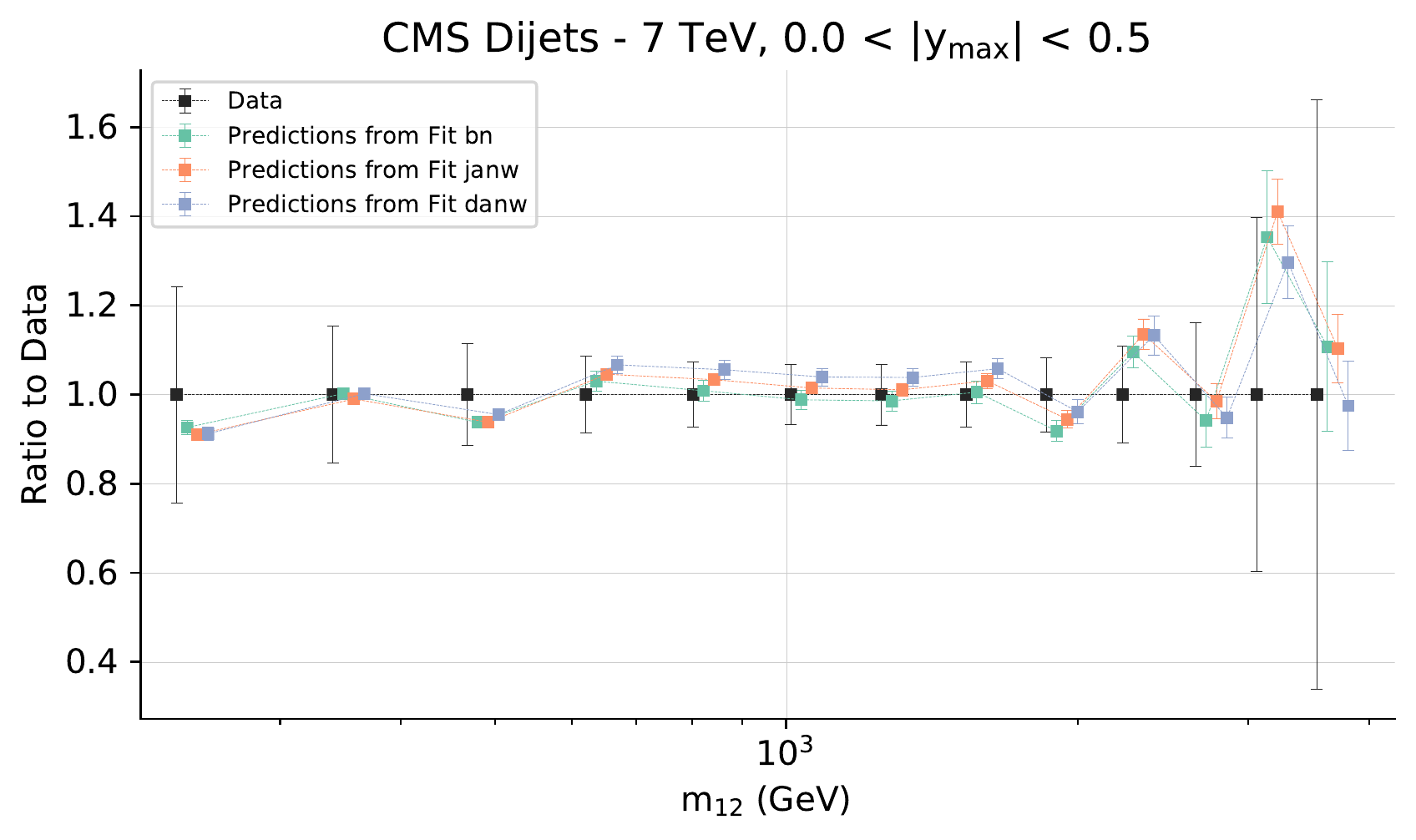}
\includegraphics[scale=0.46]{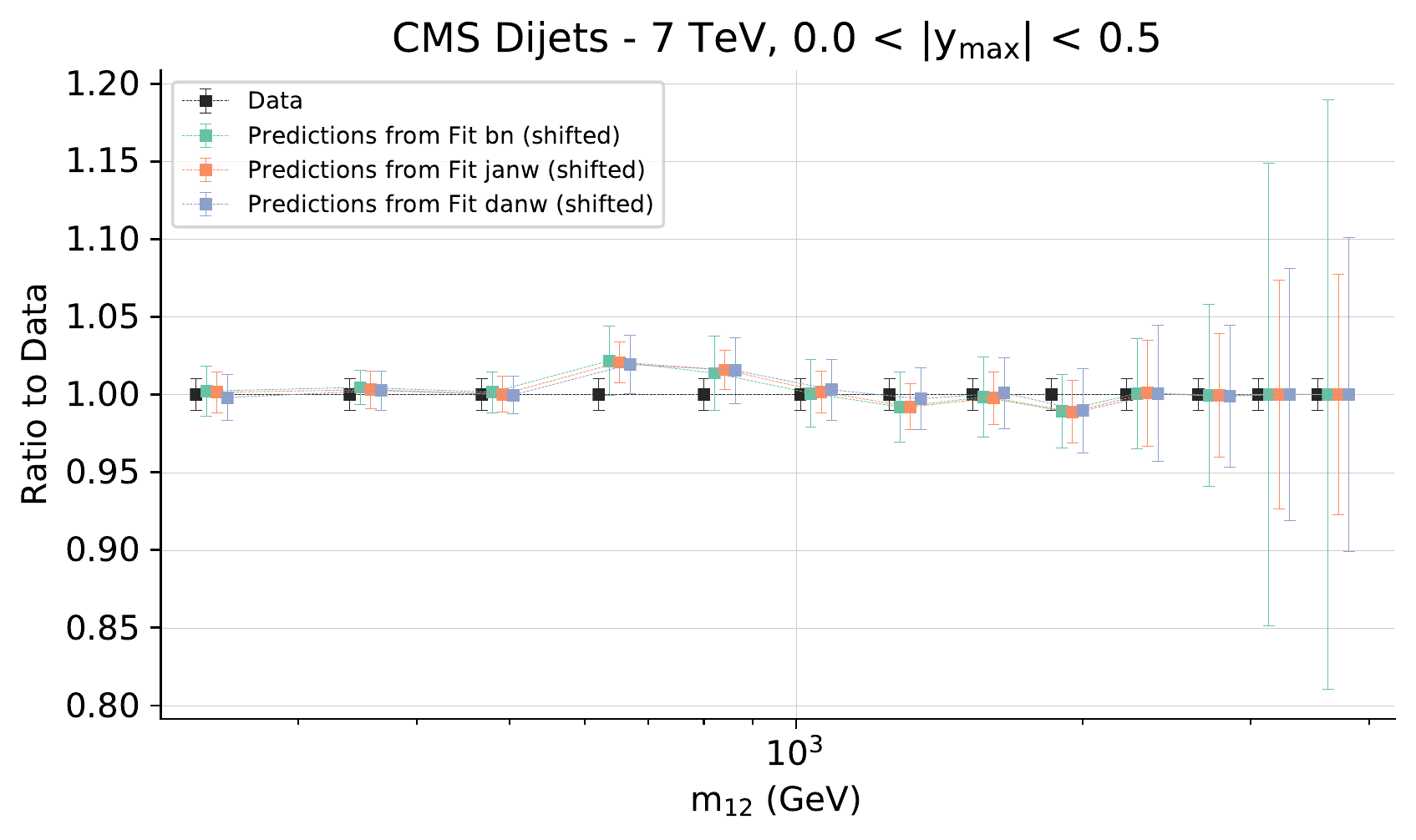}\\
\includegraphics[scale=0.46]{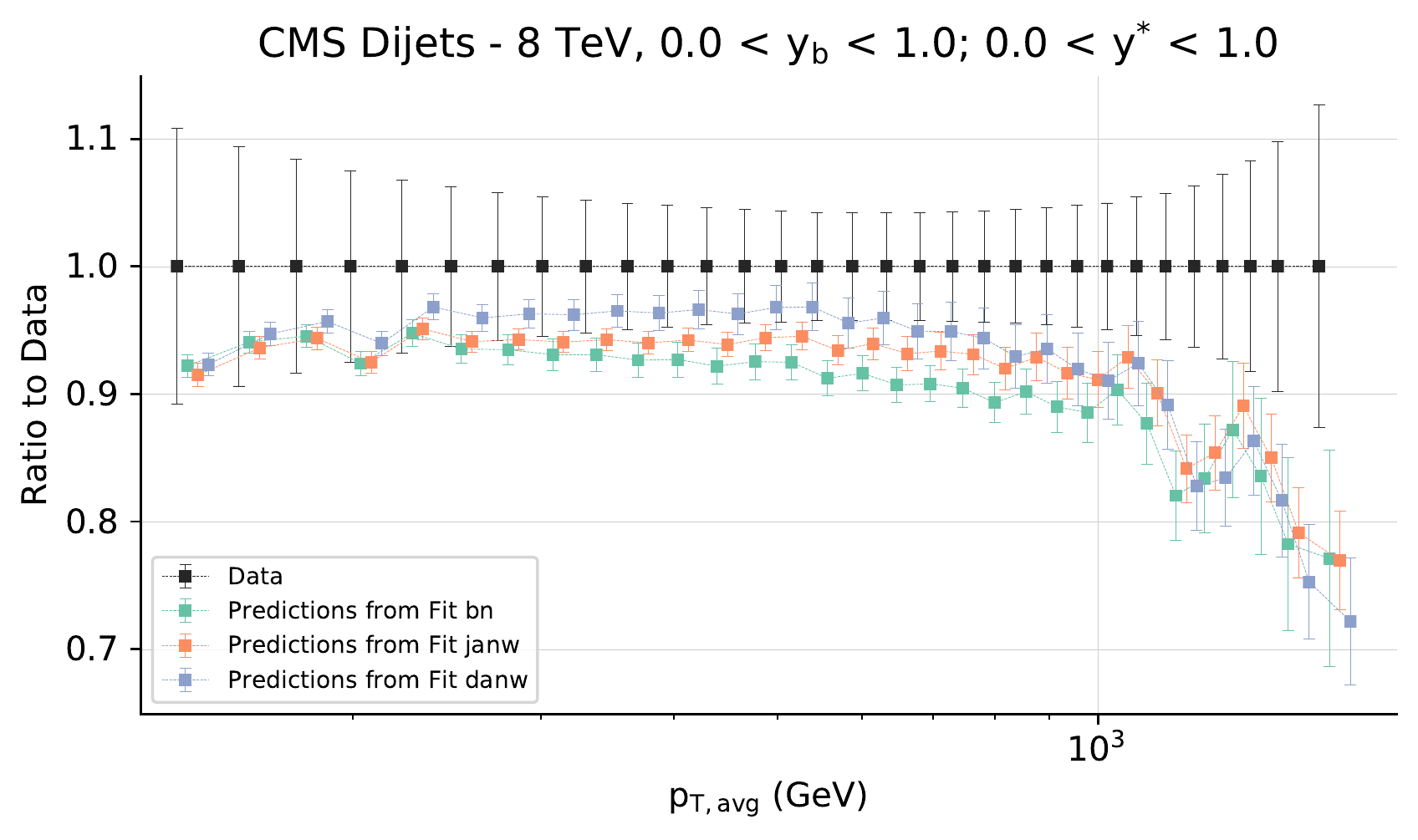}
\includegraphics[scale=0.46]{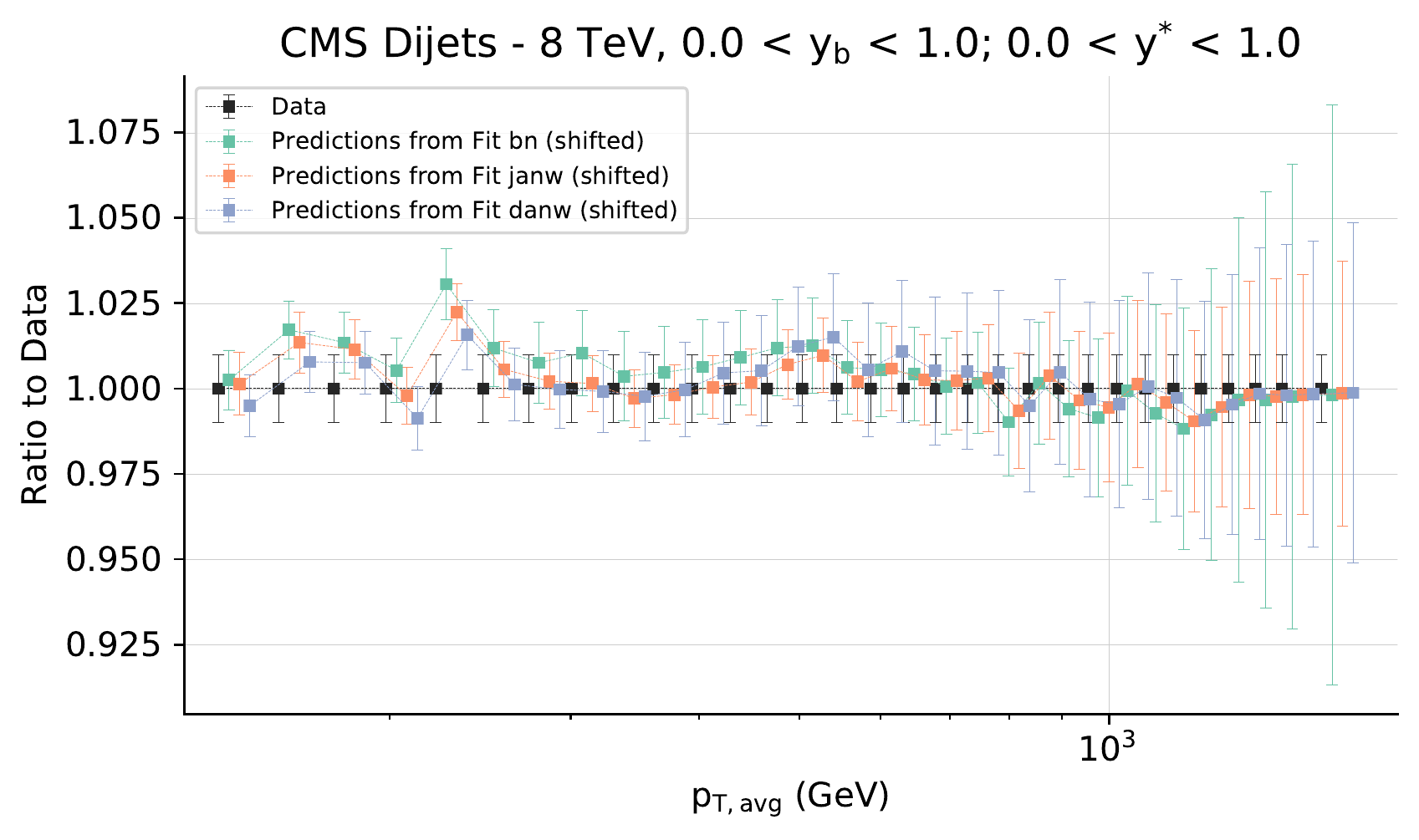}\\
\caption{Same as Fig.~\ref{fig:datatheory_jets}, but for dijets. For
  the ATLAS data only the unshifted data are shown because the breakdown
  of the statistical covariance matrix into fully uncorrelated components
  is not available.}
\label{fig:datatheory_dijets}
\end{figure}

Having assessed the impact on PDFs of jet and dijet datasets
separately, we now assess them comparatively, in terms of perturbative
stability, fit quality, and impact on PDFs. Specifically,  we  compare
directly PDFs obtained in fits to all single-inclusive (\#janw) and
dijet (\#danw) datasets with the most accurate NNLO+EW theory and
default settings in Figs.~\ref{fig:jet_dijet_1}-\ref{fig:jet_dijet_2},
where the baseline fit (with no jet data) and, in the latter case, the
CT18 PDF fit~\cite{Hou:2019efy} are also shown for reference. Also, in
Fig.~\ref{fig:datatheory_jets}-\ref{fig:datatheory_dijets} we compare to a
representative set of
datapoints from each of the single-inclusive jet and dijet datasets
predictions obtained using PDFs from
the baseline fit, the fit with single-inclusive jets, and the fit with
dijets. Predictions are shown as a ratio to the experimental data,
which are shown either with full uncertainties, or with uncorrelated
uncertainties only, with the correlated uncertainties kept into
account as a shift of the datapoint (see e.g. Eqs.~(85-86) of Ref.~\cite{Gao:2017yyd}).

Based on the $\chi^2$ values from Tables~\ref{tab:chi2s}-\ref{tab:chi2sD} and
the PDF comparisons in 
Figs.~\ref{fig:jet_dijet_1}-\ref{fig:datatheory_dijets}, our conclusions
are the following.

\begin{enumerate}

\item Concerning the relative impact on PDFs of single-inclusive jets and
dijets:
 \begin{enumerate}
\item The effect on PDFs of the inclusion of jet and dijet data in the
  NNPDF3.1 global dataset is qualitatively the
  same. Namely, they only affect the gluon, by leading to an enhancement
of its central value in the region $0.1\lesssim x \lesssim 0.4$, accompanied by a 
suppression in the region $0.01\lesssim x \lesssim 0.1$. The
suppression is by about  1\%, while the enhancement at the peak,
localized at  $x\simeq 0.3$  is by about 2.5\%
for single-inclusive jets, but stronger, by about 7.5\% for dijets. An
enhanced gluon is also present in the CT18 PDF determination, which, as mentioned,
includes the 8~TeV CMS single-inclusive jet data, and whose gluon PDF
is consistent with our result within its rather larger uncertainty.

\item The inclusion of either single-inclusive or dijets leads to a
  reduction in the gluon uncertainty, with a somewhat stronger
  reduction observed for single-inclusive jets. It should be noted
  in this respect that for the most accurate 8~TeV dijet dataset, which as
  shown in Sects.~\ref{sec:impcondata}-\ref{sec:impcondatad} is mostly
  responsible for the shift in central value (though not on the
  uncertainty),
  only CMS data
  are currently available. The constraining power of the dijet dataset
  is consequently at present more limited than that of the
  single-inclusive jet dataset.

\item The inclusion of single-inclusive jet or dijet data does not lead to a 
deterioration in the description of the rest of the data in comparison to 
the baseline fit: almost all $\chi^2$ values for other datasets are
unchanged. This shows that the single-inclusive and dijet data are not
only consistent with each other, but also with the rest of the global
dataset, and their impact on the gluon central value, accompanied by a
reduction in uncertainty, corresponds to a genuine addition of new
information in the fit. Indeed, a comparative
assessment of the impact of jet, $Z$ $p_T$ and top production data on
the gluon distribution in Ref.~\cite{Nocera:2017zge} showed good
consistency, specifically wih the top data also leading to an
enhancement of the gluon in the $x\gtrsim 0.1$ region. 
An exception is  the ATLAS top rapidity distributions, which seem to
be in tension with the ATLAS 8~TeV single-inclusive jet data, as
discussed in Sect.~\ref{sec:impcondata}. The quality of the fit to
this data also deteriorates, though by a smaller amount, when dijet
data are fitted; note however that in this case the quality of the
fit to CMS top rapidity data improves. 
\end{enumerate}

\item Concerning relative fit quality:

\begin{enumerate}

\item The quality of the fit to single-inclusive jet data and dijet
  data when each of them is fitted is comparable, though somewhat
  better for dijets ($\chi^2=1.65$ vs. $\chi^2=1.88$). The quality of
  the fit to dijets when single inclusive jets are fitted and
  conversely are almost identical ($\chi^2=2.10$ for dijets when
  fitting single-inclusive jets vs. $\chi^2=2.06$ for single-inclusive
  jets when fitting dijets), and only marginally worse than the
  quality of the fit to each dataset when it is fitted. This confirms
  the full consistency of the two datasets, with a marginal preference
  for dijets.

\item The fit including dijet data is also  somewhat more
  internally consistent than the fit including single-inclusive jet
  data. Indeed, the $\chi^2$ per datapoint of the global fit is closer to one
  (1.22 vs 1.28), and also, the $\chi^2$ for individual datasets is generally  
  better. In particular, this happens for top production data,
  also sensitive to the large-$x$ gluon. It is unclear whether
  this is due to a greater theoretical accuracy of the NNLO dijet
  observable, or to better quality of the dijet data (specifically a
  better control of correlated systematics). However, the issue is
  phenomenologically immaterial, given that  the shape 
  and size of the data to theory ratio are qualitatively comparable for all of 
  the jet and dijet data (including for the rapidity bins not displayed in 
  Figs.~\ref{fig:datatheory_jets}-\ref{fig:datatheory_dijets}), regardless of
  which dataset is actually fitted.

\end{enumerate}

\item Concerning relative perturbative stability:

\begin{enumerate}

 \item When fitting  the dijet data, fit quality to the fitted data
   improves significantly from NLO to NNLO ($\chi^2=2.44$ at NLO vs.
   $1.65$ at NNLO), but the fit quality to the single-inclusive jet
   data actually deteriorates from NLO to NNLO (from $\chi^2=1.54$ to
   $2.06$). When fitting the
   single-inclusive jet data, the fit quality to the fitted data does
   not improve and actually deteriorates from NLO to NNLO (from $\chi^2=1.25$
   to $\chi^2=1.88$) but, perhaps surprisingly, the fit quality to the
   dijet data, not fitted, does improve  (from $\chi^2=3.29$ at NLO
   to the NNLO $\chi^2=2.10$). Whereas this shows a good theoretical
   consistency of the dijet data, it is unclear whether the lack of
   improvement of the single-inclusive jet data is due to a less
   stable perturbative behavior of the jet observable, or to issues
   with data.

  \item As already noted in Sect.~\ref{sec:hoimpd}, the fit quality to
    all other data included in the global datasets deteriorates at NLO
    when including jet data, with a greater deterioration seen in the
    case of dijets, and more moderate for single-inclusive jets: the
    total $\chi^2$ per datapoint for the global fit goes from
    $\chi^2=1.20$ of the baseline to $1.28$ in the former case and
    $1.33$ in the latter. At NNLO, when dijets are fitted the global
    fit quality significantly improves and becomes almost the same as that of the baseline
    ($\chi^2=1.22$, in comparison to $\chi^2=1.18$ of the baseline)
    while for the fit to single-inclusive jets it does not
    improve. The greater deterioration of fit quality at NLO for dijets
    can be understood as a consequence of the fact, observed in point
    1.a above, that dijets have a greater pull on the gluon: hence
    missing NNLO corrections lead to a stronger loss of accuracy. The
    lack of improvement in the description of single-inclusive jets
    shows again that this observable seems to be somewhat less
    well-behaved, either for theoretical or experimental reasons.
\end{enumerate}
\end{enumerate}

We generally conclude that single-inclusive jets and dijets are
mutually consistent and at NNLO consistent with the global dataset and
have a similar impact on the gluon. The dijet observable has a better
behaved perturbative behavior and a stronger pull on the gluon PDF
and it appears to be marginally preferable, though 
it leads to a less pronounced decrease of the gluon uncertainty,
possibly because ATLAS dijet measurements are not yet
available at 8~TeV, while single-inclusive jet measurements are available both
from ATLAS and CMS.

\section{Summary and outlook}
\label{sec:summary}

We have presented an extensive  phenomenological
investigation of  inclusive jet production at the LHC, 
 exploiting recent theory calculations, in particular of NNLO QCD
 corrections, and studying for the first time  in a systematic and
 comparative way
 the inclusive dijet observable, along with the single-inclusive jet
 observable which is routinely used for PDF determination.

 We have found 
 full consistency between the constraints imposed on
 parton distributions, specifically the gluon, by single-inclusive jets
 and dijets, thus conclusively establishing the viability of the
 dijet observable for precision QCD phenomenology and PDF
 determination, as originally suggested twenty-five years ago~\cite{Giele:1994xd}. 
  We have also validated the conclusions of previous theoretical
 studies~\cite{Ridder:2013mf,Currie:2013dwa,Currie:2016bfm,Currie:2017ctp,Currie:2018xkj,Cacciari:2019qjx}.
 Specifically we have shown that NNLO corrections are crucial in order to
 ensure compatibility of the jet observables with the rest of the
 global dataset, and also that while at NLO the choice of central
 scale makes a significant impact (with the scale choice $\widehat{H}_T$ for
 single-inclusive jets better behaved at NLO), at NNLO perturbative
 stability appears to be achieved.

In a comparative assessment of single-inclusive jets vs. dijets,  we have found that the dijet observable has a more marked impact on
the gluon central value. Also, phenomenologically  it displays  a
better-behaved perturbative 
behavior, with a clear improvement of data-theory agreement as the
perturbative order of the theory prediction increases. However, the
single-inclusive jet observable leads to a more significant reduction
of the gluon uncertainty. Either, or both, of the latter observations
could be due to theoretical reasons but also to the nature of the
current data. Specifically, there are indications that some of the
single-inclusive datasets  are in tension with the rest of the global
datasets, which could explain the less clear perturbative behavior of
fits including this observable. Also, the current dijet dataset is
more limited than the single-inclusive dataset, thus possibly
explaining the more limited effect of dijets on the gluon uncertainty.

One of our main results is that the  more recent 8~TeV data generally have a
rather more significant impact than previous 7~TeV data;
interestingly, the dijet 8~TeV CMS data are triple-differential, and
this may enhance their impact on PDF determination. We
accordingly expect that the availability of more precise data,
possibly also  for a greater variety of kinematic observables,
including more differential measurements,
from LHC Run-II 13~TeV data, and then from data coming from
future LHC Run-III and HL-LHC~\cite{Khalek:2018mdn} runs  will
settle these issues and lead to a clear and consistent theoretical
picture. Eventually, the availability of
correlations between  single-inclusive jet
and dijets data will allow for their simultaneous inclusion in a PDF
determination, thereby allowing for  maximal use of the available information.
Indeed, 
we expect this to be a first step towards a widening of the set of jet
observables used in precision PDF studies, which include not only
multi-differential jet cross-sections~\cite{Gehrmann-DeRidder:2019ibf},
but also jet substructure observables, whose study has undergone rapid
progress recently~\cite{Marzani:2019hun}. The inclusion of dijet data in
the forthcoming NNPDF4.0 global PDF analysis will be a first step in
this direction.

\subsection*{Acknowledgments}

S.~F. is supported by the
European Research Council under the European Union's 
Horizon 2020 research and innovation Programme (grant agreement ERC-AdG-740006).
T.~G. is supported by The Scottish Funding Council, grant H14027.
E.~R.~N. is supported by the European Commission through the Marie 
Sk\l odowska-Curie Action ParDHonS FFs.TMDs (grant number 752748).
J.~P. acknowledges the financial support by ERC2018-ADG-835105 YoctoLHC, Fundação
para a Ciência e Tecnologia (FCT, Portugal) through the contract UIDP/50007/2020
and by FCT under project CERN/FIS-PAR/0024/2019.
J.~R. is supported by an European Research Council Starting
Grant ``PDF4BSM'' and by the Netherlands Organization for
Scientific Research (NWO).
This work was supported in part by the Swiss National Science Foundation (SNF)
under contract 200020-175595.
.


\begin{thebibliography}{100}

\bibitem{Ridder:2013mf}
A.~Gehrmann-De~Ridder, T.~Gehrmann, E.~Glover, and J.~Pires, {\it {Second order
  QCD corrections to jet production at hadron colliders: the all-gluon
  contribution}},  {\em Phys.Rev.Lett.} {\bf 110} (2013) 162003,
  [\href{http://arxiv.org/abs/1301.7310}{{\tt arXiv:1301.7310}}].

\bibitem{Currie:2013dwa}
J.~Currie, A.~Gehrmann-De~Ridder, E.~Glover, and J.~Pires, {\it {NNLO QCD
  corrections to jet production at hadron colliders from gluon scattering}},
  {\em JHEP} {\bf 1401} (2014) 110, [\href{http://arxiv.org/abs/1310.3993}{{\tt
  arXiv:1310.3993}}].

\bibitem{Currie:2016bfm}
J.~Currie, E.~W.~N. Glover, and J.~Pires, {\it {NNLO QCD predictions for single
  jet inclusive production at the LHC}},  {\em Phys. Rev. Lett.} {\bf 118}
  (2017), no.~7 072002, [\href{http://arxiv.org/abs/1611.01460}{{\tt
  arXiv:1611.01460}}].

\bibitem{Czakon:2019tmo}
M.~Czakon, A.~van Hameren, A.~Mitov, and R.~Poncelet, {\it {Single-jet
  inclusive rates with exact color at $ \mathcal{O} $ ($ {\alpha}_s^4 $)}},
  {\em JHEP} {\bf 10} (2019) 262, [\href{http://arxiv.org/abs/1907.12911}{{\tt
  arXiv:1907.12911}}].

\bibitem{Gao:2017yyd}
J.~Gao, L.~Harland-Lang, and J.~Rojo, {\it {The Structure of the Proton in the
  LHC Precision Era}},  {\em Phys. Rept.} {\bf 742} (2018) 1--121,
  [\href{http://arxiv.org/abs/1709.04922}{{\tt arXiv:1709.04922}}].

\bibitem{Martin:1987vw}
A.~D. Martin, R.~G. Roberts, and W.~J. Stirling, {\it {Structure Function
  Analysis and psi, Jet, W, Z Production: Pinning Down the Gluon}},  {\em Phys.
  Rev.} {\bf D37} (1988) 1161.

\bibitem{Aversa:1988fv}
F.~Aversa, P.~Chiappetta, M.~Greco, and J.~P. Guillet, {\it {Higher Order
  Corrections to QCD Jets}},  {\em Phys. Lett.} {\bf B210} (1988) 225.

\bibitem{Ellis:1988hv}
S.~D. Ellis, Z.~Kunszt, and D.~E. Soper, {\it {The One Jet Inclusive
  Cross-section at Order $\alpha_s^3$: Gluons Only}},  {\em Phys. Rev. Lett.}
  {\bf 62} (1989) 726.

\bibitem{Giele:1994xd}
W.~T. Giele, E.~W.~N. Glover, and D.~A. Kosower, {\it {The inclusive two jet
  triply differential cross-section}},  {\em Phys. Rev.} {\bf D52} (1995)
  1486--1499, [\href{http://arxiv.org/abs/hep-ph/9412338}{{\tt
  hep-ph/9412338}}].

\bibitem{Currie:2017ctp}
J.~Currie, E.~W.~N. Glover, A.~Gehrmann-De~Ridder, T.~Gehrmann, A.~Huss, and
  J.~Pires, {\it {Single jet inclusive production for the individual jet
  $p_{T}$ scale choice at the LHC}},  in {\em {23rd Cracow Epiphany Conference
  on Particle Theory Meets the First Data from LHC Run 2 Cracow, Poland,
  January 9-12, 2017}}, 2017.
\newblock \href{http://arxiv.org/abs/1704.00923}{{\tt arXiv:1704.00923}}.

\bibitem{Currie:2018xkj}
J.~Currie, A.~Gehrmann-De~Ridder, T.~Gehrmann, E.~W.~N. Glover, A.~Huss, and
  J.~Pires, {\it {Infrared sensitivity of single jet inclusive production at
  hadron colliders}},  {\em JHEP} {\bf 10} (2018) 155,
  [\href{http://arxiv.org/abs/1807.03692}{{\tt arXiv:1807.03692}}].

\bibitem{Cacciari:2019qjx}
M.~Cacciari, S.~Forte, D.~Napoletano, G.~Soyez, and G.~Stagnitto, {\it
  {Single-jet inclusive cross section and its definition}},  {\em Phys. Rev.}
  {\bf D100} (2019), no.~11 114015,
  [\href{http://arxiv.org/abs/1906.11850}{{\tt arXiv:1906.11850}}].

\bibitem{Dasgupta:2016bnd}
M.~Dasgupta, F.~A. Dreyer, G.~P. Salam, and G.~Soyez, {\it {Inclusive jet
  spectrum for small-radius jets}},  {\em JHEP} {\bf 06} (2016) 057,
  [\href{http://arxiv.org/abs/1602.01110}{{\tt arXiv:1602.01110}}].

\bibitem{Aad:2014vwa}
{\bf ATLAS} Collaboration, G.~Aad et~al., {\it {Measurement of the inclusive
  jet cross-section in proton-proton collisions at $ \sqrt{s}=7$ TeV using 4.5
  fb$^{-1}$ of data with the ATLAS detector}},  {\em JHEP} {\bf 02} (2015) 153,
  [\href{http://arxiv.org/abs/1410.8857}{{\tt arXiv:1410.8857}}].

\bibitem{Aaboud:2017dvo}
{\bf ATLAS} Collaboration, M.~Aaboud et~al., {\it {Measurement of the inclusive
  jet cross-sections in proton-proton collisions at $ \sqrt{s}=8 $ TeV with the
  ATLAS detector}},  {\em JHEP} {\bf 09} (2017) 020,
  [\href{http://arxiv.org/abs/1706.03192}{{\tt arXiv:1706.03192}}].

\bibitem{Chatrchyan:2012bja}
{\bf CMS} Collaboration, S.~Chatrchyan et~al., {\it {Measurements of
  differential jet cross sections in proton-proton collisions at $\sqrt{s}=7$
  TeV with the CMS detector}},  {\em Phys.Rev.} {\bf D87} (2013) 112002,
  [\href{http://arxiv.org/abs/1212.6660}{{\tt arXiv:1212.6660}}].

\bibitem{Khachatryan:2016mlc}
{\bf CMS} Collaboration, V.~Khachatryan et~al., {\it {Measurement and QCD
  analysis of double-differential inclusive jet cross sections in pp collisions
  at $ \sqrt{s}=8 $ TeV and cross section ratios to 2.76 and 7 TeV}},  {\em
  JHEP} {\bf 03} (2017) 156, [\href{http://arxiv.org/abs/1609.05331}{{\tt
  arXiv:1609.05331}}].

\bibitem{Aad:2013tea}
{\bf ATLAS Collaboration} Collaboration, G.~Aad et~al., {\it {Measurement of
  dijet cross sections in $pp$ collisions at 7 TeV centre-of-mass energy using
  the ATLAS detector}},  {\em JHEP} {\bf 1405} (2014) 059,
  [\href{http://arxiv.org/abs/1312.3524}{{\tt arXiv:1312.3524}}].

\bibitem{Sirunyan:2017skj}
{\bf CMS} Collaboration, A.~M. Sirunyan et~al., {\it {Measurement of the
  triple-differential dijet cross section in proton-proton collisions at
  $\sqrt{s}=8\,\text {TeV} $ and constraints on parton distribution
  functions}},  {\em Eur. Phys. J.} {\bf C77} (2017), no.~11 746,
  [\href{http://arxiv.org/abs/1705.02628}{{\tt arXiv:1705.02628}}].

\bibitem{Ball:2017nwa}
{\bf NNPDF} Collaboration, R.~D. Ball et~al., {\it {Parton distributions from
  high-precision collider data}},  {\em Eur. Phys. J.} {\bf C77} (2017), no.~10
  663, [\href{http://arxiv.org/abs/1706.00428}{{\tt arXiv:1706.00428}}].

\bibitem{Harland-Lang:2014zoa}
L.~A. Harland-Lang, A.~D. Martin, P.~Motylinski, and R.~S. Thorne, {\it {Parton
  distributions in the LHC era: MMHT 2014 PDFs}},  {\em Eur. Phys. J.} {\bf
  C75} (2015) 204, [\href{http://arxiv.org/abs/1412.3989}{{\tt
  arXiv:1412.3989}}].

\bibitem{Hou:2019efy}
T.-J. Hou et~al., {\it {New CTEQ global analysis of quantum chromodynamics with
  high-precision data from the LHC}},
  \href{http://arxiv.org/abs/1912.10053}{{\tt arXiv:1912.10053}}.

\bibitem{Khachatryan:2014waa}
{\bf CMS} Collaboration, V.~Khachatryan et~al., {\it {Constraints on parton
  distribution functions and extraction of the strong coupling constant from
  the inclusive jet cross section in pp collisions at $\sqrt{s} = 7$ $\,\text
  {TeV}$}},  {\em Eur. Phys. J.} {\bf C75} (2015), no.~6 288,
  [\href{http://arxiv.org/abs/1410.6765}{{\tt arXiv:1410.6765}}].

\bibitem{Harland-Lang:2017ytb}
L.~A. Harland-Lang, A.~D. Martin, and R.~S. Thorne, {\it {The Impact of LHC Jet
  Data on the MMHT PDF Fit at NNLO}},  {\em Eur. Phys. J.} {\bf C78} (2018),
  no.~3 248, [\href{http://arxiv.org/abs/1711.05757}{{\tt arXiv:1711.05757}}].

\bibitem{Cacciari:2008gp}
M.~Cacciari, G.~P. Salam, and G.~Soyez, {\it {The Anti-k(t) jet clustering
  algorithm}},  {\em JHEP} {\bf 0804} (2008) 063,
  [\href{http://arxiv.org/abs/0802.1189}{{\tt arXiv:0802.1189}}].

\bibitem{Aaboud:2017wsi}
{\bf ATLAS} Collaboration, M.~Aaboud et~al., {\it {Measurement of inclusive jet
  and dijet cross-sections in proton-proton collisions at $\sqrt{s}=13$ TeV
  with the ATLAS detector}},  {\em JHEP} {\bf 05} (2018) 195,
  [\href{http://arxiv.org/abs/1711.02692}{{\tt arXiv:1711.02692}}].

\bibitem{Khachatryan:2016wdh}
{\bf CMS} Collaboration, V.~Khachatryan et~al., {\it {Measurement of the
  double-differential inclusive jet cross section in proton–proton collisions
  at $\sqrt{s} = 13\,\text {TeV} $}},  {\em Eur. Phys. J.} {\bf C76} (2016),
  no.~8 451, [\href{http://arxiv.org/abs/1605.04436}{{\tt arXiv:1605.04436}}].

\bibitem{Sirunyan:2018xdh}
{\bf CMS} Collaboration, A.~M. Sirunyan et~al., {\it {Measurements of the
  differential jet cross section as a function of the jet mass in dijet events
  from proton-proton collisions at $ \sqrt{s}=13 $ TeV}},  {\em JHEP} {\bf 11}
  (2018) 113, [\href{http://arxiv.org/abs/1807.05974}{{\tt arXiv:1807.05974}}].

\bibitem{Sirunyan:2020uoj}
{\bf CMS} Collaboration, A.~M. Sirunyan et~al., {\it {Dependence of inclusive
  jet production on the anti-$k_\mathrm{T}$ distance parameter in pp collisions
  at $\sqrt{s} =$ 13 TeV}},  \href{http://arxiv.org/abs/2005.05159}{{\tt
  arXiv:2005.05159}}.

\bibitem{Aad:2013lpa}
{\bf ATLAS} Collaboration, G.~Aad et~al., {\it {Measurement of the inclusive
  jet cross section in pp collisions at $\sqrt{s}$=2.76 TeV and comparison to
  the inclusive jet cross section at $\sqrt{s}$=7 TeV using the ATLAS
  detector}},  {\em Eur.Phys.J.} {\bf C73} (2013) 2509,
  [\href{http://arxiv.org/abs/1304.4739}{{\tt arXiv:1304.4739}}].

\bibitem{Khachatryan:2015luy}
{\bf CMS} Collaboration, V.~Khachatryan et~al., {\it {Measurement of the
  inclusive jet cross section in pp collisions at $\sqrt{s} = 2.76\,\text
  {TeV}$}},  {\em Eur. Phys. J.} {\bf C76} (2016), no.~5 265,
  [\href{http://arxiv.org/abs/1512.06212}{{\tt arXiv:1512.06212}}].

\bibitem{Sirunyan:2018qel}
{\bf CMS} Collaboration, A.~M. Sirunyan et~al., {\it {Constraining gluon
  distributions in nuclei using dijets in proton-proton and proton-lead
  collisions at $\sqrt{s_{_\mathrm{NN}}} =$ 5.02 TeV}},  {\em Phys. Rev. Lett.}
  {\bf 121} (2018), no.~6 062002, [\href{http://arxiv.org/abs/1805.04736}{{\tt
  arXiv:1805.04736}}].

\bibitem{Eskola:2019dui}
K.~J. Eskola, P.~Paakkinen, and H.~Paukkunen, {\it {Non-quadratic improved
  Hessian PDF reweighting and application to CMS dijet measurements at 5.02
  TeV}},  {\em Eur. Phys. J.} {\bf C79} (2019), no.~6 511,
  [\href{http://arxiv.org/abs/1903.09832}{{\tt arXiv:1903.09832}}].

\bibitem{AbdulKhalek:2019mzd}
{\bf NNPDF} Collaboration, R.~Abdul~Khalek, J.~J. Ethier, and J.~Rojo, {\it
  {Nuclear parton distributions from lepton-nucleus scattering and the impact
  of an electron-ion collider}},  {\em Eur. Phys. J.} {\bf C79} (2019), no.~6
  471, [\href{http://arxiv.org/abs/1904.00018}{{\tt arXiv:1904.00018}}].

\bibitem{Aad:2014rma}
{\bf ATLAS} Collaboration, G.~Aad et~al., {\it {Measurement of three-jet
  production cross-sections in $pp$ collisions at 7 TeV centre-of-mass energy
  using the ATLAS detector}},  {\em Eur. Phys. J.} {\bf C75} (2015), no.~5 228,
  [\href{http://arxiv.org/abs/1411.1855}{{\tt arXiv:1411.1855}}].

\bibitem{Aad:2015nda}
{\bf ATLAS} Collaboration, G.~Aad et~al., {\it {Measurement of four-jet
  differential cross sections in $\sqrt{s}=8$ TeV proton-proton collisions
  using the ATLAS detector}},  {\em JHEP} {\bf 12} (2015) 105,
  [\href{http://arxiv.org/abs/1509.07335}{{\tt arXiv:1509.07335}}].

\bibitem{CMS:2014mna}
{\bf CMS} Collaboration, V.~Khachatryan et~al., {\it {Measurement of the
  inclusive 3-jet production differential cross section in proton–proton
  collisions at 7 TeV and determination of the strong coupling constant in the
  TeV range}},  {\em Eur. Phys. J.} {\bf C75} (2015) 186,
  [\href{http://arxiv.org/abs/1412.1633}{{\tt arXiv:1412.1633}}].

\bibitem{Aad:2011fc}
{\bf ATLAS} Collaboration, G.~Aad et~al., {\it {Measurement of inclusive jet
  and dijet production in pp collisions at $\sqrt{s}$ = 7 TeV using the ATLAS
  detector}},  {\em Phys. Rev.} {\bf D86} (2012) 014022,
  [\href{http://arxiv.org/abs/1112.6297}{{\tt arXiv:1112.6297}}].

\bibitem{Abulencia:2007ez}
{\bf CDF - Run II} Collaboration, A.~Abulencia et~al., {\it {Measurement of the
  Inclusive Jet Cross Section using the $k_{\rm T}$ algorithm in
  $p\overline{p}$ Collisions at $\sqrt{s}$=1.96 TeV with the CDF II Detector}},
   {\em Phys. Rev.} {\bf D75} (2007) 092006,
  [\href{http://arxiv.org/abs/hep-ex/0701051}{{\tt hep-ex/0701051}}].

\bibitem{Ball:2018iqk}
{\bf NNPDF} Collaboration, R.~D. Ball, S.~Carrazza, L.~Del~Debbio, S.~Forte,
  Z.~Kassabov, J.~Rojo, E.~Slade, and M.~Ubiali, {\it {Precision determination
  of the strong coupling constant within a global PDF analysis}},  {\em Eur.
  Phys. J.} {\bf C78} (2018), no.~5 408,
  [\href{http://arxiv.org/abs/1802.03398}{{\tt arXiv:1802.03398}}].

\bibitem{Bertone:2017bme}
{\bf NNPDF} Collaboration, V.~Bertone, S.~Carrazza, N.~P. Hartland, and
  J.~Rojo, {\it {Illuminating the photon content of the proton within a global
  PDF analysis}},  {\em SciPost Phys.} {\bf 5} (2018), no.~1 008,
  [\href{http://arxiv.org/abs/1712.07053}{{\tt arXiv:1712.07053}}].

\bibitem{Ball:2017otu}
R.~D. Ball, V.~Bertone, M.~Bonvini, S.~Marzani, J.~Rojo, and L.~Rottoli, {\it
  {Parton distributions with small-x resummation: evidence for BFKL dynamics in
  HERA data}},  {\em Eur. Phys. J.} {\bf C78} (2018), no.~4 321,
  [\href{http://arxiv.org/abs/1710.05935}{{\tt arXiv:1710.05935}}].

\bibitem{AbdulKhalek:2019ihb}
{\bf NNPDF} Collaboration, R.~Abdul~Khalek et~al., {\it {Parton Distributions
  with Theory Uncertainties: General Formalism and First Phenomenological
  Studies}},  {\em Eur. Phys. J.} {\bf C79} (2019), no.~11 931,
  [\href{http://arxiv.org/abs/1906.10698}{{\tt arXiv:1906.10698}}].

\bibitem{AbdulKhalek:2019bux}
{\bf NNPDF} Collaboration, R.~Abdul~Khalek et~al., {\it {A first determination
  of parton distributions with theoretical uncertainties}},  {\em Eur. Phys.
  J.} {\bf C} (2019) 79:838, [\href{http://arxiv.org/abs/1905.04311}{{\tt
  arXiv:1905.04311}}].

\bibitem{Nocera:2017zge}
E.~R. Nocera and M.~Ubiali, {\it {Constraining the gluon PDF at large x with
  LHC data}},  {\em PoS} {\bf DIS2017} (2018) 008,
  [\href{http://arxiv.org/abs/1709.09690}{{\tt arXiv:1709.09690}}].

\bibitem{Gehrmann-DeRidder:2019ibf}
A.~Gehrmann-De~Ridder, T.~Gehrmann, E.~W.~N. Glover, A.~Huss, and J.~Pires,
  {\it {Triple Differential Dijet Cross Section at the LHC}},  {\em Phys. Rev.
  Lett.} {\bf 123} (2019), no.~10 102001,
  [\href{http://arxiv.org/abs/1905.09047}{{\tt arXiv:1905.09047}}].

\bibitem{Currie:2017eqf}
J.~Currie, A.~Gehrmann-De~Ridder, T.~Gehrmann, E.~W.~N. Glover, A.~Huss, and
  J.~Pires, {\it {Precise predictions for dijet production at the LHC}},  {\em
  Phys. Rev. Lett.} {\bf 119} (2017), no.~15 152001,
  [\href{http://arxiv.org/abs/1705.10271}{{\tt arXiv:1705.10271}}].

\bibitem{Currie:2018oxh}
J.~Currie, A.~Gehrmann-De~Ridder, T.~Gehrmann, N.~Glover, A.~Huss, and
  J.~Pires, {\it {Jet cross sections at the LHC with NNLOJET}},  {\em PoS} {\bf
  LL2018} (2018) 001, [\href{http://arxiv.org/abs/1807.06057}{{\tt
  arXiv:1807.06057}}].

\bibitem{Dittmaier:2012kx}
S.~Dittmaier, A.~Huss, and C.~Speckner, {\it {Weak radiative corrections to
  dijet production at hadron colliders}},  {\em JHEP} {\bf 1211} (2012) 095,
  [\href{http://arxiv.org/abs/1210.0438}{{\tt arXiv:1210.0438}}].

\bibitem{Nagy:2001fj}
Z.~Nagy, {\it {Three jet cross-sections in hadron hadron collisions at
  next-to-leading order}},  {\em Phys.Rev.Lett.} {\bf 88} (2002) 122003,
  [\href{http://arxiv.org/abs/hep-ph/0110315}{{\tt hep-ph/0110315}}].

\bibitem{Wobisch:2011ij}
{\bf fastNLO} Collaboration, M.~Wobisch, D.~Britzger, T.~Kluge, K.~Rabbertz,
  and F.~Stober, {\it {Theory-Data Comparisons for Jet Measurements in
  Hadron-Induced Processes}},  \href{http://arxiv.org/abs/1109.1310}{{\tt
  arXiv:1109.1310}}.

\bibitem{Bertone:2016lga}
V.~Bertone, S.~Carrazza, and N.~P. Hartland, {\it {APFELgrid: a high
  performance tool for parton density determinations}},  {\em Comput. Phys.
  Commun.} {\bf 212} (2017) 205--209,
  [\href{http://arxiv.org/abs/1605.02070}{{\tt arXiv:1605.02070}}].

\bibitem{Britzger:2019kkb}
D.~Britzger et~al., {\it {Calculations for deep inelastic scattering using fast
  interpolation grid techniques at NNLO in QCD and the extraction of $\alpha_s$
  from HERA data}},  {\em Eur. Phys. J.} {\bf C79} (2019), no.~10 845,
  [\href{http://arxiv.org/abs/1906.05303}{{\tt arXiv:1906.05303}}].

\bibitem{Ridder:2016rzm}
A.~Gehrmann-De~Ridder, T.~Gehrmann, N.~Glover, A.~Huss, and T.~A. Morgan, {\it
  {NNLO QCD corrections for $Z$ boson plus jet production}},  {\em PoS} {\bf
  RADCOR2015} (2016) 075, [\href{http://arxiv.org/abs/1601.04569}{{\tt
  arXiv:1601.04569}}].

\bibitem{Carrazza:2017bjw}
S.~Carrazza, {\it {Modeling NNLO jet corrections with neural networks}},  in
  {\em {23rd Cracow Epiphany Conference on Particle Theory Meets the First Data
  from LHC Run 2 Cracow, Poland, January 9-12, 2017}}, 2017.
\newblock \href{http://arxiv.org/abs/1704.00471}{{\tt arXiv:1704.00471}}.

\bibitem{Bellm:2019yyh}
J.~Bellm et~al., {\it {Jet cross sections at the LHC and the quest for higher
  precision}},  \href{http://arxiv.org/abs/1903.12563}{{\tt arXiv:1903.12563}}.

\bibitem{Ball:2008by}
{\bf The NNPDF} Collaboration, R.~D. Ball et~al., {\it {A determination of
  parton distributions with faithful uncertainty estimation}},  {\em Nucl.
  Phys.} {\bf B809} (2009) 1--63, [\href{http://arxiv.org/abs/0808.1231}{{\tt
  arXiv:0808.1231}}].

\bibitem{Alekhin:2011sk}
S.~Alekhin et~al., {\it {The PDF4LHC Working Group Interim Report}},
  \href{http://arxiv.org/abs/1101.0536}{{\tt arXiv:1101.0536}}.

\bibitem{Arneodo:1996kd}
{\bf New Muon} Collaboration, M.~Arneodo et~al., {\it {Accurate measurement of
  $F_2^d/F_2^p$ and $R_d-R_p$}},  {\em Nucl. Phys.} {\bf B487} (1997) 3--26,
  [\href{http://arxiv.org/abs/hep-ex/9611022}{{\tt hep-ex/9611022}}].

\bibitem{Arneodo:1996qe}
{\bf New Muon} Collaboration, M.~Arneodo et~al., {\it {Measurement of the
  proton and deuteron structure functions, $F_2^p$ and $F_2^d$, and of the
  ratio $\sigma_L/\sigma_T$}},  {\em Nucl. Phys.} {\bf B483} (1997) 3--43,
  [\href{http://arxiv.org/abs/hep-ph/9610231}{{\tt hep-ph/9610231}}].

\bibitem{Whitlow:1991uw}
L.~W. Whitlow, E.~M. Riordan, S.~Dasu, S.~Rock, and A.~Bodek, {\it {Precise
  measurements of the proton and deuteron structure functions from a global
  analysis of the SLAC deep inelastic electron scattering cross-sections}},
  {\em Phys. Lett.} {\bf B282} (1992) 475--482.

\bibitem{Benvenuti:1989rh}
{\bf BCDMS} Collaboration, A.~C. Benvenuti et~al., {\it {A High Statistics
  Measurement of the Proton Structure Functions $F_2(x, Q^2)$ and $R$ from Deep
  Inelastic Muon Scattering at High $Q^2$}},  {\em Phys. Lett.} {\bf B223}
  (1989) 485.

\bibitem{Onengut:2005kv}
{\bf CHORUS} Collaboration, G.~Onengut et~al., {\it {Measurement of nucleon
  structure functions in neutrino scattering}},  {\em Phys. Lett.} {\bf B632}
  (2006) 65--75.

\bibitem{Goncharov:2001qe}
{\bf NuTeV} Collaboration, M.~Goncharov et~al., {\it {Precise measurement of
  dimuon production cross-sections in $\nu_{\mu}$Fe and $\bar{\nu}_{\mu}$Fe
  deep inelastic scattering at the Tevatron}},  {\em Phys. Rev.} {\bf D64}
  (2001) 112006, [\href{http://arxiv.org/abs/hep-ex/0102049}{{\tt
  hep-ex/0102049}}].

\bibitem{Mason:2006qa}
D.~A. Mason, {\em {Measurement of the strange - antistrange asymmetry at NLO in
  QCD from NuTeV dimuon data}}.
\newblock PhD thesis, Oregon U., 2006.

\bibitem{Abramowicz:2015mha}
{\bf ZEUS, H1} Collaboration, H.~Abramowicz et~al., {\it {Combination of
  measurements of inclusive deep inelastic ${e^{\pm }p}$ scattering cross
  sections and QCD analysis of HERA data}},  {\em Eur. Phys. J.} {\bf C75}
  (2015), no.~12 580, [\href{http://arxiv.org/abs/1506.06042}{{\tt
  arXiv:1506.06042}}].

\bibitem{Abramowicz:1900rp}
{\bf H1 , ZEUS} Collaboration, H.~Abramowicz et~al., {\it {Combination and QCD
  Analysis of Charm Production Cross Section Measurements in Deep-Inelastic ep
  Scattering at HERA}},  {\em Eur.Phys.J.} {\bf C73} (2013) 2311,
  [\href{http://arxiv.org/abs/1211.1182}{{\tt arXiv:1211.1182}}].

\bibitem{Aaron:2009af}
{\bf H1} Collaboration, F.~D. Aaron et~al., {\it {Measurement of the Charm and
  Beauty Structure Functions using the H1 Vertex Detector at HERA}},  {\em Eur.
  Phys. J.} {\bf C65} (2010) 89--109,
  [\href{http://arxiv.org/abs/0907.2643}{{\tt arXiv:0907.2643}}].

\bibitem{Abramowicz:2014zub}
{\bf ZEUS} Collaboration, H.~Abramowicz et~al., {\it {Measurement of beauty and
  charm production in deep inelastic scattering at HERA and measurement of the
  beauty-quark mass}},  {\em JHEP} {\bf 09} (2014) 127,
  [\href{http://arxiv.org/abs/1405.6915}{{\tt arXiv:1405.6915}}].

\bibitem{Webb:2003ps}
{\bf NuSea} Collaboration, J.~C. Webb et~al., {\it {Absolute Drell-Yan dimuon
  cross sections in 800-GeV/c p p and p d collisions}},
  \href{http://arxiv.org/abs/hep-ex/0302019}{{\tt hep-ex/0302019}}.

\bibitem{Webb:2003bj}
J.~C. Webb, {\it {Measurement of continuum dimuon production in 800-GeV/c
  proton nucleon collisions}},  \href{http://arxiv.org/abs/hep-ex/0301031}{{\tt
  hep-ex/0301031}}.

\bibitem{Towell:2001nh}
{\bf FNAL E866/NuSea} Collaboration, R.~S. Towell et~al., {\it {Improved
  measurement of the anti-d/anti-u asymmetry in the nucleon sea}},  {\em Phys.
  Rev.} {\bf D64} (2001) 052002,
  [\href{http://arxiv.org/abs/hep-ex/0103030}{{\tt hep-ex/0103030}}].

\bibitem{Moreno:1990sf}
G.~Moreno et~al., {\it {Dimuon production in proton - copper collisions at
  $\sqrt{s}$ = 38.8-GeV}},  {\em Phys. Rev.} {\bf D43} (1991) 2815--2836.

\bibitem{Aaltonen:2010zza}
{\bf CDF} Collaboration, T.~A. Aaltonen et~al., {\it {Measurement of
  $d\sigma/dy$ of Drell-Yan $e^+e^-$ pairs in the $Z$ Mass Region from
  $p\bar{p}$ Collisions at $\sqrt{s}=1.96$ TeV}},  {\em Phys. Lett.} {\bf B692}
  (2010) 232--239, [\href{http://arxiv.org/abs/0908.3914}{{\tt
  arXiv:0908.3914}}].

\bibitem{Abazov:2007jy}
{\bf D0} Collaboration, V.~M. Abazov et~al., {\it {Measurement of the shape of
  the boson rapidity distribution for $p \bar{p} \to Z/\gamma^* \to e^{+}
  e^{-}$ + $X$ events produced at $\sqrt{s}$=1.96-TeV}},  {\em Phys. Rev.} {\bf
  D76} (2007) 012003, [\href{http://arxiv.org/abs/hep-ex/0702025}{{\tt
  hep-ex/0702025}}].

\bibitem{Abazov:2013rja}
{\bf D0} Collaboration, V.~M. Abazov et~al., {\it {Measurement of the muon
  charge asymmetry in $p\bar{p}$ $\to$ W+X $\to$ $\mu$$\nu$ + X events at
  $\sqrt{s}$=1.96 TeV}},  {\em Phys.Rev.} {\bf D88} (2013) 091102,
  [\href{http://arxiv.org/abs/1309.2591}{{\tt arXiv:1309.2591}}].

\bibitem{D0:2014kma}
{\bf D0} Collaboration, V.~M. Abazov et~al., {\it {Measurement of the electron
  charge asymmetry in $\boldsymbol{p\bar{p}\rightarrow W+X \rightarrow e\nu
  +X}$ decays in $\boldsymbol{p\bar{p}}$ collisions at
  $\boldsymbol{\sqrt{s}=1.96}$ TeV}},  {\em Phys. Rev.} {\bf D91} (2015), no.~3
  032007, [\href{http://arxiv.org/abs/1412.2862}{{\tt arXiv:1412.2862}}].
  [Erratum: Phys. Rev.D91,no.7,079901(2015)].

\bibitem{Aad:2013iua}
{\bf ATLAS} Collaboration, G.~Aad et~al., {\it {Measurement of the high-mass
  Drell--Yan differential cross-section in pp collisions at $\sqrt{s}$=7 TeV
  with the ATLAS detector}},  {\em Phys.Lett.} {\bf B725} (2013) 223,
  [\href{http://arxiv.org/abs/1305.4192}{{\tt arXiv:1305.4192}}].

\bibitem{Aad:2014qja}
{\bf ATLAS} Collaboration, G.~Aad et~al., {\it {Measurement of the low-mass
  Drell-Yan differential cross section at $\sqrt{s}$ = 7 TeV using the ATLAS
  detector}},  {\em JHEP} {\bf 06} (2014) 112,
  [\href{http://arxiv.org/abs/1404.1212}{{\tt arXiv:1404.1212}}].

\bibitem{Aad:2011dm}
{\bf ATLAS} Collaboration, G.~Aad et~al., {\it {Measurement of the inclusive
  $W^{\pm}$ and $Z/\gamma^*$ cross sections in the electron and muon decay
  channels in pp collisions at $\sqrt{s}$= 7 TeV with the ATLAS detector}},
  {\em Phys.Rev.} {\bf D85} (2012) 072004,
  [\href{http://arxiv.org/abs/1109.5141}{{\tt arXiv:1109.5141}}].

\bibitem{Aaboud:2016btc}
{\bf ATLAS} Collaboration, M.~Aaboud et~al., {\it {Precision measurement and
  interpretation of inclusive $W^+$, $W^-$ and $Z/\gamma^*$ production cross
  sections with the ATLAS detector}},
  \href{http://arxiv.org/abs/1612.03016}{{\tt arXiv:1612.03016}}.

\bibitem{Aad:2015auj}
{\bf ATLAS} Collaboration, G.~Aad et~al., {\it {Measurement of the transverse
  momentum and $\phi ^*_{\eta }$ distributions of Drell–Yan lepton pairs in
  proton–proton collisions at $\sqrt{s}=8$ TeV with the ATLAS detector}},
  {\em Eur. Phys. J.} {\bf C76} (2016), no.~5 291,
  [\href{http://arxiv.org/abs/1512.02192}{{\tt arXiv:1512.02192}}].

\bibitem{Aad:2014kva}
{\bf ATLAS} Collaboration, G.~Aad et~al., {\it {Measurement of the $t\bar{t}$
  production cross-section using $e\mu $ events with b-tagged jets in pp
  collisions at $\sqrt{s}$ = 7 and 8 $\,\mathrm{TeV}$ with the ATLAS
  detector}},  {\em Eur. Phys. J.} {\bf C74} (2014), no.~10 3109,
  [\href{http://arxiv.org/abs/1406.5375}{{\tt arXiv:1406.5375}}]. [Addendum:
  Eur. Phys. J.C76,no.11,642(2016)].

\bibitem{Aaboud:2016pbd}
{\bf ATLAS} Collaboration, M.~Aaboud et~al., {\it {Measurement of the
  $t\bar{t}$ production cross-section using $e\mu$ events with b-tagged jets in
  pp collisions at $\sqrt{s}$=13 TeV with the ATLAS detector}},  {\em Phys.
  Lett.} {\bf B761} (2016) 136--157,
  [\href{http://arxiv.org/abs/1606.02699}{{\tt arXiv:1606.02699}}].

\bibitem{Aad:2015mbv}
{\bf ATLAS} Collaboration, G.~Aad et~al., {\it {Measurements of top-quark pair
  differential cross-sections in the lepton+jets channel in $pp$ collisions at
  $\sqrt{s}=8$ TeV using the ATLAS detector}},  {\em Eur. Phys. J.} {\bf C76}
  (2016), no.~10 538, [\href{http://arxiv.org/abs/1511.04716}{{\tt
  arXiv:1511.04716}}].

\bibitem{Chatrchyan:2012xt}
{\bf CMS} Collaboration, S.~Chatrchyan et~al., {\it {Measurement of the
  electron charge asymmetry in inclusive W production in pp collisions at
  $\sqrt{s}$ = 7 TeV}},  {\em Phys.Rev.Lett.} {\bf 109} (2012) 111806,
  [\href{http://arxiv.org/abs/1206.2598}{{\tt arXiv:1206.2598}}].

\bibitem{Chatrchyan:2013mza}
{\bf CMS} Collaboration, S.~Chatrchyan et~al., {\it {Measurement of the muon
  charge asymmetry in inclusive pp to WX production at $\sqrt{s}$ = 7 TeV and
  an improved determination of light parton distribution functions}},  {\em
  Phys.Rev.} {\bf D90} (2014) 032004,
  [\href{http://arxiv.org/abs/1312.6283}{{\tt arXiv:1312.6283}}].

\bibitem{Chatrchyan:2013tia}
{\bf CMS} Collaboration, S.~Chatrchyan et~al., {\it {Measurement of the
  differential and double-differential Drell-Yan cross sections in
  proton-proton collisions at $\sqrt{s} =$ 7 TeV}},  {\em JHEP} {\bf 1312}
  (2013) 030, [\href{http://arxiv.org/abs/1310.7291}{{\tt arXiv:1310.7291}}].

\bibitem{Khachatryan:2016pev}
{\bf CMS} Collaboration, V.~Khachatryan et~al., {\it {Measurement of the
  differential cross section and charge asymmetry for inclusive $\mathrm
  {p}\mathrm {p}\rightarrow \mathrm {W}^{\pm }+X$ production at ${\sqrt{s}} =
  8$ TeV}},  {\em Eur. Phys. J.} {\bf C76} (2016), no.~8 469,
  [\href{http://arxiv.org/abs/1603.01803}{{\tt arXiv:1603.01803}}].

\bibitem{Khachatryan:2015oaa}
{\bf CMS} Collaboration, V.~Khachatryan et~al., {\it {Measurement of the Z
  boson differential cross section in transverse momentum and rapidity in
  proton–proton collisions at 8 TeV}},  {\em Phys. Lett.} {\bf B749} (2015)
  187--209, [\href{http://arxiv.org/abs/1504.03511}{{\tt arXiv:1504.03511}}].

\bibitem{Khachatryan:2016mqs}
{\bf CMS} Collaboration, V.~Khachatryan et~al., {\it {Measurement of the t-tbar
  production cross section in the e-mu channel in proton-proton collisions at
  sqrt(s) = 7 and 8 TeV}},  {\em JHEP} {\bf 08} (2016) 029,
  [\href{http://arxiv.org/abs/1603.02303}{{\tt arXiv:1603.02303}}].

\bibitem{Khachatryan:2015uqb}
{\bf CMS} Collaboration, V.~Khachatryan et~al., {\it {Measurement of the top
  quark pair production cross section in proton-proton collisions at $\sqrt(s)
  =$ 13 TeV}},  {\em Phys. Rev. Lett.} {\bf 116} (2016), no.~5 052002,
  [\href{http://arxiv.org/abs/1510.05302}{{\tt arXiv:1510.05302}}].

\bibitem{Khachatryan:2015oqa}
{\bf CMS} Collaboration, V.~Khachatryan et~al., {\it {Measurement of the
  differential cross section for top quark pair production in pp collisions at
  $\sqrt{s} = 8\,\text {TeV} $}},  {\em Eur. Phys. J.} {\bf C75} (2015), no.~11
  542, [\href{http://arxiv.org/abs/1505.04480}{{\tt arXiv:1505.04480}}].

\bibitem{Aaij:2012vn}
{\bf LHCb} Collaboration, R.~Aaij et~al., {\it {Inclusive $W$ and $Z$
  production in the forward region at $\sqrt{s} = 7$ TeV}},  {\em JHEP} {\bf
  1206} (2012) 058, [\href{http://arxiv.org/abs/1204.1620}{{\tt
  arXiv:1204.1620}}].

\bibitem{Aaij:2012mda}
{\bf LHCb} Collaboration, R.~Aaij et~al., {\it {Measurement of the
  cross-section for $Z \to e^+e^-$ production in $pp$ collisions at
  $\sqrt{s}=7$ TeV}},  {\em JHEP} {\bf 1302} (2013) 106,
  [\href{http://arxiv.org/abs/1212.4620}{{\tt arXiv:1212.4620}}].

\bibitem{Aaij:2015gna}
{\bf LHCb} Collaboration, R.~Aaij et~al., {\it {Measurement of the forward $Z$
  boson production cross-section in $pp$ collisions at $\sqrt{s}=7$ TeV}},
  {\em JHEP} {\bf 08} (2015) 039, [\href{http://arxiv.org/abs/1505.07024}{{\tt
  arXiv:1505.07024}}].

\bibitem{Aaij:2015zlq}
{\bf LHCb} Collaboration, R.~Aaij et~al., {\it {Measurement of forward W and Z
  boson production in $pp$ collisions at $ \sqrt{s}=8 $ TeV}},  {\em JHEP} {\bf
  01} (2016) 155, [\href{http://arxiv.org/abs/1511.08039}{{\tt
  arXiv:1511.08039}}].

\bibitem{Ball:2009qv}
{\bf The NNPDF} Collaboration, R.~D. Ball et~al., {\it {Fitting Parton
  Distribution Data with Multiplicative Normalization Uncertainties}},  {\em
  JHEP} {\bf 05} (2010) 075, [\href{http://arxiv.org/abs/0912.2276}{{\tt
  arXiv:0912.2276}}].

\bibitem{zahari_kassabov_2019_2571601}
Z.~Kassabov, ``{Reportengine: A framework for declarative data analysis}.''
  https://doi.org/10.5281/zenodo.2571601, Feb., 2019.

\bibitem{Ball:2012wy}
R.~D. Ball, S.~Carrazza, L.~Del~Debbio, S.~Forte, J.~Gao, et~al., {\it {Parton
  Distribution Benchmarking with LHC Data}},  {\em JHEP} {\bf 1304} (2013) 125,
  [\href{http://arxiv.org/abs/1211.5142}{{\tt arXiv:1211.5142}}].

\bibitem{Ball:2014uwa}
{\bf NNPDF} Collaboration, R.~D. Ball et~al., {\it {Parton distributions for
  the LHC Run II}},  {\em JHEP} {\bf 04} (2015) 040,
  [\href{http://arxiv.org/abs/1410.8849}{{\tt arXiv:1410.8849}}].

\bibitem{Khalek:2018mdn}
R.~Abdul~Khalek, S.~Bailey, J.~Gao, L.~Harland-Lang, and J.~Rojo, {\it {Towards
  Ultimate Parton Distributions at the High-Luminosity LHC}},  {\em Eur. Phys.
  J. C} {\bf 78} (2018), no.~11 962,
  [\href{http://arxiv.org/abs/1810.03639}{{\tt arXiv:1810.03639}}].

\bibitem{Marzani:2019hun}
S.~Marzani, G.~Soyez, and M.~Spannowsky, {\em {Looking inside jets: an
  introduction to jet substructure and boosted-object phenomenology}},
  vol.~958.
\newblock Springer, 2019.

\end{thebibliography}
\providecommand{\href}[2]{#2}\begingroup\raggedright\endgroup

\end{document}